\documentclass[submitting]{nst}

\usepackage{subfigure,dcolumn}
\usepackage[T2A,T1]{fontenc}
\usepackage[russian,english]{babel}

\usepackage{color}
\usepackage{soul}

\newcommand{\Jpsi}{J/$\psi$ }
\newcommand{\pT}{$p_{_T}$ }

\newcommand{\sNN}{$\sqrt{s_{_\mathrm{NN}}}$ }
\newcommand{\s}{$\sqrt{s}$ }
\newcommand{\pp}{$p$+$p$ }
\newcommand{\cucu}{Cu+Cu }
\newcommand{\auau}{Au+Au }
\newcommand{\raa}{$R_{\mathrm{AA}}$ }
\newcommand{\npart}{$N_{\textrm{part}}$ }
\newcommand{\cte}{\rm c\rightarrow e}
\newcommand{\bte}{\rm b\rightarrow e}

\usepackage{listings}
\begin{document}

\title{An experimental review of open heavy flavor and quarkonium production at RHIC}
\thanks{This work is supported in part by the National Key R\&D Program of China under Grand Nos. 2018YFE0104900 and 2018YFE0205200, the National Natural Science Foundation of China under Grand Nos. 11675168, 11890712 and 11720101001.}

\author{Zebo Tang}
\affiliation{State Key Laboratory of Particle Detection and Electronics, University of Science and Technology of China, Hefei, China}
\affiliation{Department of Modern Physics, University of Science and Technology of China, Hefei, China}
\author{Wangmei Zha}
\affiliation{State Key Laboratory of Particle Detection and Electronics, University of Science and Technology of China, Hefei, China}
\affiliation{Department of Modern Physics, University of Science and Technology of China, Hefei, China}
\author{Yifei Zhang}
\email[Corresponding author, ]{ephy@ustc.edu.cn}
\affiliation{State Key Laboratory of Particle Detection and Electronics, University of Science and Technology of China, Hefei, China}
\affiliation{Department of Modern Physics, University of Science and Technology of China, Hefei, China}

\begin{abstract}
Open heavy flavor and quarkonium are unique probes of the hot-dense medium produced in heavy-ion collisions. Their production in $p$+$p$ collisions also provide important test of QCD. In this paper, we review the selected results on open heavy flavor and quarkonium in $p$+$p$ and heavy-ion collisions achieved at RHIC. Physics implications are also discussed.
\end{abstract}


\maketitle

\section{Introduction}

In the ultra-relativistic heavy-ion collisions, the smash of two colliding nucleus creates an extremely hot and dense medium, in which the quarks and gluons are liberated from confinement inside hadrons and form a new state of matter called Quark-Gluon Plasma (QGP)~\cite{Gyulassy:2004vg,Bass:1998vz}. Since past twenty years, many experiment evidences from Relativistic Heavy Ion Collider (RHIC) and Large Hadron Collider (LHC), such as the jet quenching and strong particle flow of light flavor hadrons (consists of light quarks $u$, $d$, $s$), show that the QGP matter is strongly coupled and behaves like a liquid with small viscosity over entropy density~\cite{Adams:2005dq,Adcox:2004mh,Muller:2012zq,Chen:2018tnh}. However, most of the light flavor hadrons are produced late in the collisions with final state effects and the information of the QGP created in early stage of the collisions may be smeared. 

Heavy quark masses ($m_{c} \sim$ 1.3 GeV/$c^{2}$, $m_{b} \sim$ 4.8 GeV/$c^{2}$) are much larger than light quarks and QCD energy scale ($\Lambda_{\rm QCD}$). They are predominately produced via initial hard processes at early stage in the ultra-relativistic heavy-ion collisions and the probability of thermal production is negligible, especially at RHIC energies. Thus heavy quarks experience the whole evolution of the QCD matter created in heavy-ion collisions and are ideal probes to study the QGP matter properties. Most of the heavy quarks hadronize into open heavy flavor mesons (e.g. $D^0$, $D^{\pm}$, $D_{s}^{\pm}$, $B^0$, $B^{\pm}$, $B_{s}^{\pm}$) and baryons (e.g. $\Lambda_{c}$), while a small fraction $<$ 1\% of the total heavy quarks forms hidden heavy flavor, quarkonium states (e.g. J/$\psi$, $\Upsilon$) and their families.

Theoretic model predicts that heavy quarks loose less energy compared with light quarks due to the suppression of gluon radiation angle by their large masses~\cite{Dokshitzer:2001zm}. The measurement of open charm/bottom nuclear modification factor ($R_{AA}$), which is defined as the yield measured in Au+Au collisions divided by that in $p$+$p$ collisions and scaled by the average number of binary collisions ($N_{\rm coll}$), is commonly used to measure the medium effect characterizing as any deviation from unity. Strong suppression of open charm hadron $R_{AA}$ in central heavy-ion collisions at high transverse momentum ($p_T$) from recent measurements of STAR and ALICE experiments~\cite{Adamczyk:2014uip,ALICE:2012ab} indicates strong interactions between charm and hot-dense medium. The similar suppression level of charmed and light hadrons can be explained by model calculations including both elastic and inelastic energy loss~\cite{Djordjevic:2004nq,Buzzatti:2011vt}. In the mean time, the open bottom hadron measurements are very challenging due to poor production rate and small hadronic decay branching ratios. The effective way to measure bottom is via its decay products. 

On the other hand, the second order coefficient of Fourier expansion of particle azimuth distribution in the momentum space, $v_2$, is commonly used to measure medium bulk properties and how the medium transports the partons~\cite{Poskanzer:1998yz}. Heavy quarks are more difficult to participate in the partonic collectivity due to their large masses. Recent measurements of large elliptic flow of $D$ mesons indicate that charm quark has similar flow as light quarks and may reach thermalization~\cite{D0_v2_STAR,Abelev:2013lca,Abelev:2014ipa}.

By comparison of experimental data and theoretical model calculations, the transport diffusion coefficient of charm quark traversing in the medium can be obtained with large uncertainties~\cite{D0_v2_STAR}. To well understand the interactions between heavy quarks and medium, experiments keep devoting efforts in upgrading detectors to pursue high spatial resolution and fast response for precise measurement of heavy flavor hadron production in new era.

Quarkonium is a tightly bound state of a heavy quark and its anti-quark. Charmonium (bottomonium) refers the bound state of charm (bottom) quark and its anti-quark. Table~\ref{tab:quarkonium} shows the mass, binding energy and radius of various quarkonium states~\cite{quarkonium_size_Satz06}. If QGP is formed, the potential of heavy quark and its anti-quark is expected to be modified by the deconfined medium. The real part of the potential can get color-screened statically in the medium, resulting in a broadening of the wave function of the pair of heavy quark and its anti-quark. While the imaginary part of the potential is related to the dissociation of quarkonium arising from the scattering of quarkonium with the medium constituents such as gluons. The suppression of \Jpsi due to color screening was proposed as a signature of the QGP formation~\cite{colorscreen} and was considered as a strong experimental evidence of deconfinement in the medium produced in Pb+Pb collisions at SPS~\cite{SPS_deconfinement_Jpsi}.

The temperature required to dissociate a quarkonium state (dissociation temperature, $T_d$) depends on the binding energy of the quarkonium state. More loosely bounded state has lower $T_d$. In both charmonium and bottomonium sectors, $T_d$ decreases with increasing quarkonium mass and the excited states have lower $T_d$ than the 1S state. Based on the radius of the quarkonium as shown in Tab.~\ref{tab:quarkonium}, it is expected that $T_d^{\Upsilon(1S)} > T_d^{\chi_b} \sim T_d^{J/\psi} \sim T_d^{\Upsilon(2S)} > T_d^{\chi_b'} \sim T_d^{\chi_c} \sim T_d^{\Upsilon(3S)} > T_d^{\psi(2S)}$. The systematical measurements of quarkonium suppression can also help to constrain the temperature profile and the dynamic evolution of the fireball produced in ultra-relativistic heavy-ion collisions.

\begin{table*}[tb]\centering
\caption{The mass, binding energy and radius of charmonium and bottomonium states~\cite{quarkonium_size_Satz06}.}
\begin{tabular}{c|ccc|ccccc}
\hline
State & \multicolumn{3}{c|}{Charmonium} & \multicolumn{5}{c}{Bottomonium} \\
{} & \Jpsi & $\chi_c$ & $\psi(2S)$ 
& $\Upsilon(1S)$ & $\chi_b$ & $\Upsilon(2S)$ & $\chi_b'$  & $\Upsilon(3S)$  \\
\hline
Mass (GeV/$c^2$) & 3.10 & 3.53 & 3.68 & 9.46 & 9.99 & 10.02 & 10.26 & 10.36 \\
$\Delta E$ (GeV/$c^2$) & 0.64 & 0.20 & 0.05 & 1.10 & 0.67 & 0.54 & 0.31 & 0.20 \\
Radius (fm) & 0.25 & 0.36 & 0.45 & 0.14 & 0.22 & 0.28 & 0.34 & 0.39 \\
\hline
\end{tabular}
\label{tab:quarkonium}
\end{table*}

In contrast to the color-screening, quarknoium production yield could be enhanced due to (re)combination of (un)associated heavy quark and its anti-quark during QGP evolution and/or hadronization. The dissociation rate and/or the recombination probability depend on the properties of QGP, such as the temperature profile and the evolution of the fireball etc, as well as the size of the quarknium.  Although the (re)combination effect is competing with the QGP melting effect, both of them require deconfinement and can be used to search for QGP and study its properties. 

In additional to these two hot nuclear matter effects, quarkonium production in heavy-ion collisions is also affected by cold nuclear matter (CNM) effects, including modification of parton distribution function in nucleus (nPDF), breakup by hadrons, the scattering and/or energy loss of the parton evolved in quarkonium production etc. The CNM effects can be experimentally studied via the collisions of $p$ or light nucleus and heavy nucleus. There are other effects need to be taken into account when interpret the experimental results. One important effect is the feeddown contribution of the quarkonium production. 

Since the relative contributions of these effects have different dependences on various variables, such as initial energy density, system size, total heavy-quark cross-section, size and transverse momentum ($p_T$) of the quarkonium state etc, a comprehensive study of the quarkonium yield as a function of the collision energy, collision system, quarkonium $p_T$ and rapidity of different quarkonium states, as well as collectivity of heavy flavor hadron and quarkonium, is essential for a complete understanding of quarkonium production in heavy-ion collisions.

In the following sections, we imply the average of particle and anti-particle when using the term of particle unless otherwise specified.

\section{Open heavy flavor production}\label{sec.II}

\subsection{Open charm production}\label{sec.IIA}
\begin{figure*}[!htb]
\centering
\includegraphics[width=12 cm]{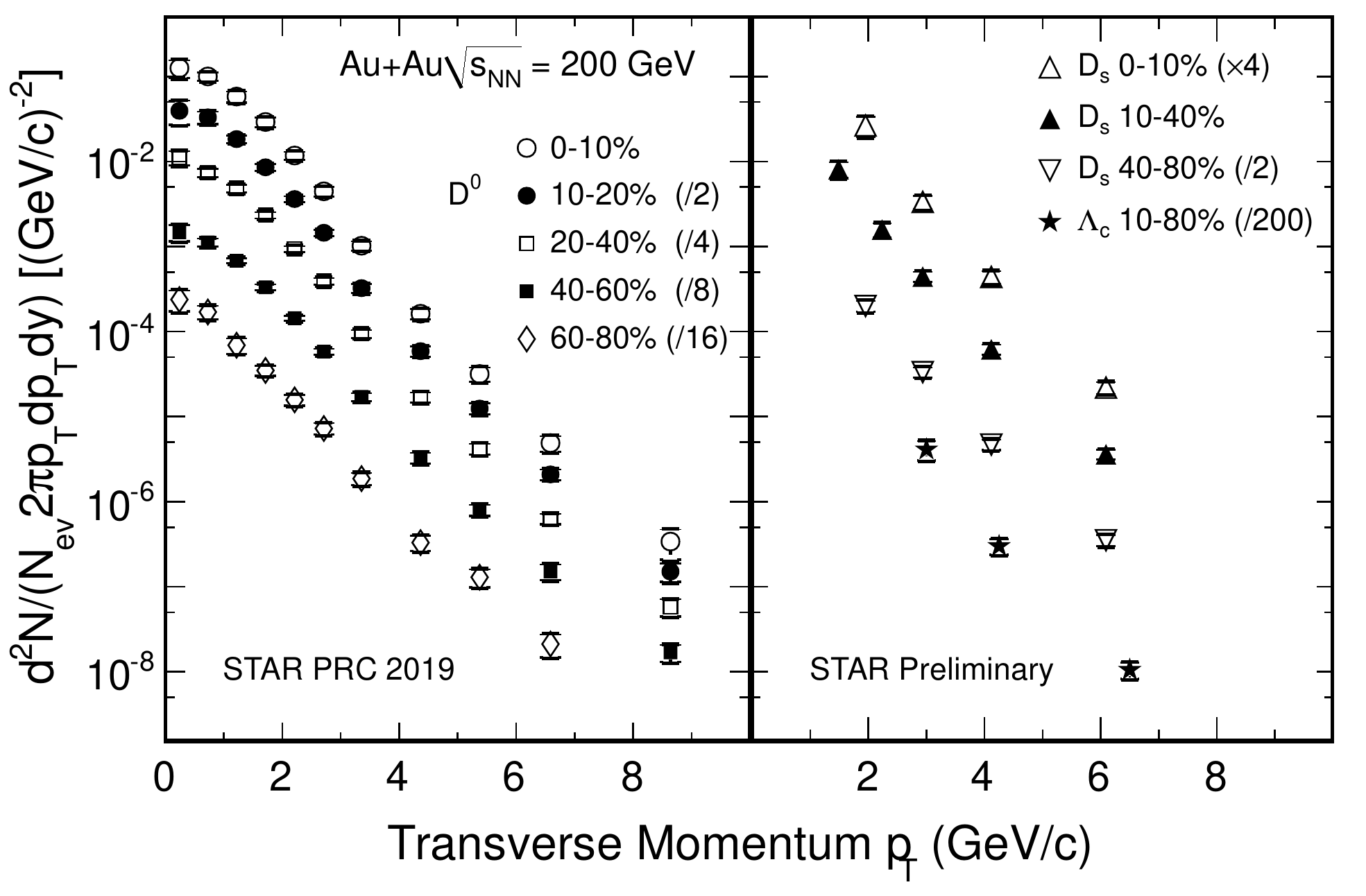}
\caption{Left Panel: The $D^0$ $p_{T}$  spectra at $|y|<1$ in 0--10\%, 10--20\%, 20--40\%, 40--60\% and 60--80\% Au+Au collisions~\cite{D0_Spectra_STAR_PRC2019}. Right Panel: The $D_s$ $p_{T}$  spectra (triangles) in 0--10\%, 10--40\%, 40--80\% and $\Lambda_c$ spectrum (stars) in 60--80\% Au+Au collisions at $|y|<1$~\cite{Zhou:2017ikn,Adam:2019hpq}. The statistic and systematic uncertainties are represented by vertical error bars and brackets, respectively.}
\label{chadronspectra}
\end{figure*}

The charm production cross sections in high energy $p$+$p$ collisions can be evaluated by perturbative Quantum Chromodynamics (QCD)~\cite{FONLL_online, Vogt:2007aw}. The differential transverse momentum ($p_{T}$ ) spectra of $D^0$ mesons in a wide energy range from $\sqrt{s} = $ 200 GeV upto 7 TeV in $p$+$p$ collisions measured by STAR~\cite{STARDpp, STARD500}, CDF~\cite{CDF03} and ALICE~\cite{ALICE12_2.76TeV, ALICE12_7TeV, ALICE17_7TeV} experiments respectively are in good agreement with the upper limit of Fixed-Order-Next-to-Leading-Logarithm (FONLL) calculations~\cite{FONLL1, FONLL2, FONLL_online, Luo:2020pef}. In heavy-ion collisions, charm quarks interact with the hot-dense medium and their transverse momenta are modified via energy loss, collective flow or cold nuclear matter (CNM) effects. The charmed hadrons are formed via charm hadronization from fragmentation, coalescence or recombination until chemical freeze-out. After kinetic freeze-out, the final state interactions stop,  the charmed hadrons spectra are fixed as what we measure. Figure~\ref{chadronspectra} shows the centrality dependence of charmed hadron $p_{T}$  spectra measured by STAR with the help of identifying secondary decay vertices of charmed hadrons utilizing recent developed state-of-the-art silicon pixel detector, Heavy Flavor Tracker (HFT)~\cite{Qiu:2014dha,Zhang:2014vea}. The $D^0$ $p_{T}$  spectra at mid-rapidity ($|y|<1$) in 0--10\%, 10--20\%, 20--40\%, 40--60\% and 60--80\% Au+Au collisions~\cite{D0_Spectra_STAR_PRC2019} are shown in the left panel. The $D_s$ $p_{T}$  spectra (triangles) in 0--10\%, 10--40\%, 40--80\% and $\Lambda_c$ spectrum (stars) in 60--80\% Au+Au collisions at $|y|<1$~\cite{Zhou:2017ikn,Adam:2019hpq} are shown in the right panel.  The spectra in some centrality bins are scaled with arbitrary factors indicated on the figure for clarity. 

\begin{figure}[!htb]
\centering
\includegraphics[width=8 cm]{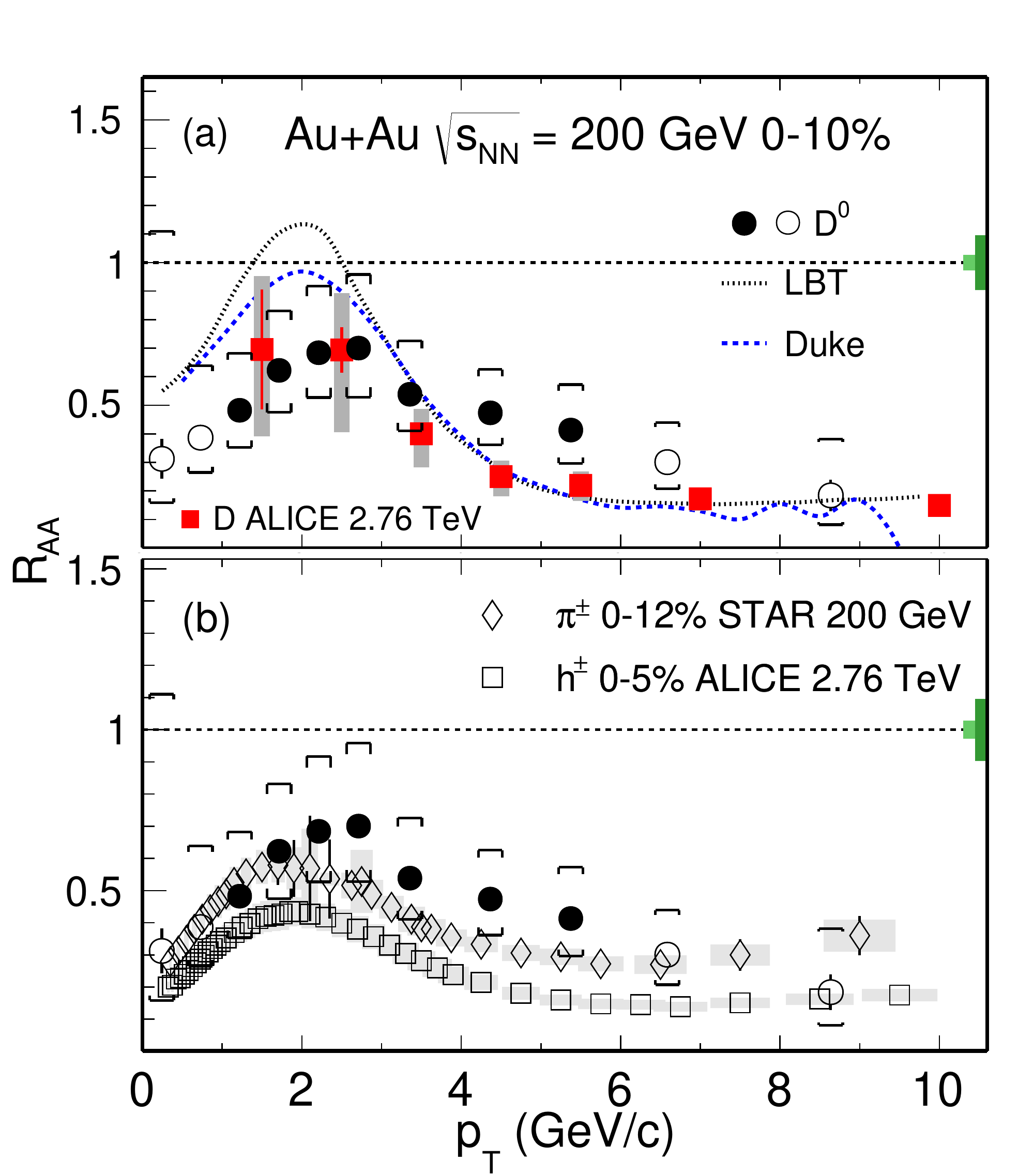}
\caption{(color online) $D^{0}$ $R_{\rm AA}$ in 0--10\% Au+Au collisions at ${\sqrt{s_{\rm NN}} = \rm{200\,GeV}}$ compared to the ALICE $D$-meson result in 0--10\% Pb + Pb collisions at $\sqrt{s_{_{\rm NN}}}$ = 2.76\,TeV (a) and charged hadrons from ALICE and $\pi^{\pm}$ from STAR (b). Also shown in panel (a) are the model calculations from the LBT and Duke groups~\cite{Cao:2016gvr,Xu:2017obm}. Notations for statistical and systematic uncertainties are the same as in previous figures.}
\label{D0RAA}
\end{figure}

The nuclear modification factor $R_{\rm AA}$ is calculated as the ratio of $N_{\rm bin}$--normalized yields between Au+Au and $p$+$p$ collisions. The $R_{\rm AA}$ of $D^0$ mesons in 0-10\% central Au+Au collisions at ${\sqrt{s_{\rm NN}} = \rm{200\,GeV}}$~\cite{D0_Spectra_STAR_PRC2019} is compared to that of (a) average $D$-meson from ALICE~\cite{Adam:2015sza} and (b) charged hadrons from ALICE and $\pi^{\pm}$ from STAR~\cite{Abelev:2012hxa,Abelev:2007ra}, shown in Fig.~\ref{D0RAA}. The $D^0$ $R_{\rm AA}$ from this measurement is comparable to that from the LHC measurements in Pb+Pb collisions at $\sqrt{s_{_{\rm NN}}}$ = 2.76\,TeV despite the large energy difference between these measurements. A significant suppression is seen at $p_{T}>$ 5\,GeV/$c$. The suppression level is similar to that of light flavor mesons, indicating strong interactions of charm with medium and energy loss. At $p_{T}$\,$<$\,5\,GeV/$c$, the $D^0$ $R_{\rm AA}$ shows a characteristic bump structure. The Duke model and the Linearized Boltzmann Transport (LBT) model~\cite{Cao:2016gvr,Xu:2017obm} calculations that predict sizable collective motion for charm quarks during the medium evolution can qualitatively describe the STAR data. The uncertainties from the $p$+$p$ reference~\cite{STARDpp} dominates the systematic uncertainty for STAR $R_{\rm AA}$.

\begin{figure*}[!htb]
\centering
\includegraphics[width=7 cm]{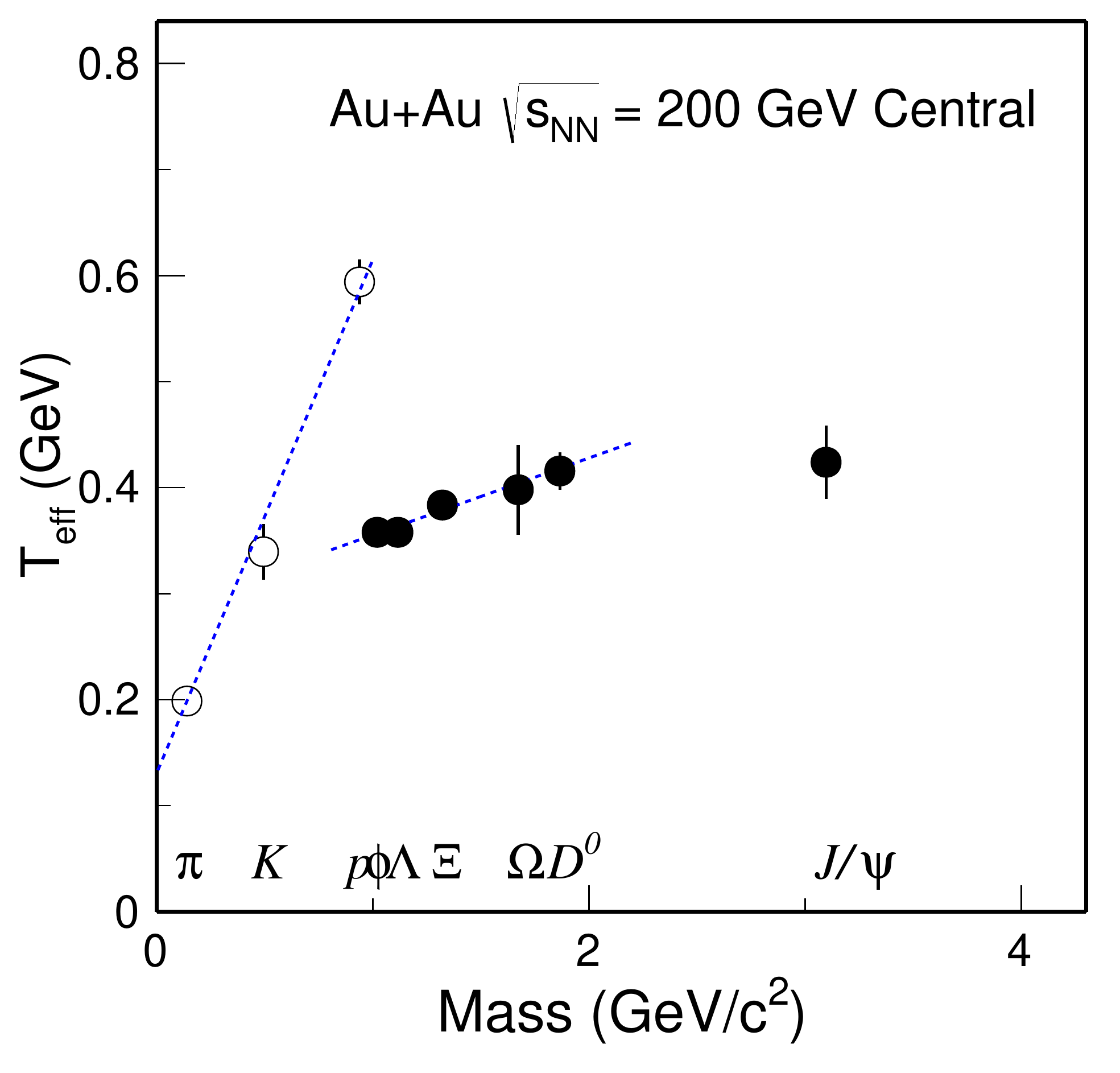}
\includegraphics[width=7 cm]{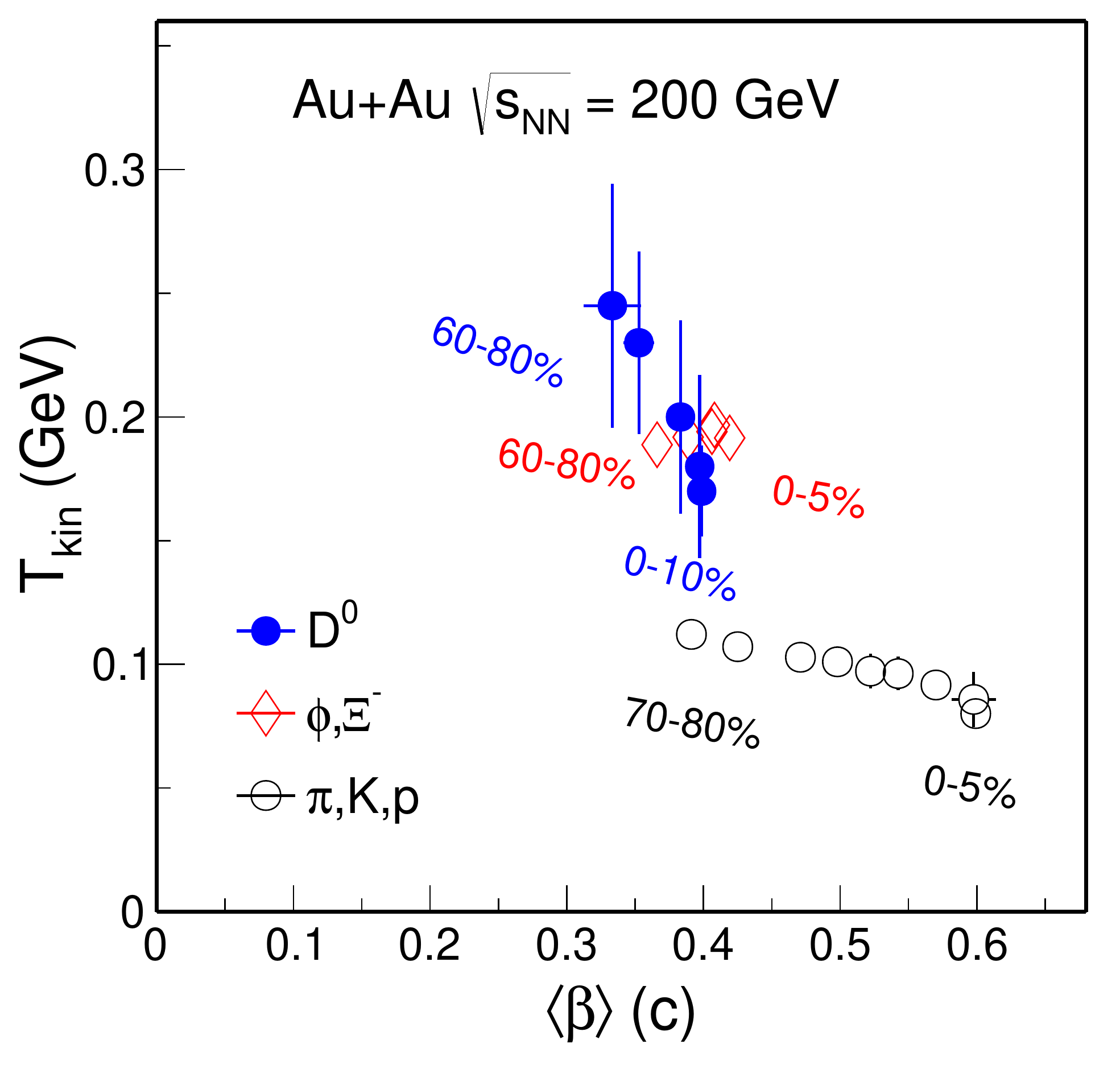}
\caption{(color online) Left Panel: Slope parameter $T_{\rm eff}$ for different particles in 0-10\% central Au+Au collisions~\cite{Adams:2003xp,Abelev:2007rw,Adams:2006ke,Adamczyk:2013tvk}. The dashed lines depict linear function fits to $\pi,K,p$ and $\phi,\Lambda,\Xi^{-},\Omega^{-},D^0$ respectively. Right Panel: Results of $T_{\rm kin}$ vs. $\langle\beta\rangle$ from the Blast-Wave model fits to different groups of particles. The data points for each group of particles present the results from different centrality bins. The one in most central collisions is at the largest $\langle\beta\rangle$.}
\label{thermalpar}
\end{figure*}

Several functions, such as Levy, power-law, $m_T$ exponential, Blast-wave~\cite{Schnedermann:1993ws}, Tsallis Blast-wave~\cite{Tsallis09}, are used to fit the $D^0$ data in above centrality bins to extract the collectivity and thermal properties. In this paper, only the physics results and conclusions are discussed. The analysis details can be found in Ref.~\cite{D0_Spectra_STAR_PRC2019}. The obtained slope parameter $T_{\rm eff}$ for $D^0$ mesons is compared to other light and strange hadrons measured at RHIC. Left panel of Fig.~\ref{thermalpar} summarizes the slope parameter $T_{\rm eff}$ for various identified hadrons ($\pi^{\pm}$, $K^{\pm}$, $p$/$\bar{p}$, $\phi$, $\Lambda$, $\Xi^-$, $\Omega$, $D^0$ and J/$\psi$) in central Au+Au collisions at $\sqrt{s_{_{\rm NN}}} {\rm = 200\,GeV}$~\cite{Adams:2003xp,Abelev:2007rw,Adams:2006ke,Adamczyk:2013tvk}. Point-by-point statistical and systematic uncertainties are added as a quadratic sum when performing these fits. All fits are performed up to $m_{T} - m_{0} <1\,\rm{GeV}/c^2$ for $\pi,\ K,\ p$, $<2$\,GeV/$c^2$ for $\phi,\ \Lambda,\ \Xi$, and $<3$\,GeV/$c^2$ for $\Omega,\ D^{0},\ J/\psi$, respectively. 

The slope parameter $T_{\rm eff}$ in a thermalized medium can be characterized by the random (generally interpreted as a kinetic freeze-out temperature $T_{\rm fo}$) and collective (radial flow velocity $\langle\beta_{T}\rangle$) components with a simple relation~\cite{Adams:2005dq,Csorgo:1995bi,Kolb:2003dz}:
\begin{equation}
  \begin{aligned}
T_{\rm eff} = T_{\rm fo} + m_0 \langle\beta_{T}\rangle^2.
  \end{aligned}
\label{equ:equation7}
\end{equation}
Therefore, $T_{\rm eff}$ will show a linear dependence as a function of particle mass $m_0$ with a slope that can be used to characterize the radial flow collective velocity.

The data points of $\phi,\ \Lambda,\ \Xi^{-},\ \Omega^{-},\ D^0$ follow a linear dependence with different slope compared to that of $\pi,\ K,\ p$, as represented by the dashed lines shown in left panel of Fig.~\ref{thermalpar}. Light flavor hadrons, such as, $\pi,\ K,\ p$ gain radial collectivity through the whole system evolution, therefore the linear dependence exhibits a larger slope. On the other hand hadrons contains strangeness or heavy quarks, such as, $\phi,\ \Lambda,\ \Xi^{-},\ \Omega^{-},\ D^0$ may freeze out from the system earlier, and therefore receive less radial collectivity, resulting in a smaller slope of the linear dependence of $T_{\rm eff}$ versus mass.

Figure~\ref{thermalpar} right panel summarizes the fit parameters $T_{\rm kin}$ vs. $\langle\beta\rangle$ from the Blast-Wave model fits to different groups of particles: black markers for the simultaneous fit to $\pi,\ K,\ p$; red markers for the simultaneous fit to $\phi,\ \Xi^-$ and blue markers for the fit to $D^0$.\,\,\,The data points for each group of particles represent the fit results from different centrality bins with the most central data point at the largest $\langle\beta\rangle$ value.\,\,\,Similar as in the fit to the $m_{T}$ spectra, point-by-point statistical and systematic uncertainties are added in quadrature when performing the fit. The fit results for $\pi,\ K,\ p$ are consistent with previously published results~\cite{Tsallis09}. The fit results for multi-strangeness particles $\phi,\ \Xi^{-}$, and for $D^0$ show much smaller mean transverse velocity $\langle\beta\rangle$ and larger kinetic freeze-out temperature, suggesting these particles decouple from the system earlier and gain less radial collectivity compared to light hadrons. The resulting $T_{\rm kin}$ parameters for $\phi,\ \Xi^-$ and for $D^0$ are close to the pseudocritical temperature $T_{c}$ calculated from a lattice QCD calculation at zero baryon chemical potential~\cite{Bazavov:2011nk}, indicating negligible contribution from the hadronic stage to the observed radial flow of these particles.\,\,Therefore, the collectivity that $D^0$ mesons obtain is mostly through the partonic stage re-scatterings in the QGP phase. 

\begin{figure}[!htb]
\centering
\includegraphics[width=8 cm]{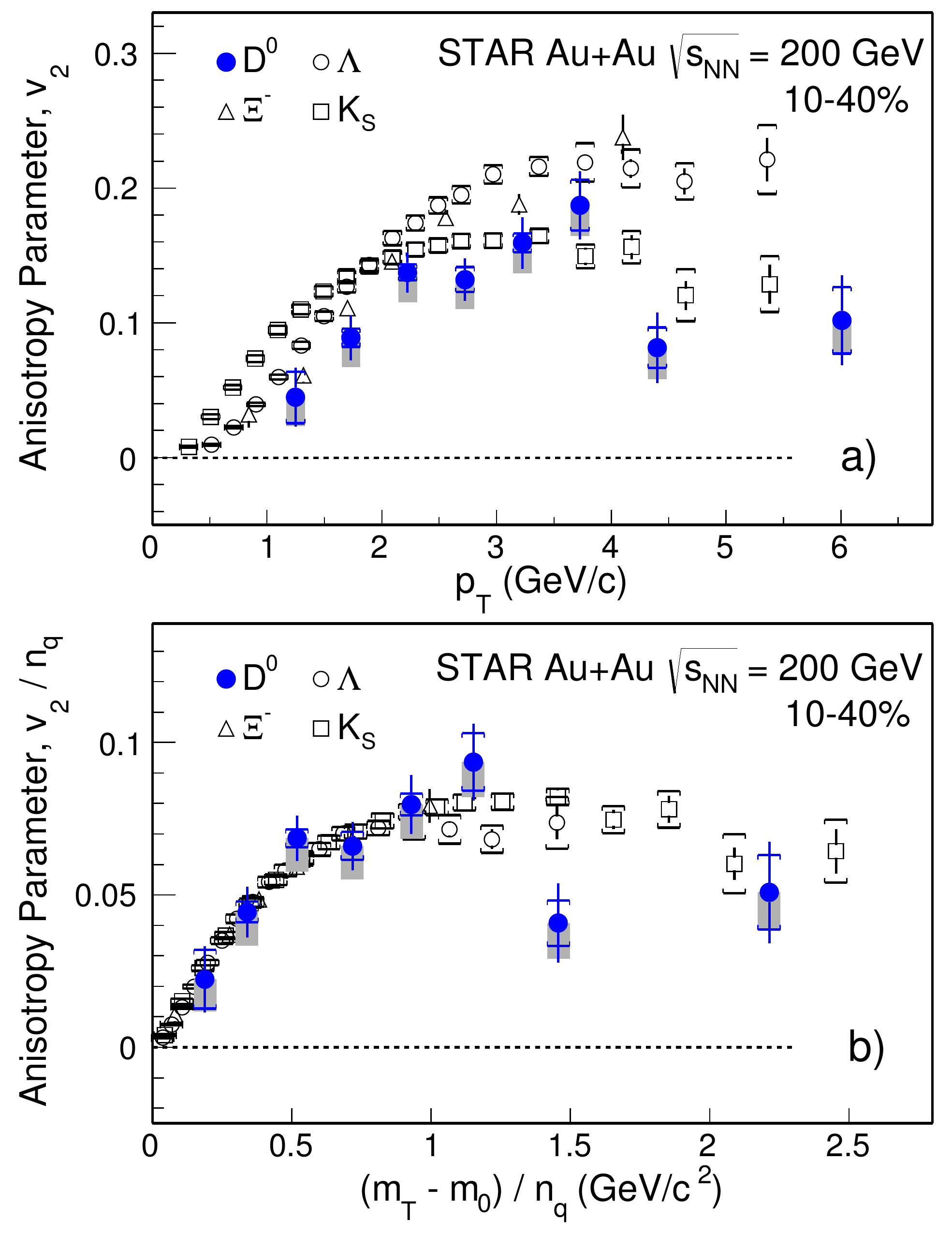}
\caption{(color online) (a) $v_2$ as a function of $p_{T}$  and (b) $v_2/n_q$ as a function of $(m_{\rm T}-m_0)/n_q$ for $D^0$ in 10--40$\%$ centrality Au+Au collisions compared with $K^0_S$, $\Lambda$, and $\Xi^-$~\cite{Abelev:2008ae}. 
The vertical bars and brackets represent statistical and systematic uncertainties, and the grey bands represent the estimated non-flow contribution.}
\label{fig:v2CompareWithData}
\end{figure}

Other observable to measure bulk collectivity is the elliptic flow characterized by the second order coefficient of particle azimuth distribution in the momentum space, $v_2$. The elliptic flow measurement of multi-strange hadrons and $\phi$ mesons indicates the partonic collectivity has been built up at the top energy heavy ion collisions of RHIC~\cite{Adamczyk:2015ukd}. Recently, with the help of the silicon vertex detector HFT, STAR experiment measured the $D^0$ $v_2$~\cite{D0_v2_STAR} in Au+Au collisions at $\sqrt{s_{_{\rm NN}}}$ 200 GeV. Figure~\ref{fig:v2CompareWithData} compares the measured $D^0$ $v_2$ from the event plane method in 10--40$\%$ centrality bin with $v_2$ of $K_{s}^0$, $\Lambda$, and $\Xi^-$~\cite{Abelev:2008ae}. The comparison between $D^0$ and light hadrons needs to be done in a narrow centrality bin 
to avoid the bias caused by the fact that the $D^0$ yield scales with number of binary collisions while the yield of light hadrons scales approximately with number of the participants~\cite{Esha:2016svw}.
Panel (a) shows $v_2$ as a function of $p_{T}$  where a clear mass ordering for $p_{T}$ \,$<$\,2\,GeV/$c$ including $D^0$ mesons is observed. For $p_{T}$ \,$>$\,2\,GeV/$c$, the $D^0$ meson $v_2$ follows that of other light mesons indicating significant charm quark flow at RHIC~\cite{Molnar:2003ff,Abelev:2008ae,Adamczyk:2015ukd}. 
Recent ALICE measurements show that the $D^0$ $v_2$ is comparable to that of charged hadrons in 0-50\% Pb+Pb collisions at $\sqrt{s_{_{\rm NN}}}$ = 2.76\,TeV~\cite{Abelev:2013lca,Abelev:2014ipa} suggesting sizable charm flow at the LHC.
Panel (b) shows $v_2/n_q$ as a function of scaled transverse kinetic energy, $(m_{\rm T}-m_0)/n_q$, where $n_q$ is the number of constituent quarks in the hadron, $m_0$ is the rest mass, and $m_{\rm T}=\sqrt{p_{\rm T}^2+m_0^2}$. We find that the $D^0$ $v_2$ falls into the same universal trend as all other light hadrons~\cite{Molnar:2003ff,Abelev:2008ae}, in particular for $(m_{\rm T}-m_0)/n_q$\,$<$\,1\,GeV/$c^2$. This suggests that charm quarks have gained significant flow through interactions with the sQGP medium in 10--40\% Au+Au collisions at $\sqrt{s_{_{\rm NN}}}$ = 200\,GeV.

\begin{figure*}[!htb]
\centering
\includegraphics[width=13 cm]{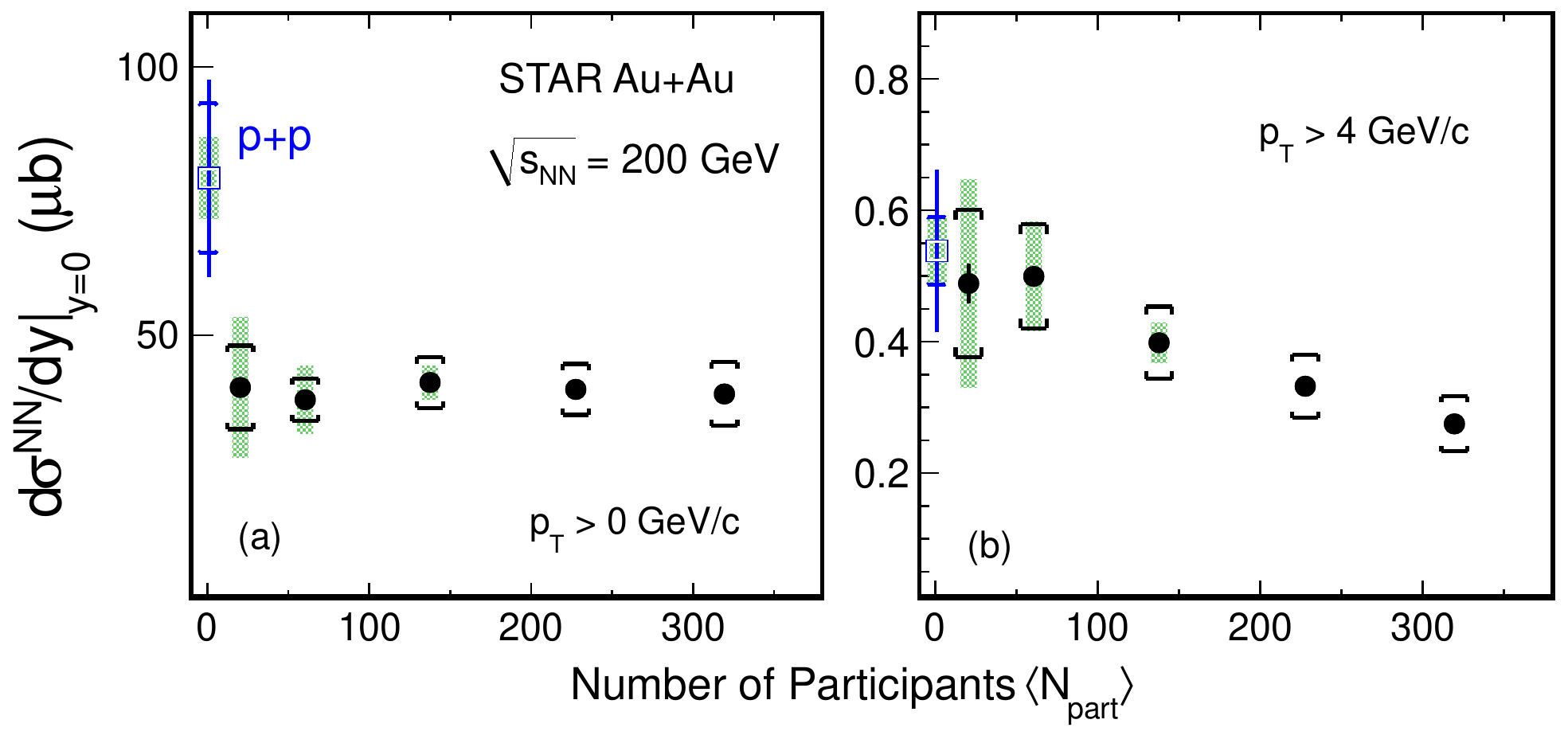}
\caption{(color online) Integrated $D^0$ cross section per nucleon-nucleon collision at mid-rapidity in $\sqrt{s_{\rm NN}} = $ 200 GeV Au+Au collisions for $p_{T}$  > 0 (a) and $p_{T}$  > 4 GeV/c (b) as a function of centrality Npart. The statistical and systematic uncertainties are shown as error bars and brackets on the data points. The green boxes on the data points depict the overall normalization uncertainties in $p$+$p$ and Au+Au data respectively. }
\label{cxsec}
\end{figure*}  

In heavy-ion collisions, charm quark interacts with the QGP matter when traversing in the medium. The transverse momentum of charm quark is modified by the medium via energy loss or collective flow. However, the total number of charm quarks may keep conserved since they are produced in initial hard processes before the QGP formation and there is no more charm quark created later via thermal production at RHIC energies. Figure~\ref{cxsec} (a) and (b) shows the $p_{T}$--integrated cross section for $D^0$ production per nucleon-nucleon collision $d\sigma^{\rm NN}/dy|_{y=0}$ from different centrality bins in $\sqrt{s_{_{\rm NN}}} {\rm = 200\,GeV}$ 200 GeV Au+Au collisions for the full $p_{T}$ range and for $p_{T}$\,$>$\,4\,GeV/$c$, respectively~\cite{D0_Spectra_STAR_PRC2019}. The result from the $p$+$p$ measurement at the same collision energy is also shown in both panels~\cite{STARDpp}.

The high $p_{T}$  ($>$\,4\,GeV/$c$) $d\sigma^{\rm NN}/dy|_{y=0}$ shows a clear decreasing trend from peripheral to mid-central and central collisions and the result in peripheral collisions is consistent with $p$+$p$ collisions within uncertainties. This is consistent with charm loses more energy in more central collisions at high $p_{T}$ . However, the $d\sigma^{\rm NN}/dy|_{y=0}$ integrated over full $p_{T}$  range shows approximately a flat distribution as a function of $N_{\rm part}$. The values for the full $p_{T}$  range in mid-central to central Au+Au collisions are smaller than that in $p$+$p$ collisions with $\sim1.5\sigma$ effect considering the large uncertainties from the $p$+$p$ measurements. The total charm quark yield in heavy-ion collisions is expected to follow the number-of-binary-collision scaling since charm quarks are conserved at RHIC energies. However, the cold nuclear matter (CNM) effect including shadowing could also play an important role. In addition, hadronization through coalescence could alter the hadrochemistry distributions of charm quark in various charm hadron states which may lead to the reduction in the observed $D^0$ yields in Au+Au collisions~\cite{GRECO2004202}. For instance, hadronization through coalescence can lead to an enhancement of the charmed baryon $\Lambda_{c}^+$ yield over $D^0$ yield~\cite{Oh2009,Zhao:2018jlw,Plumari:2017ntm}, and together with the strangeness enhancement in the hot QCD medium and sequential hadronization, can also lead to an enhancement in the charmed strange meson $D_{s}^+$ yield relative to $D^0$~\cite{He2013,He2019,Zhao:2018jlw,Plumari:2017ntm}. 

\begin{figure*}[!htb]
\centering
\includegraphics[width=7 cm]{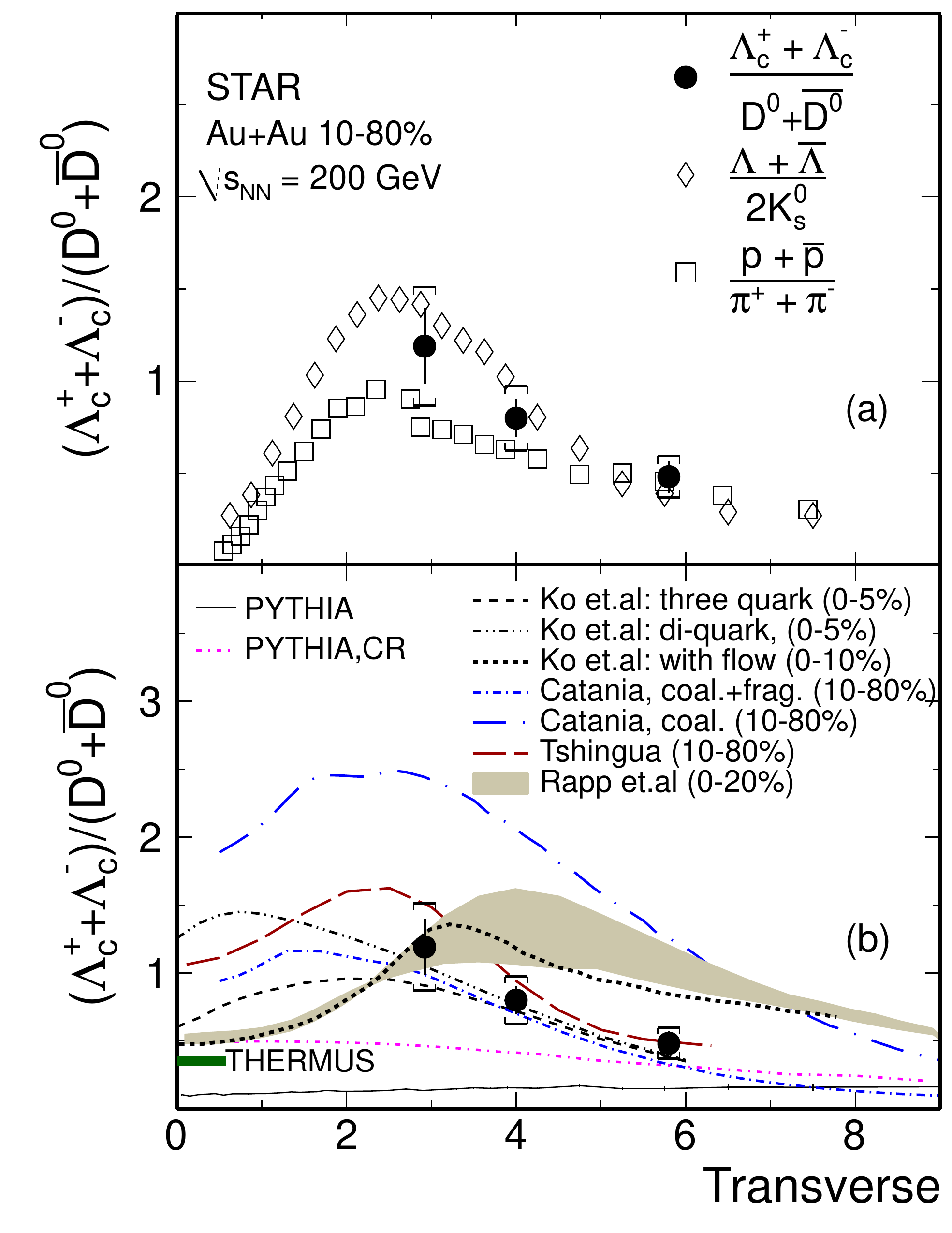}
\includegraphics[width=7 cm]{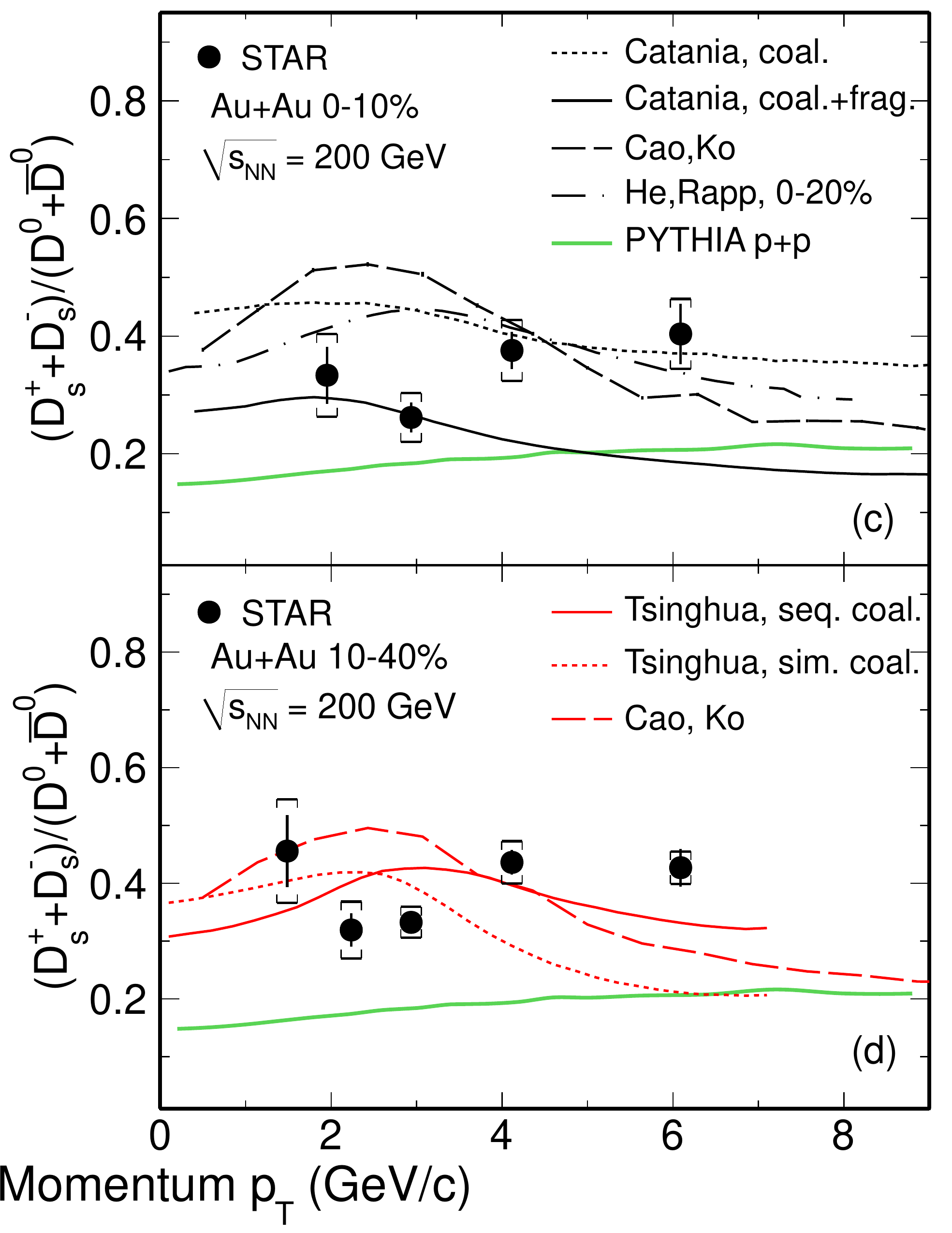}
\caption{(color online) Left Panels: The measured $\Lambda_{c}/D^0$ ratio at mid-rapidity ($|y|<$ 1) as a function of $p_{T}$  for Au+Au collisions at $\sqrt(s_{\rm NN}) = $ 200\,GeV in 10-80\% centrality, compared to the baryon-to-meson ratios for light and strange hadrons (a) and various model calculations (b). The $p_{T}$  integrated $\Lambda_{c}/D^0$ ratio from the THERMUS~\cite{Wheaton:2004qb} model calculation with a freeze-out temperature of $T_{\rm ch}=160$\,MeV is shown as a horizontal bar on the left axis of the plot. Right panels: (c) The integrated $D_{s}/D^{0}$ ratio (black solid circles) of 1.5 $<$ $p_{T}$ $<$ 8\,GeV/$c$ as a function of $p_{T}$ compared to model calculation (curves) in 0-10\% Au+Au collisions at $\sqrt{s_{_{\rm NN}}}$ = 200\,GeV. (d) Same $D_{s}/D^{0}$ ratio as (c) but with 10-40\% centrality. The vertical lines and brackets on the data points indicate statistical and systematic uncertainties respectively.}
\label{chadronization}
\end{figure*}  

The STAR Heavy Flavor Tracker (HFT) with a silicon pixel detector achieved $\sim$30 $\mu m$ spacial resolution of the track impact parameter to the primary vertex allows a topological reconstruction of the decay vertices of open charm hadrons. Figure~\ref{chadronization} left panels show the charmed baryon over meson ratio compared with light and strange baryon over meson ratios~\cite{Agakishiev:2011ar,Abelev:2006jr} (a) and various models (b). The $\Lambda_c/D^0$ ratio is comparable in magnitude to the $\Lambda/K^0_s$ and $p$/$\pi$ ratios and shows a similar $p_{T}$  dependence in the measured region. A significant enhancement is seen compared to the calculations from the latest PYTHIA 8.24 release (Monash tune~\cite{Skands:2014pea}) without (green solid curve) and without (magenta dot-dashed curve) color reconnections (CR)~\cite{Bierlich:2015rha}. The implementation with CR is found to enhance the baryon production with respect to mesons. However, both calculations fail to fully describe the data and its $p_{T}$  dependence. Figure~\ref{chadronization} (b) also shows the comparison to various models with coalescence hadronization of charm quarks~\cite{Oh2009,Plumari:2017ntm,Zhao:2018jlw,He2013,He2019}. The comparisons suggest coalescence hadronization plays an important role in charm-quark hadronization in the presence of QGP. Also, the data can be used to constrain the coalescence model calculations and their model parameters. 

Figure~\ref{chadronization} right panel shows the $D_{s}/D^{0}$ ratio as a function of $p_{T}$  compared to coalescence model calculations for 0-10\% (c) and 10-40\% (d) collision centralities. Several models incorporating coalescence hadronization of charm quarks and strangeness enhancement are used to describe the $p_{T}$  dependence of $D_{s}/D^{0}$ ratio. 
Those models assume that $D_s^{\pm}$ mesons are formed by recombination of charm quarks with equilibrated strange quarks in the QGP~\cite{Oh2009,Plumari:2017ntm,Zhao:2018jlw,He2013,He2019}. In particular, the sequential coalescence model together with charm quark conservation~\cite{Zhao:2018jlw} considers that more charm quarks are hadronized to $D_s^{\pm}$ mesons than $D^{0}$ since the former is created earlier in the QGP, which results in further enhancement of $D_{s}/D^{0}$ ratio in Au+Au collisions relative to $p$+$p$ collisions.

STAR experiment extracted the total charm production cross section per binary nucleon collision at midrapidity in 200 GeV Au+Au collisions by summing all yields of the open charm hadron states~\cite{STARccXesc}, which is consistent with that in $p$+$p$ collisions~\cite{STARDpp} within uncertainties. The numbers are reported as,

\begin{eqnarray} \label{eq:cxsAA}
AuAu:  d\sigma^{\rm NN}/dy|_{y=0} = 152 \pm 13 (stat) \pm 29 (sys) \mu b, \\
pp:  d\sigma/dy|_{y=0} = 130 \pm 30 (stat) \pm 26 (sys) \mu b
\end{eqnarray}

This result is consistent with charm quark conservation in heavy-ion collisions at RHIC top energy.

\subsection{Open bottom production}

Theoretical calculations predict that the heavy quark energy loss is less than that of light quarks due to suppression of the gluon radiation angle by the quark mass. Bottom quark mass is a factor of three larger than charm quark mass, thus the less bottom quark energy loss is expected compared to charm quark when they traverse the hot-dense medium created in the heavy-ion collisions~\cite{Dokshitzer:2001zm,Djordjevic:2004nq,Buzzatti:2011vt}. However, the low production cross section of bottom quark in RHIC energy and very small hadronic decay branching ratio prevent direct measurement of open bottom hadrons in experiments at RHIC. Fortunately, different life time of open charm hadrons and open bottom hadrons allow us to separate their decay products utilizing STAR HFT to distinguish their decay vertices and provide the impact parameter (or the distance of closest approach to primary collision vertex, DCA) distributions. 

\begin{figure}[!htb]
\centering
\includegraphics[width=8 cm]{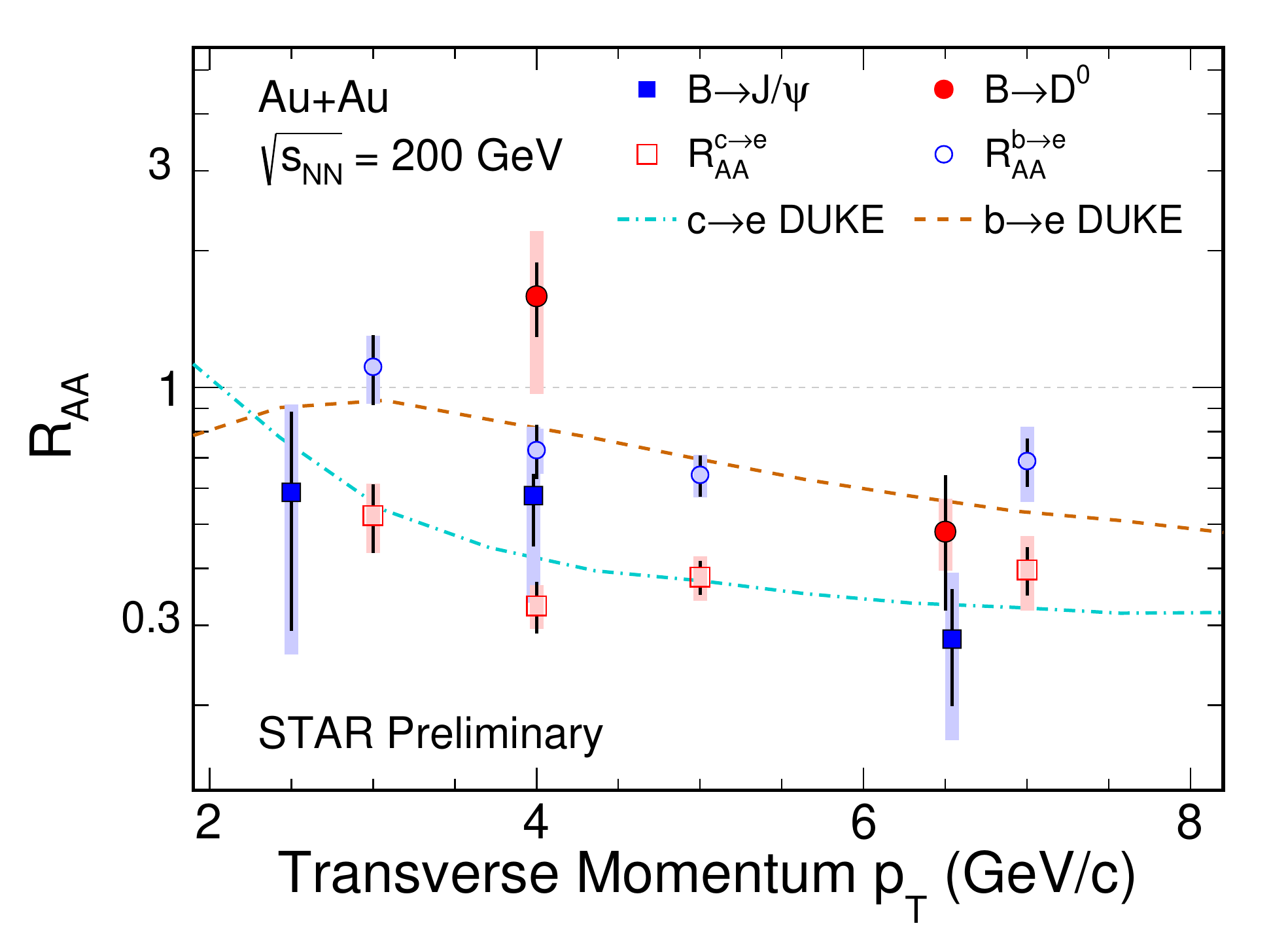}
\caption{(color online) $R_{\rm AA}$ of $B{\rightarrow}J/\psi$ (blue solid squares), $B{\rightarrow}D^0$ (red solid circles), $b{\rightarrow}e$ (blue open circles) and $c{\rightarrow}e$ (red open squares) at mid-rapidity in $\sqrt{s_{\rm NN}} = $ 200 GeV Au+Au collisions from STAR experiment~\cite{Chen:2018tqb,Zhang:2019dxw,Zhang:2018iva}. Vertical bars and bands represent statistic and systematic uncertainties, respectively. Dashed and dot-dashed curves are DUKE model calculations~\cite{Cao:2015hia} for $b{\rightarrow}e$ and $c{\rightarrow}e$, respectively.}
\label{b2X}
\end{figure}   

Recently the non-prompt products from open bottom decays, $B{\rightarrow}J/\psi$, $B{\rightarrow}D^0$ and $b{\rightarrow}e$, were measured by the STAR experiment at mid-rapidity in $\sqrt{s_{\rm NN}} = $ 200 GeV Au+Au collisions via a template fit method using the different shape of impact parameters between signal and background~\cite{Chen:2018tqb,Zhang:2019dxw,Zhang:2018iva}. The results of $R_{AA}$ as a function of $p_{T}$  are shown in Fig.~\ref{b2X}. The data of $B{\rightarrow}J/\psi$, represented by blue solid squares, are observed suppressed in the whole $p_{T}$  region from 2 to 8 GeV/c. The similar suppression is also observed for $B{\rightarrow}D^0$ (red solid circles) and $b{\rightarrow}e$ (blue open circles) at high $p_{T}$ . These results indicate that interactions between bottom quark and the hot-dense nuclear matter lead to bottom quark energy loss in the medium. It should be noted that the non-prompt $J/\psi$, $D^0$ and electrons shown here are in 0-80\%, 0-10\% and 0-80\% Au+Au collisions, respectively. Comparing with $c{\rightarrow}e$, shown as red open squares, $b{\rightarrow}e$ is systematically less suppressed, which indicates bottom looses less energy than charm. The calculations of a transport model from Duke group~\cite{Cao:2015hia} reproduce the data within uncertainties. The non-prompt $D^0$ at 4 GeV/c shows no suppression as well, again consistent with less energy loss of bottom quark due to its heavier mass comparing with charm and light quarks. These two independent measurements provide important evidence of mass-dependent parton energy loss in the hot QCD medium in high energy heavy-ion collisions. 

\begin{figure}[!htb]
\centering
\includegraphics[width=8 cm]{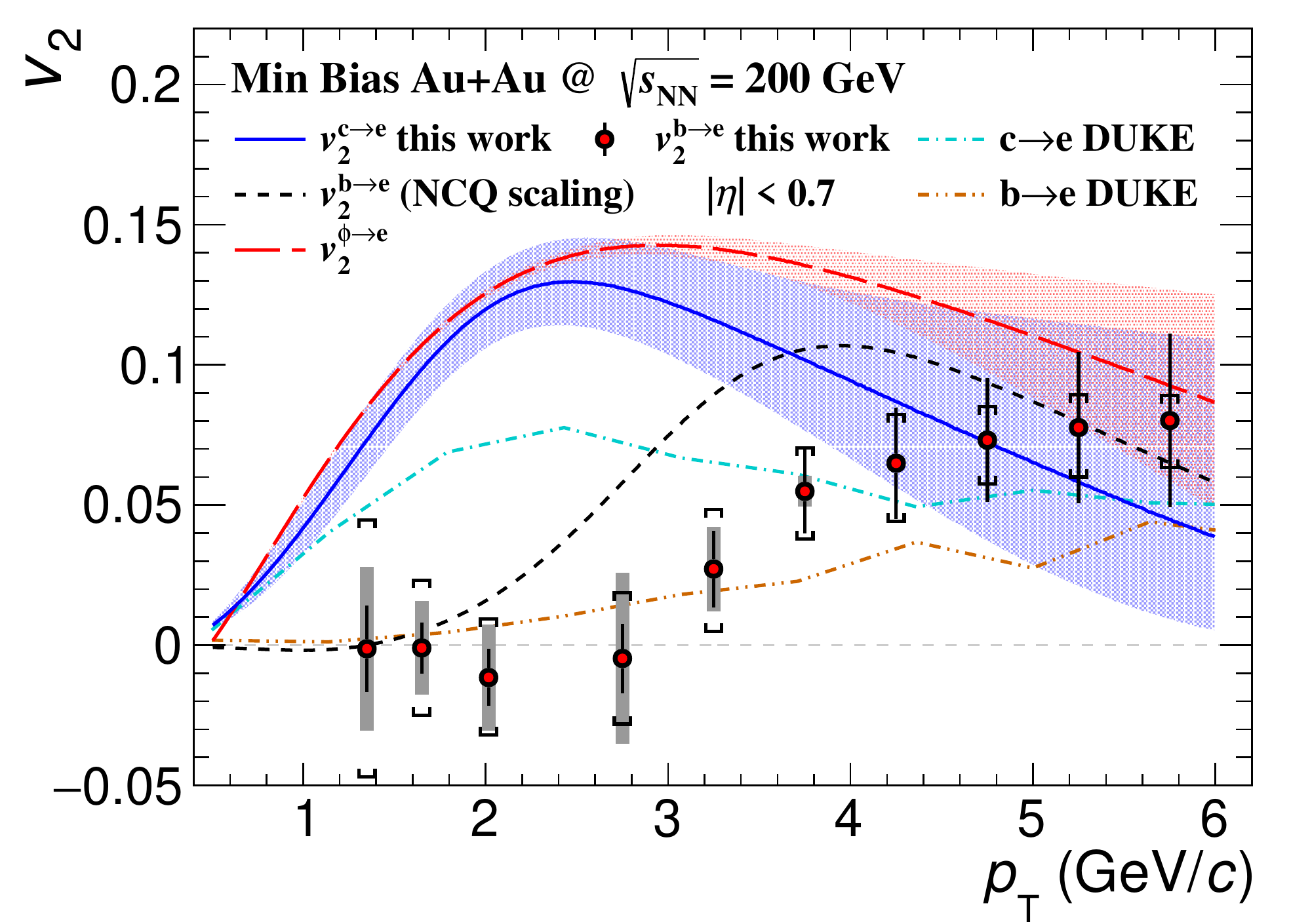}
\caption{(color online) The elliptic flows ($v_2$) of electrons from open charm (blue band) and open bottom (red circles) decays at mid-rapidity ($\left|\eta\right|$ $<$ 0.7) in minimum bias Au+Au collisions at \sNN = 200 GeV. The $v_2^{\rm b\rightarrow e}$ with B-meson $v_2$ NCQ scaling assumption and the $v_2^{\rm \phi\rightarrow e}$ are shown as the dashed curve and open squares, respectively. Results from DUKE~\cite{Cao:2015hia} model predictions are shown for comparison. Figure is taken from Ref.~\cite{Si:2019esg}.}
\label{b2ev2}
\end{figure}

Recently an experimental data driven approach was applied to extract bottom elliptic flow from heavy flavor semileptonic decay channels~\cite{Si:2019esg}. Taking the advantage of silicon vertex detectors, high precision open charm hadrons were measured by STAR experiment, which allows bottom contribution can be extracted by subtracting the contributions of open charm decays from the inclusive heavy flavor electron spectrum. Figure~\ref{b2ev2} shows the $v_2$ results of electrons from open charm ($v_2^{\cte}$) and open bottom ($v_2^{\bte}$) decays as the blue solid curve with an uncertainty band and red circles, respectively. The $v_2$ of $\rm\phi\rightarrow e$ ($v_2^{\rm\phi\rightarrow e}$ is shown as red long-dashed curve with band. DUKE model predictions~\cite{Cao:2015hia} are also shown as dot-dashed curves for comparison. The electron $v_2$ from beauty hadron decays at $p_T$ $>$ 3.0 GeV/$c$ is observed with an average of 4-sigma significance ($\chi^2/ndf$ = 29.7/6) deviating from zero. And it is consistent with electrons from charmed or strange hadron decays within uncertainties at $p_T$ $>$ 4.5 GeV/$c$. This flavor independent $v_2$ at high $p_T$ could be attributed to the initial geometry anisotropy or the path length dependence of the energy loss in the medium. A smaller $v_2^{\bte}$ compared with $v_2^{\cte}$ is observed at $p_T$ $<$ 4.0 GeV/$c$, which may be driven by the larger mass of beauty quark than that of charm quark. The $v_2^{\bte}$ deviates from the hypothesis of that B-meson $v_2$ follows the NCQ scaling (black curve) at 2.5 GeV/$c$ $<$ $p_T$ $<$ 4.5 GeV/$c$ with a confidence level of 99\% ($\chi^2/ndf$ = 14.3/4), which favors that the beauty quark elliptic flow is smaller than that of light quarks, unlike the $D^0$ $v_2$ scaled with that of light flavor hadrons by dividing number of constituent quarks in both $v_2$ and $\left(m_{\rm T}-m_0\right)$ as presented in Fig.~\ref{fig:v2CompareWithData} previously. This suggests that beauty quark is unlikely thermalized and too heavy to be moved following the collective flow of lighter partons in heavy-ion collisions at RHIC energy. 

\section{Quarkonium production in \pp collisions}
Quarkonium production in $p$+$p$ collisions is a crucial baseline of the study of quarkonium production in medium for various reasons. First of all, to quantify the modification of quarkonium production in heavy-ion collisions, we usually compare the production yield of quarkonium in heavy-ion collisions to that in $p$+$p$ collisions by calculating the nuclear modification factor
\begin{equation} \label{eq:RAA}
R_{\textrm{AA}}=\frac{y_{\textrm{AA}}} {y_{pp}~N_{\textrm{coll}}},
\end{equation}
where $y_\textrm{AA}$ and $y_{pp}$ is the yield of quarkonium in heavy-ion collisions and $p$+$p$ collisions, respectively, and $N_\textrm{coll}$ is number of binary nucleon-nucleon collisions in heavy-ion collisions. If the heavy-ion collision is only superposition of nucleon-nucleon collisions, the yield of quarkonium should follow the $N_\textrm{coll}$-scaling and $R_\textrm{AA}=1$. Deviation of \raa from unity indicates modification of quarkonium production in heavy-ion collisions. Secondly, understanding the production mechanism of quarkonium in \pp collisions is essential to interpret the \raa measurements in heavy-ion collisions. For example, quarkonium has relative large formation time, whether the $q\bar{q}$ interstate (before the quarkonium is fully formed) is color-singlet or color-octet may result in different modification of quarkonium yield in heavy-ion collisions when it traverses the medium. 

In fact, quarkonium production in hadron collisions also provides important test of QCD. The production of the pair of heavy quark and its anti-quark are dominantly from the initial hard scattering and can be calculated in the framework of perturbative QCD down to low $p_T$. However, when the heavy-quark pair forms a physical quarkonium bound state, the process involves long distances and soft momentum scales and thus is a non-perturbative. The latter relies on modeling. The detailed study of quarkonium production with hadron collider and the comparison to theoretical calculations provides an important test ground of both perturbative and non-perturbative aspects of QCD calculations.

Experimental study of the quarkonium production mechanism is usually conduced by measuring the production yield and polarization against different kinematic variables such as rapidity, \pT and compare to different theoretical calculations. 

It is also important to note that not all of the quarkonium are produced directly, but a large fraction of quarkonium are produced via the decay of other hadrons such as higher quarkonium states and $B$-hadrons for charmonium. The contribution from the decay of other hadrons is called feeddown contribution. Because of different properties such as binding energy, the directly produced quarkonium and those from the decay of different hadrons should have different modification in heavy-ion collisions, the feeddown contribution has to be taken into account when extracting physics information from the measurements of quarkonium suppression in heavy-ion collisions. The feeddown contribution is usually studied in \pp collisions and/or small systems such as $p$+A collisions. 

\subsection{J/$\psi$ production cross-section} 

\begin{figure*}[!t] 
\includegraphics
  [width=0.7\hsize]
  {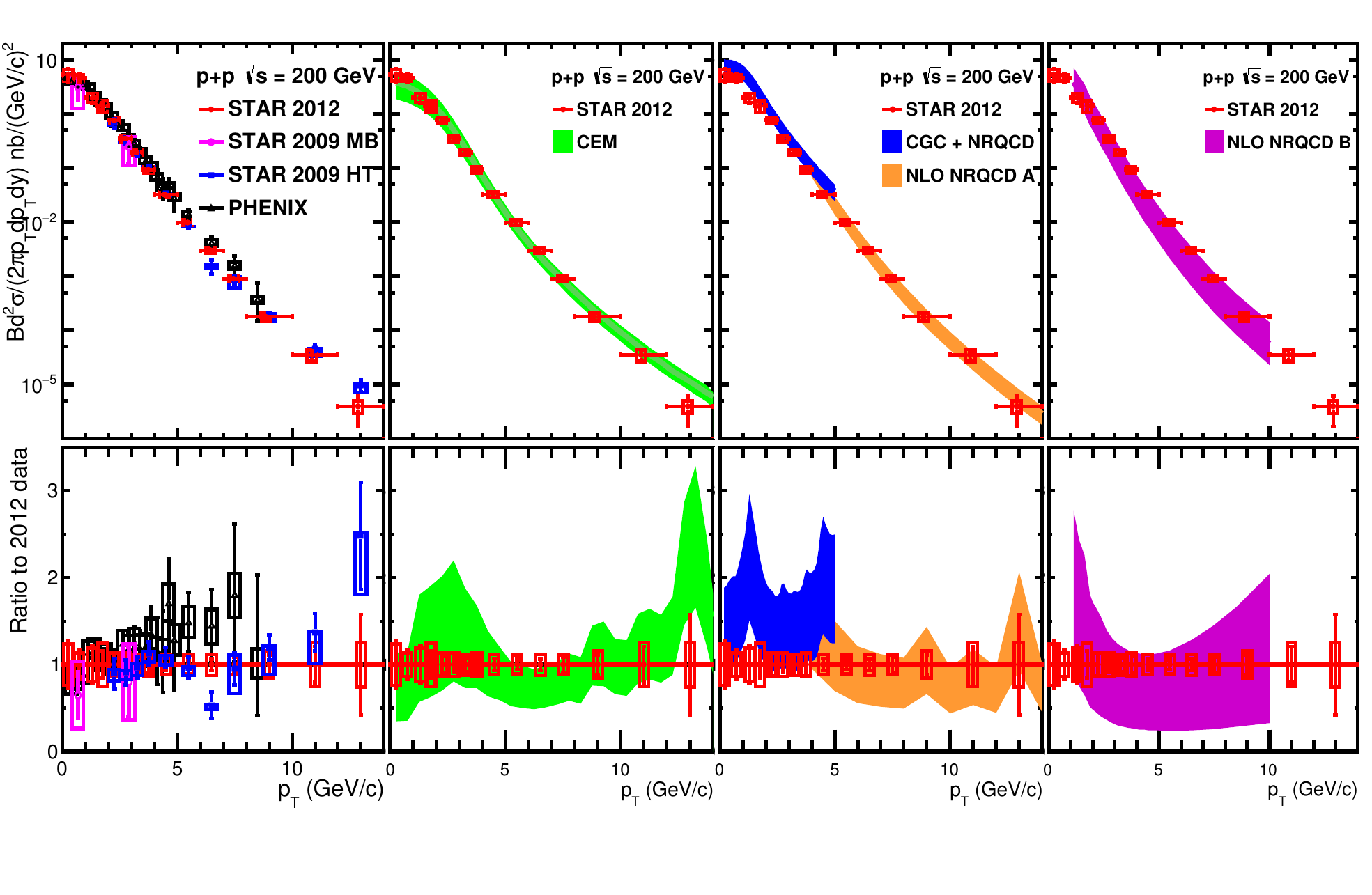}
\caption{Top panels: Inclusive \Jpsi production cross section times decay branching ratio as a function of \pT at mid-rapidity in non-single diffractive \pp collisions at \s = 200GeV measured by the STAR~\cite{Jpsi_STARpp2012,Jpsi_STARpp2009HT,Jpsi_STARpp2009MB} and PHENIX~\cite{Jpsi_PHENIXpp} Collaborations. The bars and boxes on the data points depict statistical and systematical uncertainties, respectively. The bands show theoretical calculations from CEM for direct J/$\psi$~\cite{Frawley:2008kk} (green), NRQCD for prompt J/$\psi$~\cite{NRQCD_PKU} (orange), NRQCD for direct J/$\psi$~\cite{NRQCD_Europe} (magenta) and CGC+NRQCD~\cite{CGC_NRQCD} (blue). Bottom panels: The ratios with respect to the central value of STAR measurements from the 2012 dataset.The figure is taken from~\cite{Jpsi_STARpp2012}.}
\label{fig:Jpsi_pt_200GeV_STAR}
\end{figure*}

Figure~\ref{fig:Jpsi_pt_200GeV_STAR} shows the inclusive \Jpsi production cross-section at mid-rapidity ($|y|<1$) as a function of \pT in non-single-diffractive \pp collisions at \s = 200 GeV measured by the STAR Collaboration~\cite{Jpsi_STARpp2012} via the di-electron decay channel by combining the events from minimum-bias trigger and triggered by the electromagnetic calorimeter with various thresholds, taken in 2012. The results are found to be consistent with STAR previously published results~\cite{Jpsi_STARpp2009MB, Jpsi_STARpp2009HT} and PHENIX published results ($|y|<0.35$)~\cite{Jpsi_PHENIXpp} but with improved precision at $p_T>2 ~\textrm{GeV}/c$. The decay branching ratio is not corrected for. The total production cross-section per unit rapidity for inclusive \Jpsi is extracted to be 
\begin{equation}
B_{ee} \frac{d\sigma}{dy}|_{y=0} = 43.2 \pm 3.0(\textrm{stat.}) \pm 7.5(\textrm{syst.})~\textrm{nb}.
\end{equation}
The precise \pT spectrum is compared to theoretical calculations. The theory-to-data ratios are shown in the lower panels in Fig.~\ref{fig:Jpsi_pt_200GeV_STAR}. The green band represents the calculation from Color Evaporation Model (CEM) for prompt \Jpsi at $|y|<0.35$~\cite{Frawley:2008kk}. The orange and magenta bands represent the calculation in the framework of Next-to-Leading Order (NLO) Non-Relativistic QCD (NRQCD) with different treatments, labeled as NRQCD A ~\cite{NRQCD_PKU} and NRQCD B~\cite{NRQCD_Europe}. NRQCD A is for prompt \Jpsi and NRQCD B is for direct J/$\psi$. The blue band shows the calculation from NRQCD A incorporating a Color-Glass Condensate (CGC) effective theory for small-$x$ resummation for prompt \Jpsi ~\cite{CGC_NRQCD}. The CEM and NRQCD calculations describe the data in the applicable \pT ranges within uncertainties. The CGC+NRQCD calculations at low \pT are systematically higher than the data but the lower boundary touches the data. It is noted that the contribution from $B$-hadron decay is not included in the theoretical calculations. As will be discussed in the next subsection, its contribution increases with increasing $p_T$, but less than 20\% at \pT below 5 GeV/$c$.

\begin{figure}[htb]
\includegraphics
  [width=0.8\hsize]
  {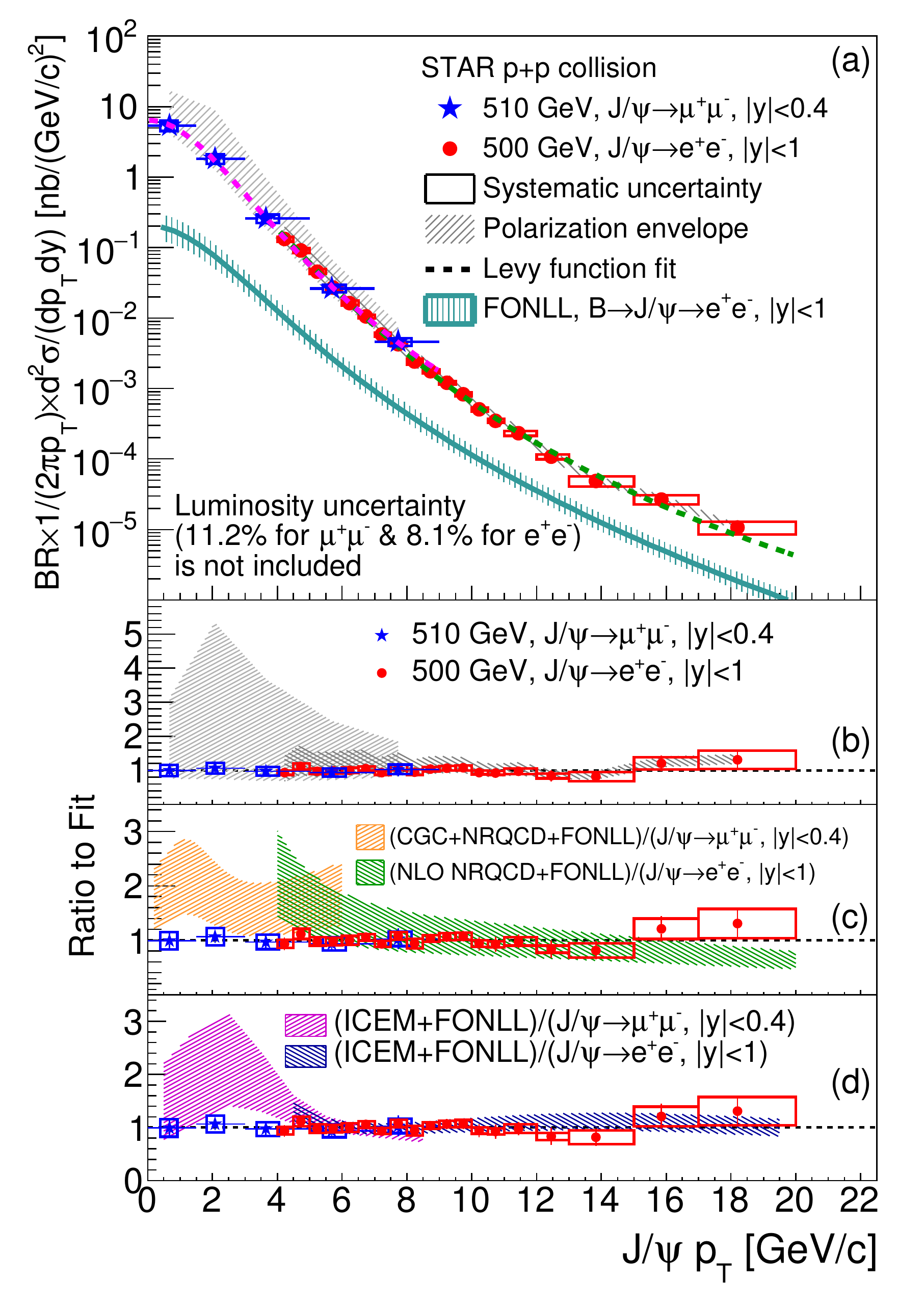}
\caption{(a) Inclusive \Jpsi production cross section times decay branching ratio as a function of \pT at mid-rapidity in non-single diffractive \pp collisions at \s = 500 and 510 GeV measured by the STAR Collaboration~\cite{Jpsi_STARpp500GeV}. (b-d), Ratios of data and different model calculations to a fit to the data. See text for details. The figure is taken from~\cite{Jpsi_STARpp500GeV}. }
\label{fig:Jpsi_pt_500GeV_STAR}
\end{figure}

Figure~\ref{fig:Jpsi_pt_500GeV_STAR} shows inclusive \Jpsi production cross-section in mid-rapidity as a function of \pT in non-single-diffractive \pp collisions at \s = 500 and 510 GeV measured by the STAR Collaboration~\cite{Jpsi_STARpp500GeV} via both dimuon and di-electron decay channels. It covers a broad \pT range ($0<p_T<20~\textrm{GeV}/c$). The data points depict the results from unpolarized-assumption and the gray band denote the polarization envelope. The total production cross-section per unit rapidity for inclusive \Jpsi within $0<p_T<9~\textrm{GeV}/c$ is
\begin{equation}
B_{\mu\mu} \frac{d\sigma}{dy}|_{y=0} = 67 \pm 6(\textrm{stat.}) \pm 10(\textrm{syst.}) ^{+100}_{-18}(\textrm{pol.}) \pm 7(\textrm{lumi.}) ~\textrm{nb}.
\end{equation}
The data is compared to NRQCD~\cite{NRQCD_PKU}, CGC+NRQCD~\cite{CGC_NRQCD} and Improved CEM (ICEM)~\cite{ICEM} calculations. All the model calculations are for prompt J/$\psi$, the feeddown from $B$-hadrons is not included. In order to do a fair comparison between data and theoretical calculations, the feeddown contribution from $B$-hadrons is estimated using FONLL calculations~\cite{FONLL_online, FONLL1, FONLL2} and added to the theoretical calculations. Figure~\ref{fig:Jpsi_pt_500GeV_STAR}(b-d) shows the ratio of the theoretical calculations to a fit to the data using an empirical function.  At high-$p_T$, the NRQCD and ICEM describe the data reasonably good. At low-$p_T$, both CGC+NRQCD and ICEM calculations are above the data, but within the huge uncertainty from the polarization envelope. 

\subsection{Feeddown contribution of J/$\psi$}

The inclusive \Jpsi production includes prompt \Jpsi and non-prompt J/$\psi$.  The former includes the directly produced \Jpsi and the contribution from the decay of excited charmonium states such as $\psi(2S)$, $\chi_{c0,1,2}$ etc. While non-prompt \Jpsi refers to the contribution from the decay of $B$-hadrons. It is important to understand the fraction of different components of inclusive \Jpsi to interpret the measurements of inclusive \Jpsi production mechanism in both \pp and A+A collisions.

\begin{figure}[!htb]\centering{
 \includegraphics
  [width=0.65\hsize]
  {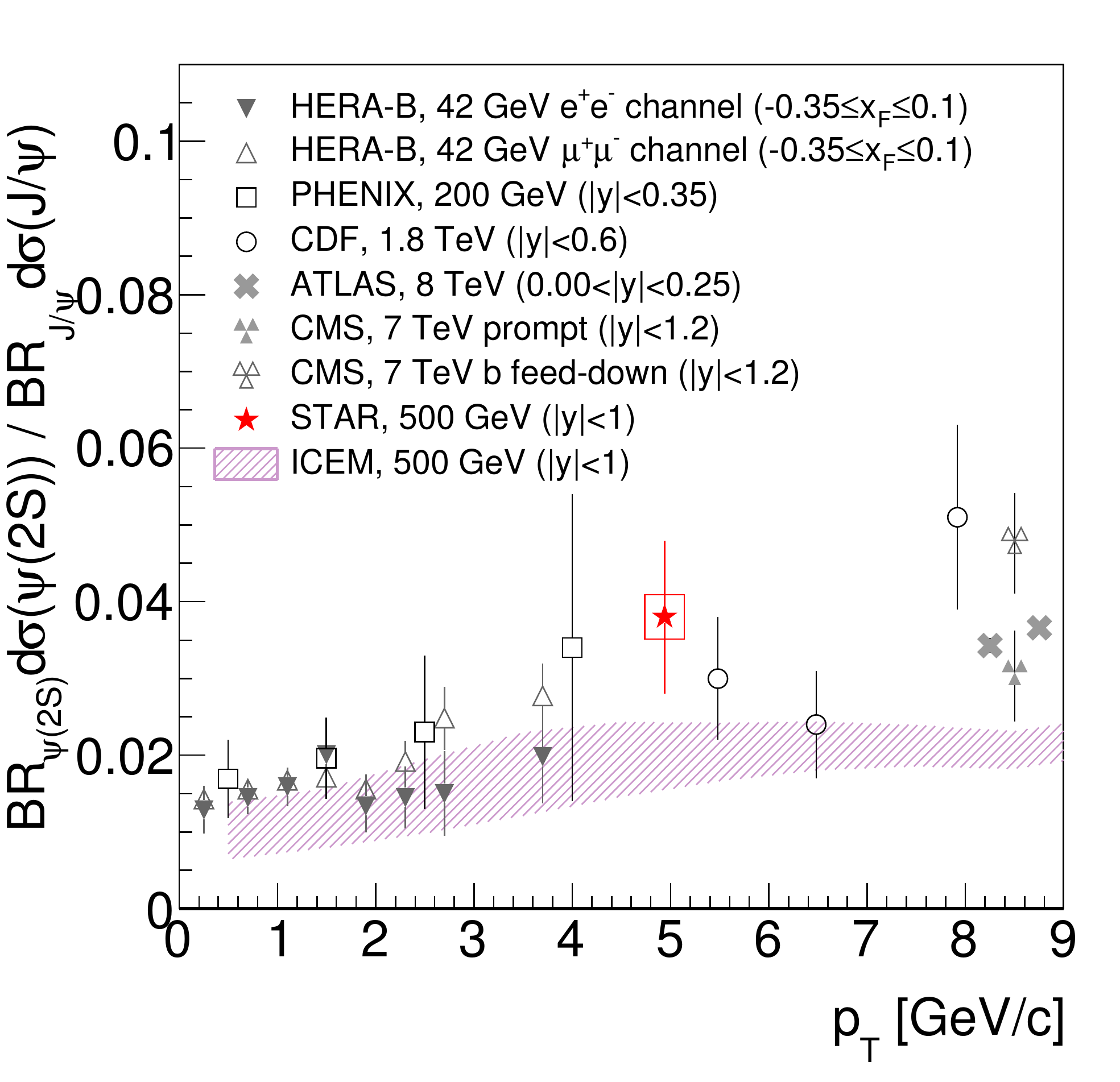}
  \includegraphics
  [width=0.65\hsize]
  {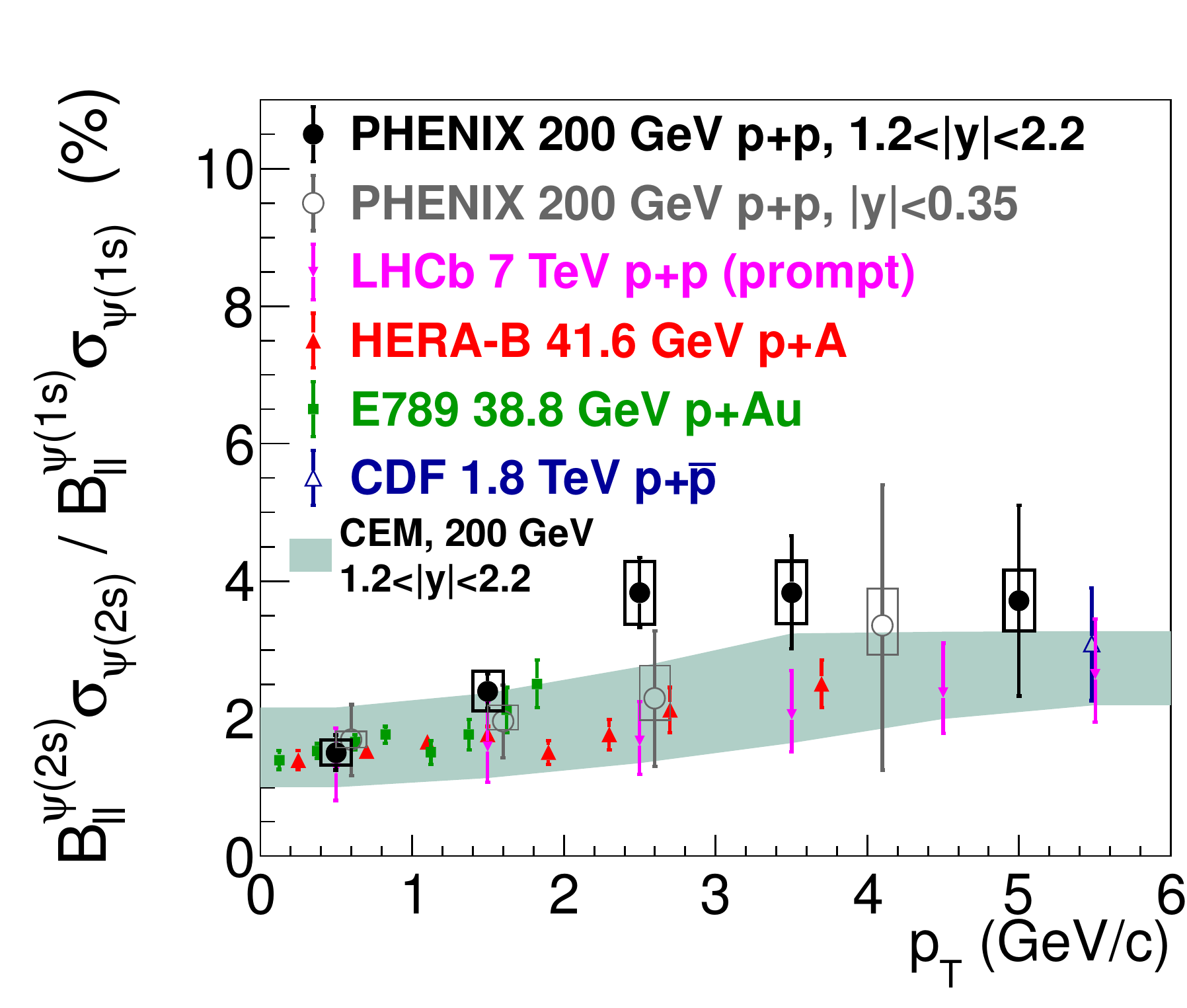}}
\caption{The ratio of the production cross section times branching ratio of $\psi(2S)$ and \Jpsi as a function of \pT in $p+p(\bar{p}$) or $p$ + A collisions from various experiments. The two panels of this figure are respectively taken from~\cite{Jpsi_STARpp500GeV} and \cite{Jpsi_psi2S_PHENIX_PRC2017}.}
\label{fig:psi2S-2-Jpsi}
\end{figure}

$\psi(2S)$ and \Jpsi is usually reconstructed in the same di-lepton decay channel. The systematic uncertainties can be largely cancelled out when calculating the ratio of their yields. The measurement is actually very challenging because the study of \Jpsi at RHIC is already statistics limited, the yield of $\psi(2S)$ is even much lower than \Jpsi and the di-lepton decay branching ratio is also much lower. This results in about a factor of 50 reduction of reconstructed $\psi(2S)$ signal than the J/$\psi$, while the combinatorial and correlated background is similar. The yield ratios of inclusive $\psi(2S)$ and inclusive J/$\psi$, after correcting for the difference of acceptance and efficiency, measured at RHIC by the STAR and the PHENIX Collaborations are shown in Fig.~\ref{fig:psi2S-2-Jpsi} and compared to world data. The uncertainties are dominated by statistical uncertainties. Note that the decay branching ratio is not corrected for, which is about 
\begin{equation}
\frac{\textrm{BR}_{\psi(2S)}} {\textrm{BR}_{J/\psi}} \sim 7.53 \pm 0.16~(e^+e^-) \textrm{ or } 7.5 \pm 0.6~(\mu^+\mu^-) ,
\end{equation}
~\cite{PDG2018}. The ratio increases with \pT but shows no much energy dependence from center-of-mass energy of 40 GeV to 7 TeV.  The ICEM calculations~\cite{ICEM} at RHIC can describe the data. The branching ratio of $\psi(2S) \rightarrow J/\psi + X$ is $(61.4 \pm 0.6)\%$~\cite{PDG2018}. The feeddown contribution of $\psi(2S)$ to inclusive \Jpsi can be roughly estimated by multiplying the ratio shown in Fig.~\ref{fig:psi2S-2-Jpsi} by a factor of $\sim 4.6 \pm 0.5$. The fraction is $<\sim 10\%$ at low \pT and increases to about 15\% at \pT up to 10 GeV/c.

\begin{figure*}[tb]
\includegraphics
  [width=0.35\hsize]
  {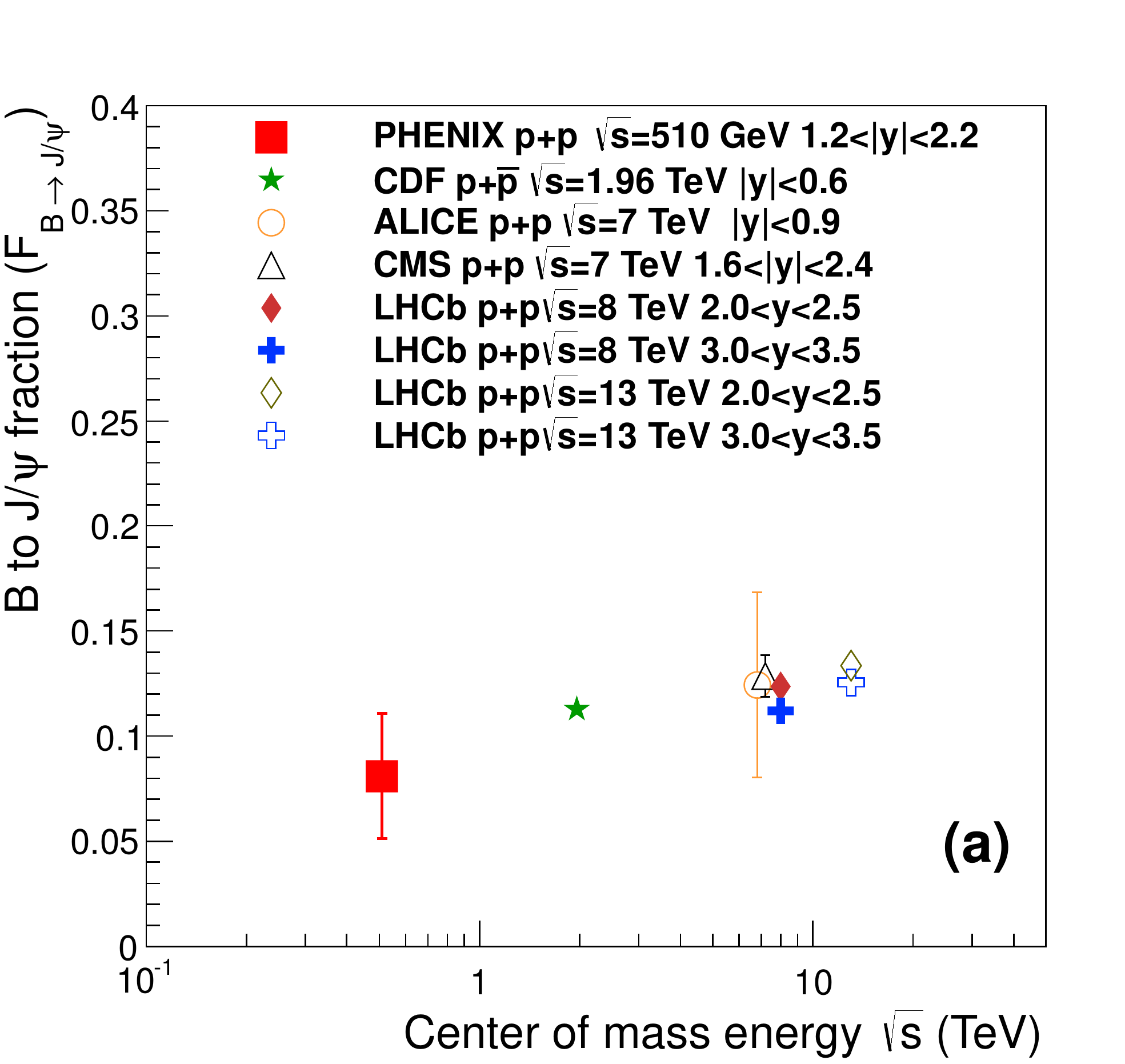}
  \includegraphics
  [width=0.35\hsize]
  {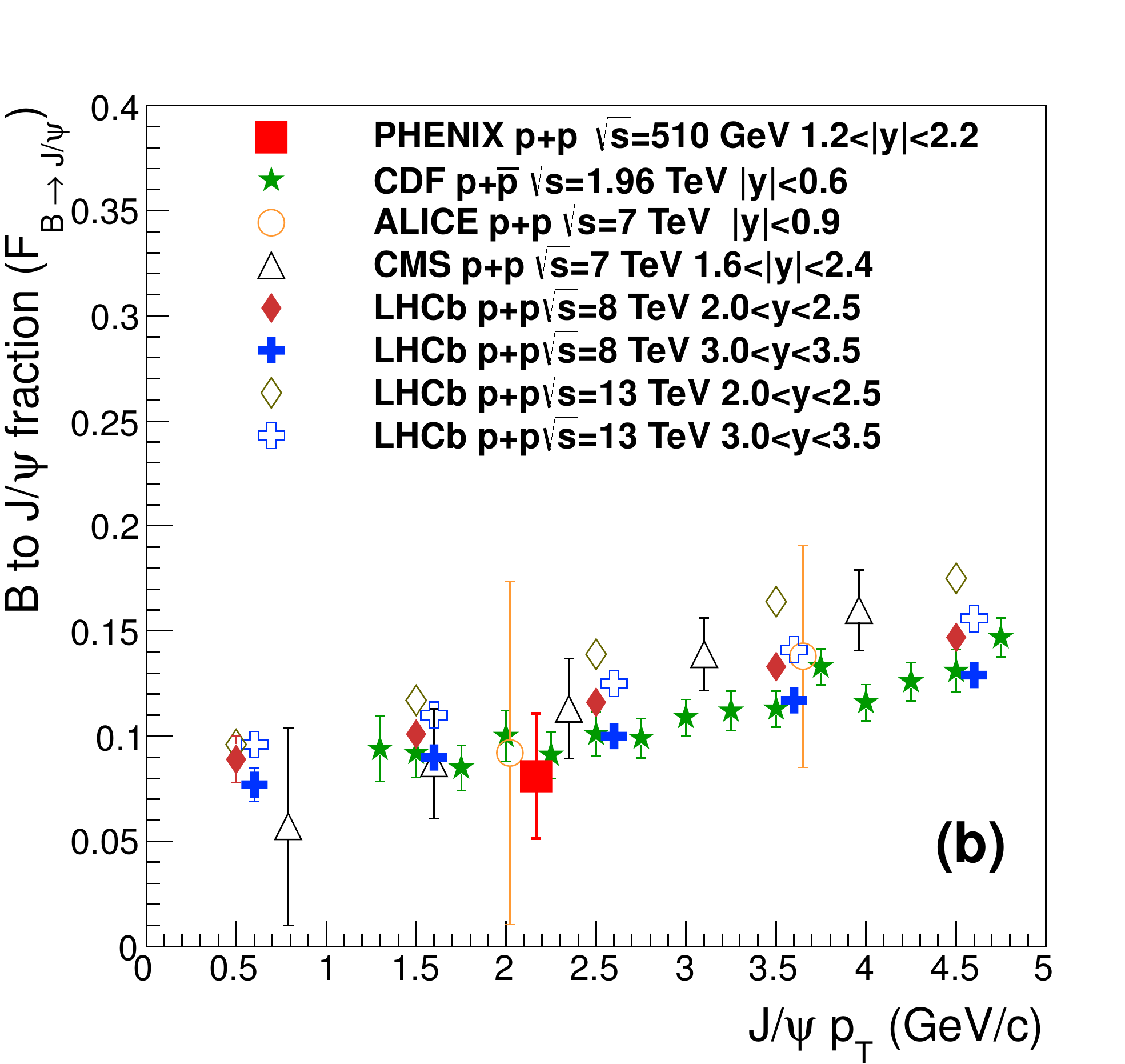}
\caption{The feeddown fraction of $B$-hadron decays to inclusive \Jpsi at a function of energy and \pT in $p+p(\bar{p})$ collisions. The figure is taken from~\cite{Jpsi_B2Jpsi_PHENIX_PRD2017}.}
\label{fig:B-2-Jpsi-pp510}
\end{figure*}

The feeddown contribution of $\chi_c$ to \Jpsi is usually studied via the radiative decay of $\chi_c$ ($\chi_c \rightarrow J/\psi + \gamma$). This measurement is also very challenging since the photon from the decay typically has very low energy, which requires the electromagnetic calorimeter having very good energy resolution and low threshold. At RHIC, so far only the PHENIX Collaboration successfully performed this measurement~\cite{excited_charmonium_PHENIX_PRD2012}. The feeddown fraction of $\chi_c$ decays in the inclusive \Jpsi is measured to be $(32 \pm 9)\%$. There is no enough statistics to study \pT dependence of the fraction. The LHCb Collaboration measured the feeddown fraction of $\chi_c$ decays in prompt \Jpsi as a function of \pT with good precision and found that the fraction is only 14\% at $p_T=2$ GeV/$c$ and monotonically increases to about 25\% at $p_T=10$ GeV/$c$ in \pp collisions at \s = 7 TeV at forward rapidity~\cite{LHCb:2012af}.

The feeddown contribution of non-prompt \Jpsi ($J/\psi \leftarrow B$) can be measured via two methods. The STAR Collaboration measured the feeddown fraction of $B$-hadron decays at mid-rapidity ($|y|<1$) at $p_T > 5$ GeV/$c$ in \pp collisions at \s = 200 GeV via the correlation function of \Jpsi and charged hadrons since the non-prompt \Jpsi is usually associated with more charged hadrons than prompt J/$\psi$~\cite{Jpsi_STARpp2009HT}. The PHENIX Collaboration measured the fraction at forward rapidity (1.2<|y|<2.2) in \pp collisions at \s = 510 GeV using the silicon vertex detector to statistically distinguish \Jpsi coming from primary vertex and secondary vertex~\cite{Jpsi_B2Jpsi_PHENIX_PRD2017}. Figure~\ref{fig:B-2-Jpsi-pp510} shows the fraction as a function of energy and $p_T$. The data from Tevatron and LHC experiments are also shown. The fraction shows no much center-of-mass energy dependence but significant \pT dependence. The fraction is $8.1\% \pm 2.3\%$ (stat.) $\pm~1.9\%$ (syst.) at low \pT ($0<p_T<5$ GeV/$c$) and $1.2<|y|<2.2$.  It increases to about 15\% at $p_T=5$ GeV/$c$ and about 25\% at $p_T=10$ GeV/$c$.

For more details of the feeddown contribution to quarkonium production, please refer to a recent review~\cite{quarkonium_feeddown}.

\subsection{J/$\psi$ polarization}


\begin{figure}[!htb]
\includegraphics
  [width=0.6\hsize]
  {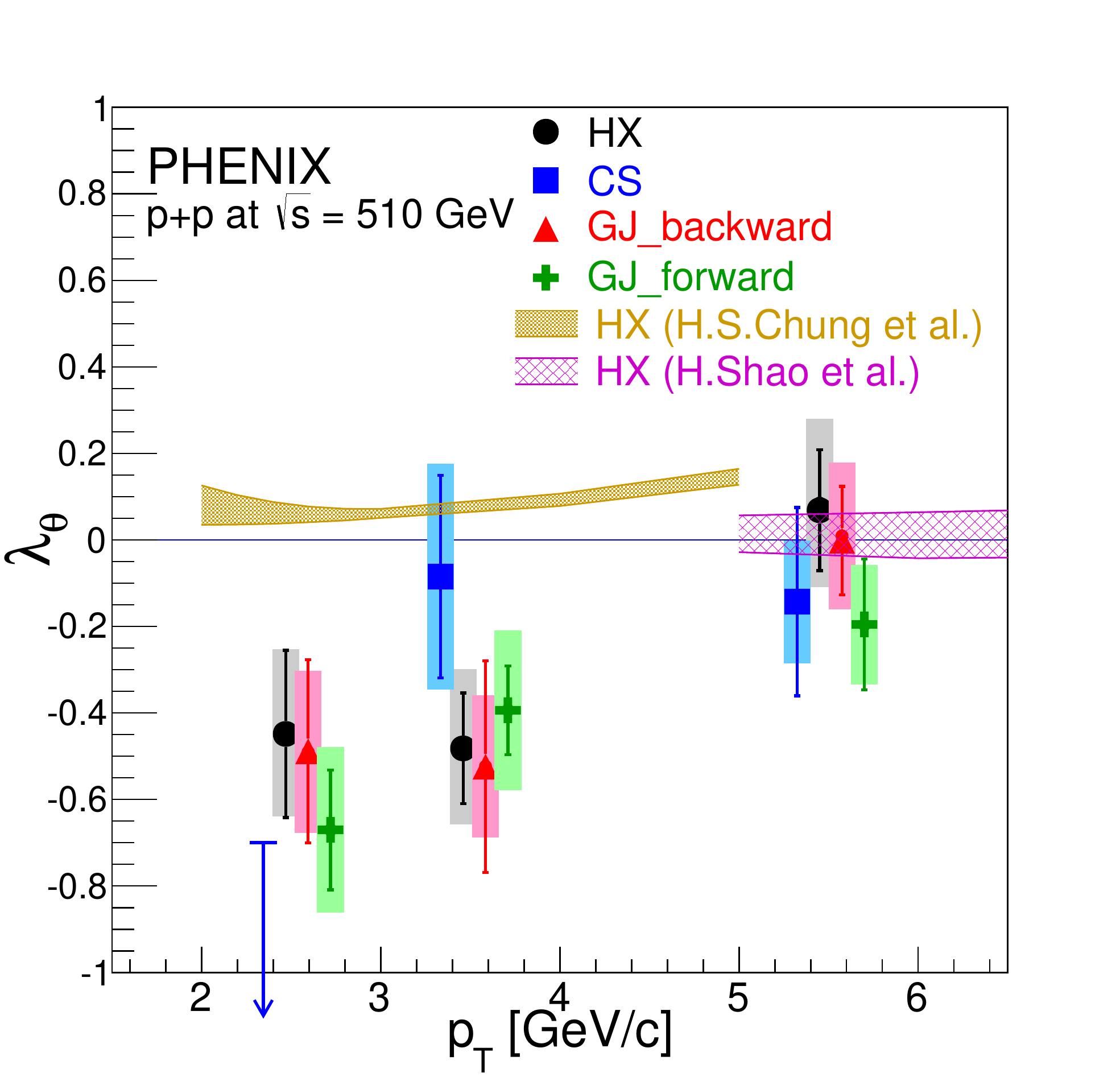}
  \includegraphics
  [width=0.6\hsize]
  {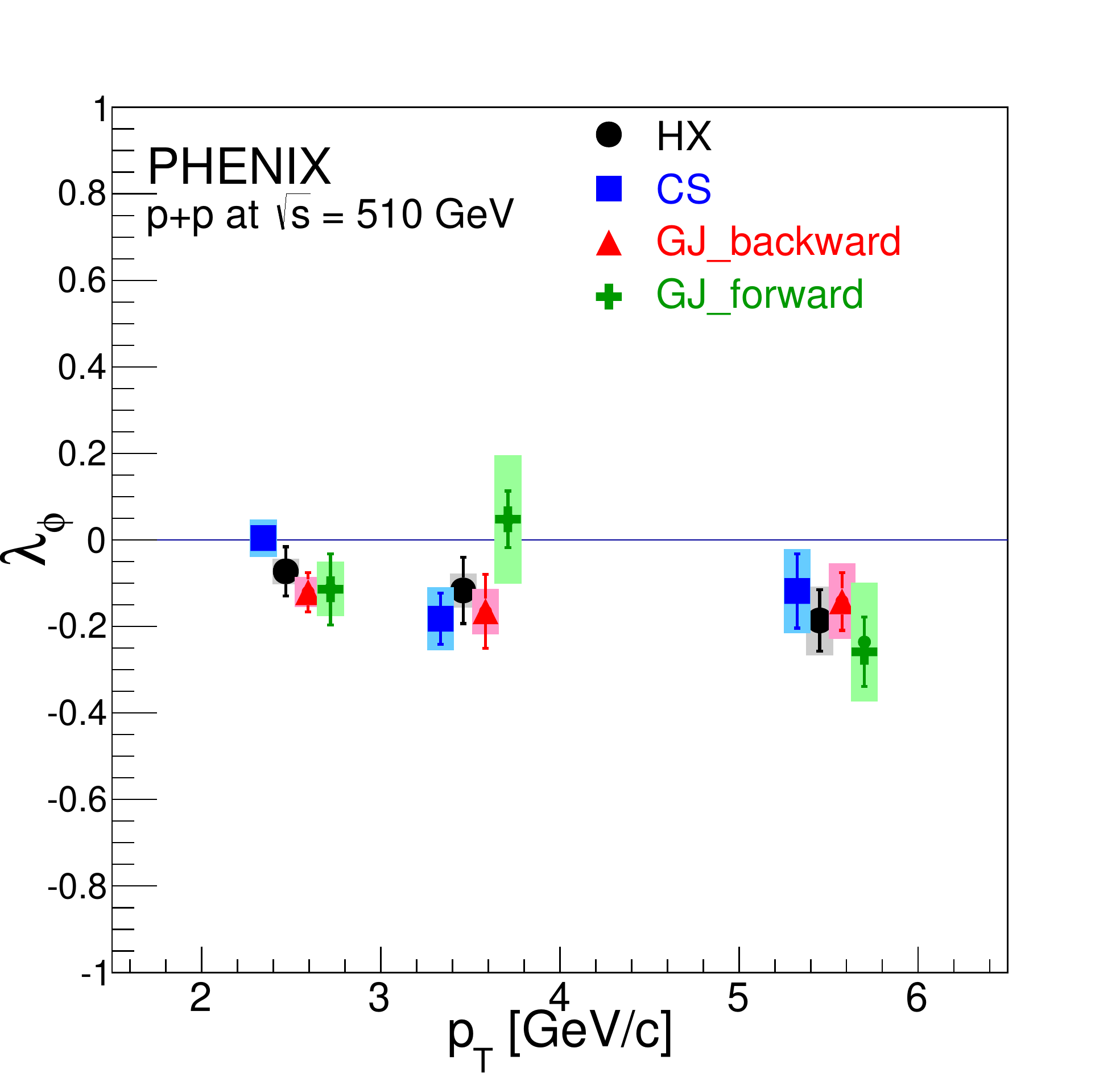}
\caption{\Jpsi polarization parameters $\lambda_\theta$ and $\lambda_\phi$ as a function of \pT with various reference frames at forward rapidity in \pp collisions at \s = 510 GeV. The figure is taken from ~\cite{Jpsi_Pol_PHENIX_PRD2017}.}
\label{fig:Jpsi_pol_PHENIX}
\end{figure}

\begin{figure*}[!htb]
\includegraphics
 [width=0.65\hsize]
 {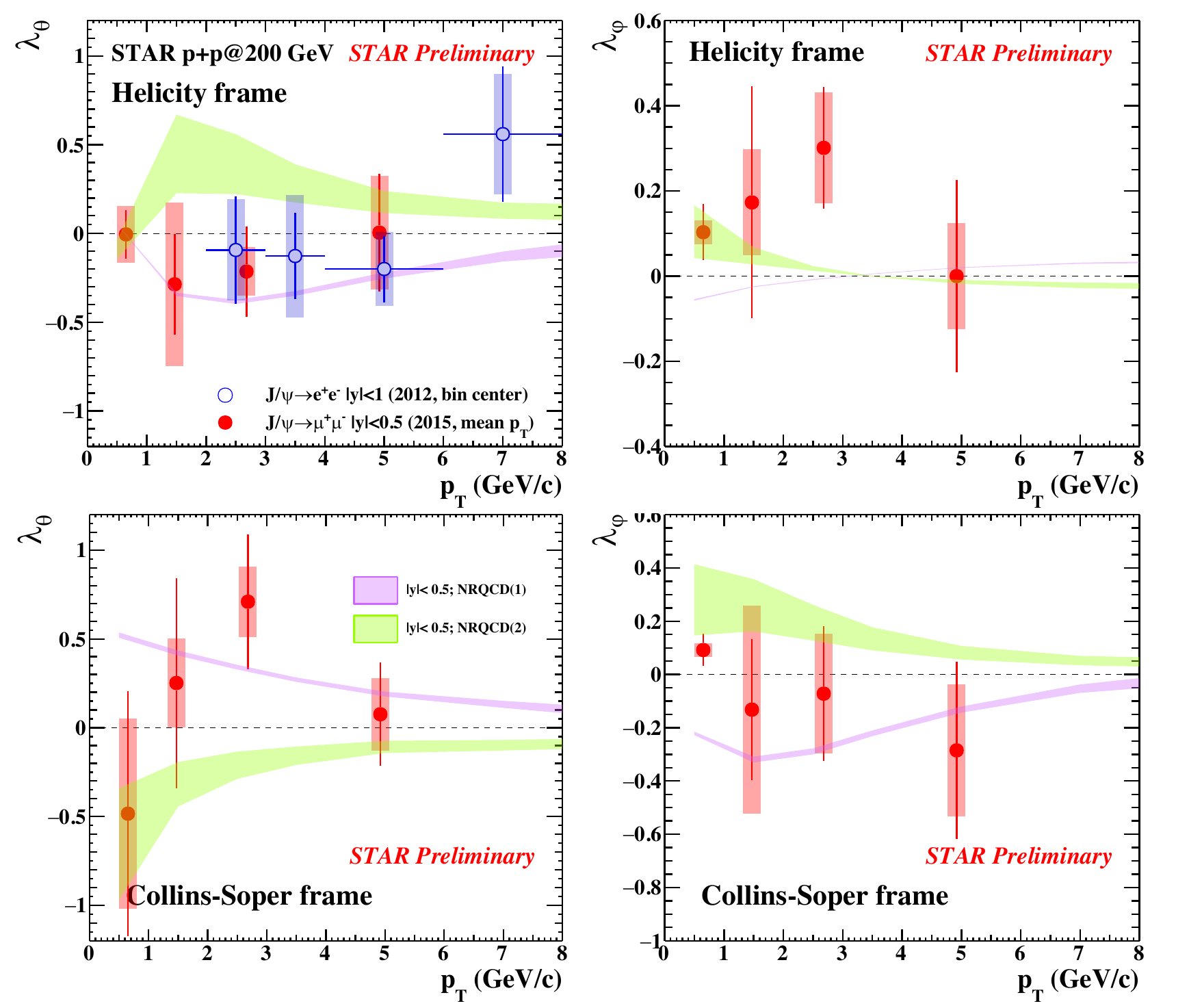}
\caption{\Jpsi polarization parameters $\lambda_\theta$ and $\lambda_\phi$ as a function of \pT in various reference frames at mid-rapidity in \pp collisions at \s = 200 GeV. The figure is taken from~\cite{ZhenLiu_HP2018Proceedings}.}
\label{fig:Jpsi_pol_STAR}
\end{figure*}

The measurements of \Jpsi polarization is important not only because the acceptance and efficiency used in the production cross section measurements depend on the polarization, but more important is that it helps to distinguish or constraint different theoretical models. Currently there is no model in the market can simultaneously describe \Jpsi production cross section and polarization. \Jpsi polarization (spin alignment) can be measured through the angular distribution of the di-lepton decay from J/$\psi$. It can be parameterized as
\begin{equation}
\begin{split}
 W(\cos\theta,\varphi) \propto \frac{1}{3+\lambda_{\theta}}  
 (1+\lambda_{\theta}\cos^{2}\theta \\ 
 +\lambda_{\varphi}\sin^{2}\theta \cos2\varphi  +\lambda_{\theta\varphi}\sin2\theta \cos\varphi).
\end{split}
\label{eq:2dPolar}
\end{equation}
Where $\lambda_{\theta}$, $\lambda_{\varphi}$ and $\lambda_{\theta \varphi}$ are the polarization parameters. $\theta$ and $\varphi$ are the polar and azimuthal angle of a lepton in the J/$\psi$ rest frame with respect to the chosen quantization axis. The coefficients depend on the chosen quantization axis. The name of the reference frames and the corresponding quantization axises are:
\begin{enumerate}
\item Helicity (HX) frame: The direction along the J/$\psi$ momentum in the center-of-mass system of the colliding beams;
\item Collins-Soper (CS) frame: The bisector of the angle formed by one beam direction and the opposite direction of the other beam in J/$\psi$ rest frame;
\item Gottfried-Jackson (GJ) frame: The direction of the beam momentum boosted into \Jpsi rest frame.
\end{enumerate}

Figure~\ref{fig:Jpsi_pol_PHENIX} shows the measurements of polarization parameters for inclusive \Jpsi in various reference frames at forward rapidity in \pp collisions at \s = 510 GeV~\cite{Jpsi_Pol_PHENIX_PRD2017}. In all frames, the polarization parameter $\lambda_\theta$ is significantly negative at low \pT and consistent with no polarization at high $p_T$. In contrast, the polarization parameter $\lambda_\phi$ is close to zero and becomes slightly negative at high $p_T$. The theoretical calculation on $\lambda_\theta$ for prompt \Jpsi in HX frame in the NRQCD factorization approach by H. S. Chung \textit{et al.}~\cite{Chung:2010iq} at $2<p_T<5$ GeV/$c$ and by H. Shao~\cite{Shao:2014yta} at \pT above 5 GeV/$c$ are also shown for comparison. Both calculations are consistent with the data at high $p_T$. However, there is discrepancy between the data and the theoretical calculation at low $p_T$. The theory expects small but positive $\lambda_\theta$, but the data is significantly negative. 

STAR recently measured \Jpsi polarization parameters $\lambda_\theta$ and $\lambda_\phi$ at mid-rapidity in \pp collisions at \s = 200 GeV, as shown in Fig.~\ref{fig:Jpsi_pol_STAR}~\cite{ZhenLiu_HP2018Proceedings}. In the HX frame, the measured $\lambda_\theta$ is slightly negative, but consistent with zero at $p_T<6$ GeV/$c$. While $\lambda_\phi$ is slightly positive. In CS frame, $\lambda_\theta$ is slightly positive, but consistent with zero within uncertainties. The bands in Fig.~\ref{fig:Jpsi_pol_STAR} depict NRQCD calculations using two sets of Long Distance Matrix Elements (LDME). The two calculations are labeled as NRQCD1~\cite{Zhang:2014ybe} and NRQCD2~\cite{Gong:2012ug}. They show significant difference especially at low $p_T$. Although with the current precision the data is consistent with both calculations, the measurements with improved precision in future should be able to constraint the LDMEs.

\subsection{$\Upsilon$ production cross section}

\begin{figure}[!htb]
\includegraphics[width=0.7\hsize]
{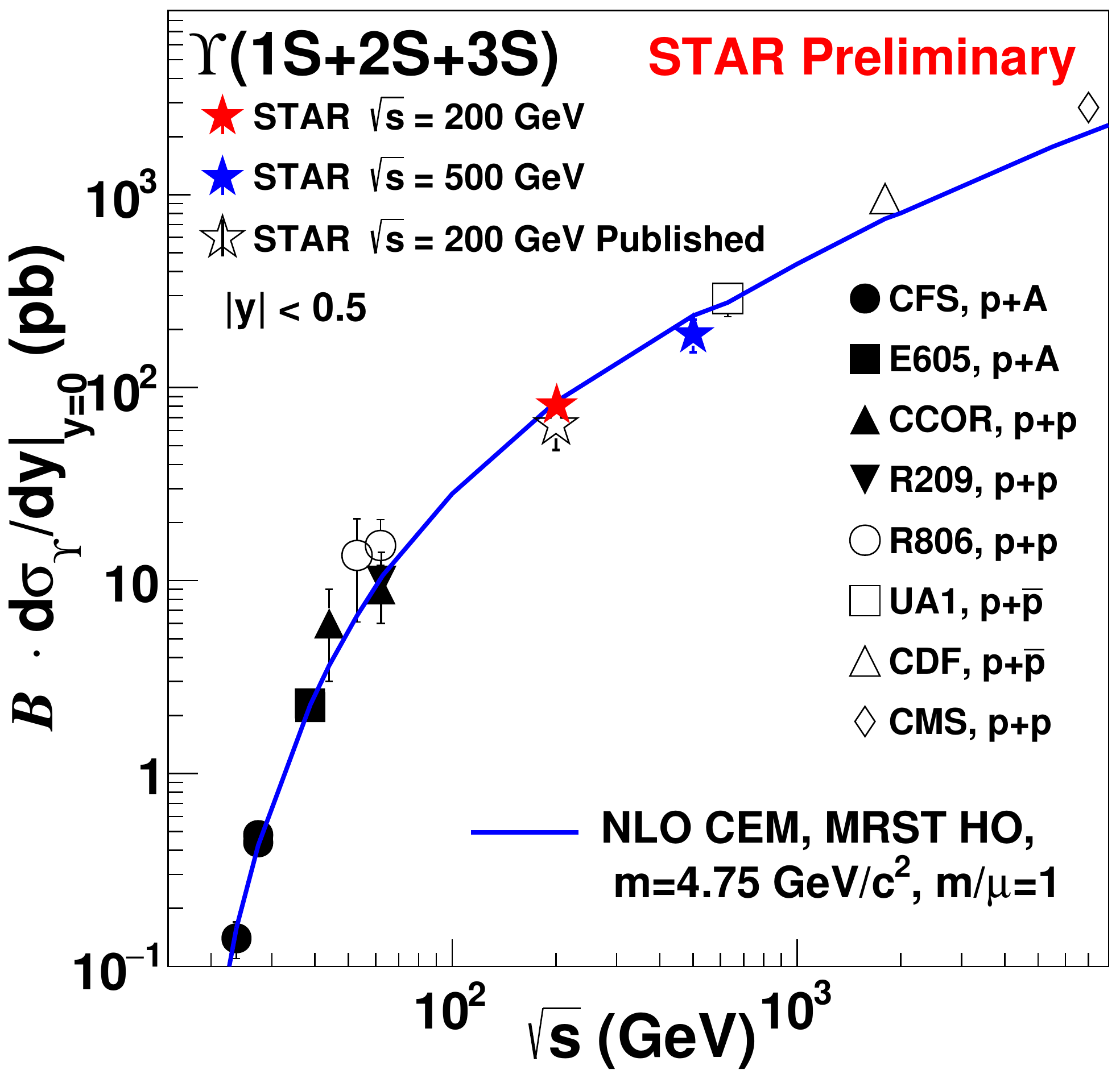}
\caption{ $\Upsilon(1S+2S+3S)$ production cross section per unity rapidity per binary nucleon-nucleon collision at mid-rapidity times $\Upsilon \rightarrow l^+l^-$ branching ratio  in $p + p(\bar{p}, \textrm{A})$ collisions. The figure is taken from~\cite{ZhenLiu_HP2018Proceedings}.}
\label{fig:Y_pp}
\end{figure}

The production cross section of $\Upsilon$ times the di-lepton decay branching ratio is about 3 orders of magnitude lower than that of \Jpsi and is 9 orders of magnitude lower than that of inelastic \pp collision at RHIC. About half billion minimum bias \pp collision events is needed to produce one $\Upsilon \rightarrow l^+l^-$ decay at mid-rapidity. In order to enhance the recorded integral luminosity for $\Upsilon$ study, a special trigger based on the barrel electromagnetic calorimeter is designed and make the measurement of $\Upsilon$ at STAR possible. However, the statistics and momentum resolution are not good enough to separate the 1S, 2S and 3S states in \pp collisions. The 1S, 2S and 3S states are measured together. Figure~\ref{fig:Y_pp} shows the measured $\Upsilon(1S+2S+3S)$ cross section per unit rapidity at mid-rapidity times the $\Upsilon \rightarrow e^+e^-$ branching ratio as a function of center-of-mass energy~\cite{ZhenLiu_HP2018Proceedings}. Theoretical calculation from the Next-to-Leading Order (NLO) Color Evaporation Model (CEM)~\cite{Frawley:2008kk} describes the cross section from the center-of-mass energy of  20 GeV to 7 TeV and the RHIC data follow the world wide trend. PHENIX also measured $\Upsilon(1S+2S+3S)$ at forward rapidity (1.2<|y|<2.2), thanks to the di-muon trigger~\cite{Upsilon_PHENIX_pp_dAu}. The rapidity distribution is obtained by combining the measurements from STAR (|y|<1) and PHENIX (1.2<|y|<2.2). It is narrower than the NLO CEM prediction~\cite{Frawley:2008kk}. 

The binding energy of $\Upsilon(1S)$, $\Upsilon(2S)$ and $\Upsilon(3S)$ are very different. The modification of the production in nuclear medium is expect to be different for different $\Upsilon$ states. It is of particular interest to separate the 1S, 2S and 3S states. Although it is not achievable at current stage in \pp collisions at 200 GeV, but it is possible to separate 1S from 2S+3S or even 1S, 2S and 3S in $p(d)$ + A and A + A collisions due to higher statistics and better momentum resolution thanks to better primary vertex resolution. However, the production cross section of 1S and 2S+3S (or 2S and 3S) state in \pp collisions is needed as a reference to study the different nuclear matter effects of different $\Upsilon$ states. The authors of Ref.~\cite{Upsilon_pp_systematic} performed a systematic study of $\Upsilon$ production in $p + p(\bar{p}, \textrm{A})$ collisions from world wide experiments and developed a way to predict $\Upsilon(2S)/\varUpsilon(1S)$ and $\Upsilon(3S)/\varUpsilon(1S)$  in \pp collisions at given center-of-mass energy. The production cross section of 1S, 2S and 3S in \pp collisions can be obtained by combining the derived ratios and the measurement of $\Upsilon(1S+2S+3S)$ production cross section. The STAR Collaboration also established a method to derive the \pT spectrum of $\Upsilon(1S)$, $\Upsilon(2S)$ and $\Upsilon(3S)$ in \pp at 200 GeV in order to study the \pT dependence of the nuclear modification factor of $\Upsilon$ states. Recently, the STAR managed to measure the \pT spectrum of $\Upsilon(1S)$ and $\Upsilon(2S+3S)$ in \pp collisions at 500 GeV~\cite{Leszek_HQProceedings}. The newly installed iTPC in STAR will improve the mass resolution of $\Upsilon$~\cite{iTPC_MWPC}. The possibility of further improving the mass resolution by changing the working gas of TPC to suppress transverse diffusion is under investigation. 

\section{J/$\psi$ production in medium}

Although the production mechanism in \pp collisions is not fully understood, quarkonium in heavy-ion collisions is one of a few most important probes of QGP. The suppression of \Jpsi production yield in relativistic heavy-ion collisions with respect to the yield in \pp collisions scaled by the number of binary nucleon-nucleon collisions has been proposed as a ``smoking gun'' signature of QGP formation by T. Matsui and H. Satz in 1986~\cite{colorscreen}. The suppression is the result of \Jpsi dissociation due to the screening of the potential between charm quark and anti-charm quark in the deconfined hot, dense medium. \Jpsi production in nucleus-nucleus collisions is extensively studied at CERN SPS since 1986. The pioneer experiment NA38 found that \Jpsi production in S+U collisions is suppressed relative to $p$+U collisions and as a function of the transverse energy $E_T$, which is related to the collision centrality. However, it is found later that the suppression pattern was compatible with the extrapolation of the trend observed in $p$+A collisions and can be accounted for normal nuclear absorption. The NA50 experiment collected high statistics data with $p$ beam with energy of 450 or 400 GeV on target ranging from Be, Al, Cu, Ag, W and Pb. The normal nuclear absorption of \Jpsi production is obtained by systematically studying the $p$+A data~\cite{CNM_NA50}. 

The NA50 experiment also collected data with Pb beam with energy per nucleon of 158 GeV on Pb-target (\sNN = 17.3 GeV) in 1995, 1996, 1998 and 2000. The analysis of these data showed that \Jpsi production, relative to Drell-Yan, is anomalously suppressed with respect to the normal nuclear absorption pattern, which is extracted by extrapolating the \Jpsi suppression in $p$+A collisions as aforementioned~\cite{SPS_deconfinement_Jpsi}. In order to take $p$+A data under the same condition as the A+A data, the NA60 experiment run with proton beam on nuclear targets (Be, Al, Cu, In, W, Pb and U) and In beam on In target with energy per nucleon of 158 GeV~\cite{Jpsi_RAA_NA60}. The nuclear absorption is found to have beam energy dependence. The anomalous \Jpsi suppression is calculated with the new nuclear absorption cross section and also take the difference of the parton distribution function in nucleus (nPDF) and nucleon (shadowing effect) into account. The observed suppression of \Jpsi is compatible with the extrapolation of cold-nuclear-matter (CNM) effects upto $N_{\textrm{part}} \sim 200$. When $N_{\textrm{part}} > 200$, there is an anomalous suppression of upto $\sim 20\%-30\%$ in the most central Pb+Pb collisions. 

\begin{figure*}[!htb]
\includegraphics[width=0.35\hsize]
{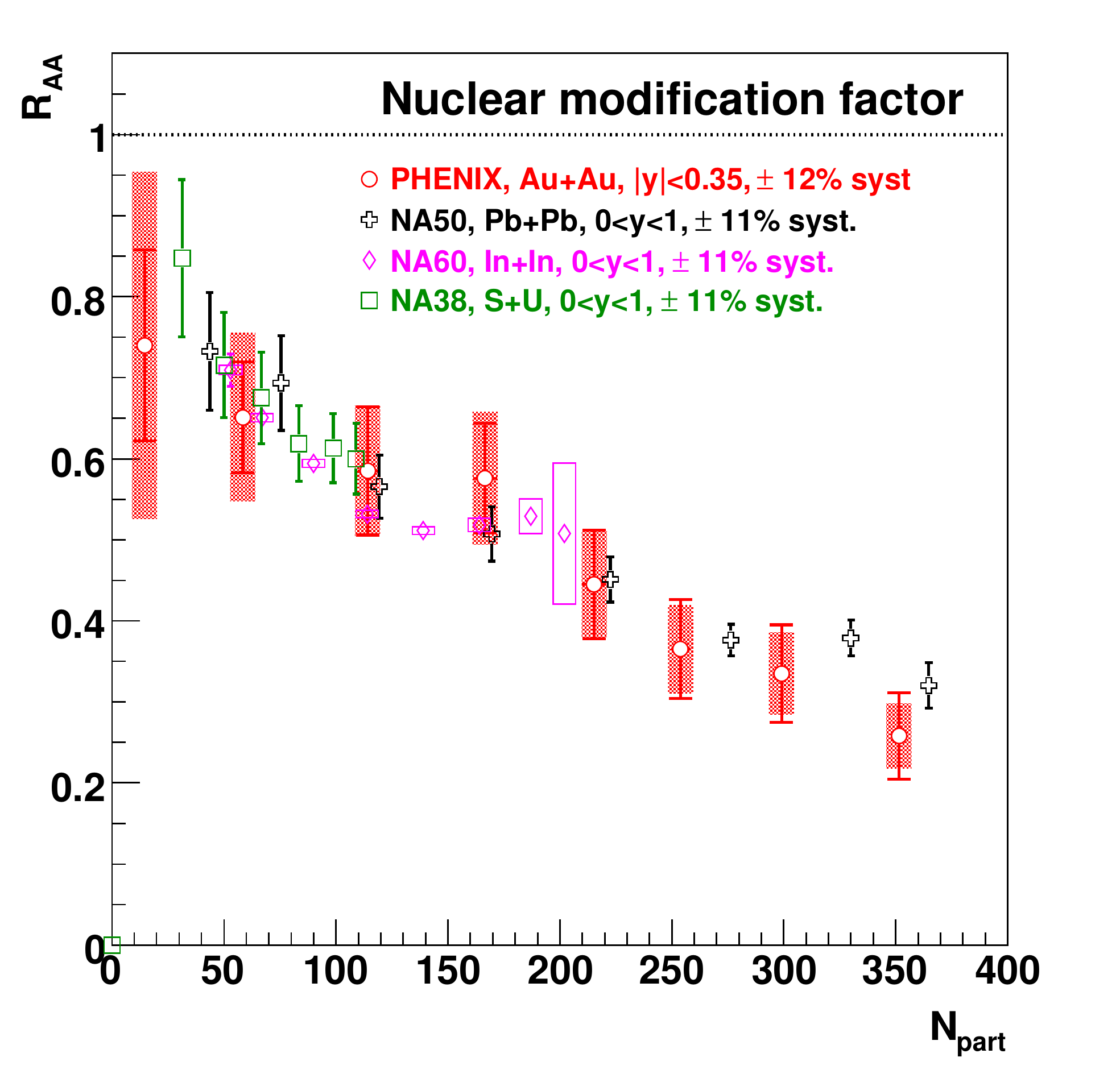}
\includegraphics[width=0.35\hsize]
{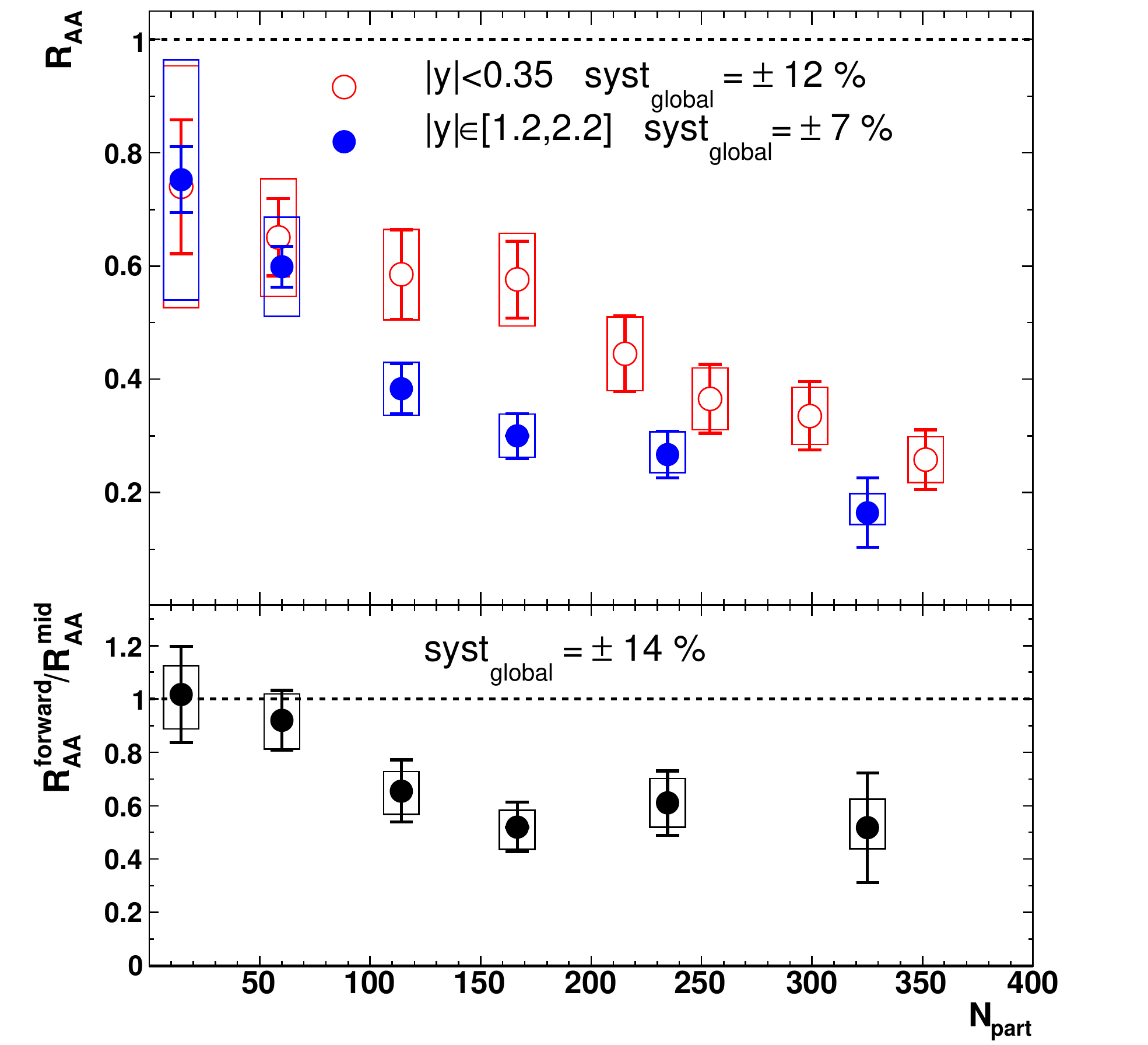}
\caption{ \raa as a function of \npart for inclusive \Jpsi at SPS and RHIC. The panels are taken from~\cite{Kluberg:2009wc,PHENIX_Jpsi_AuAu}.}
\label{fig:Jpsi_PHENIX_AuAu}
\end{figure*}

The PHENIX Collaboration at RHIC measures \Jpsi production in \auau collisions at \sNN = 200 GeV, more than a factor of 10 higher than that of SPS, for rapidities |y|<0.35 and 1.2<|y|<2.2, through the di-electron and di-muon decay channel, respectively.  The left panel of Fig.~\ref{fig:Jpsi_PHENIX_AuAu} shows the nuclear modification factor \raa as a function of $N_{\textrm{part}}$ at mid-rapidity in \auau collisions at \sNN = 200 GeV measured by the PHENIX Collaboration~\cite{PHENIX_Jpsi_AuAu} and in Pb+Pb, In+In and S+U collisions at \sNN = 17.3 GeV measured by the NA50, NA60 and NA38 Collaborations~\cite{Jpsi_NA38, SPS_deconfinement_Jpsi, Jpsi_RAA_NA60}, respectively. Even the center-of-mass energies differ by one order of magnitude, the amount of suppression is very similar at RHIC and SPS energies. This was puzzling as stronger suppression iss expected at RHIC due to high energy density and/or initial temperature. Another puzzle was that the suppression is observed to be much stronger at forward rapidity than at mid-rapidity as shown in the right panel of Fig.~\ref{fig:Jpsi_PHENIX_AuAu}. It was expected that the suppression at mid-rapidity where the energy density is higher to be stronger. An additional \Jpsi production mechanism, (re)combination of charm quark and anti-charm quark, is introduced~\cite{BraunMunzinger:2000px, Andronic:2003zv, Grandchamp:2001pf, Thews:2000rj, Gorenstein:2002hc}.  The idea of the (re)combination mechanism is that when the charm and anti-charm quark not initial produced as a bound state get close enough in space and momentum after transportation in QGP, they may form a bound state such as J/$\psi$. The production yield of \Jpsi is only about 1\% of the total number of charm and anti-charm quark pair in \pp collisions, the (re)combination mechanism may have sizable effect even though the (re)combination probability is small. The probability is proportional to the square of number of charm and anti-charm quark pair produced in one event. The probability is negligible in \pp collisions at RHIC energy. But since the number of charm and anti-charm quark pair is roughly scaled by the number of binary nucleon-nucleon collisions which can reach as large as 1000 in central \auau collisions, the probability is much larger in heavy-ion collisions than in \pp collisions. Unlike the color-screening (or QGP melting) mechanism, the (re)combination mechanism could result in enhancement of \raa in heavy-ion collisions. Since the charm and anti-quark charm pair production cross section increases dramatically with center-of-mass energy, the (re)combination mechanism play more important role in heavy-ion collisions at higher center-of-mass energy. Based on a theoretical calculation~\cite{ccbar_vs_energy_MNR} on charm and anti-charm quark pair production cross section in \pp collisions, we estimated that the number of charm and anti-charm quark pair in a 0-10\% Pb+Pb collisions at SPS energy (\sNN = 17.3 GeV) is $0.13 \pm 0.03$. The number increases to $18 \pm 4$ in 0-10\% \auau collisions at RHIC energy (\sNN = 200) and over 100 in 0-10\% Pb+Pb collisions at LHC energies. The contribution of (re)combination is negligible at SPS but could have sizable effect at RHIC and LHC. The theoretical models (such as transport models) including both QGP melting and (re)combination can explain the \raa observed at RHIC at both mid-rapidity and forward rapidity, as well as at SPS. 

The \Jpsi production in heavy-ion collisions at RHIC and LHC is the interplay of QGP melting, CNM effects as well as (re)combination effects. To study the properties of the QGP via J/$\psi$, we need good understanding of each of the three effects. The CNM effects are usually be studied experimentally in $p$+A collisions or the collisions of light ions where the QGP melting and/or (re)combination effect is unlikely existed at least at RHIC energy. But the separation of QGP melting and (re)combination effects is very difficult. Fortunately these two have very different collision energy, collision system and \pT dependence. A systematically study of \Jpsi production in heavy-ion collisions is helpful to understand the \Jpsi production mechanism in heavy-ion collisions and study the properties of QGP using J/$\psi$.

\subsection{Collision energy dependence}

\begin{figure*}[!htb]
\includegraphics[width=0.38\hsize]
{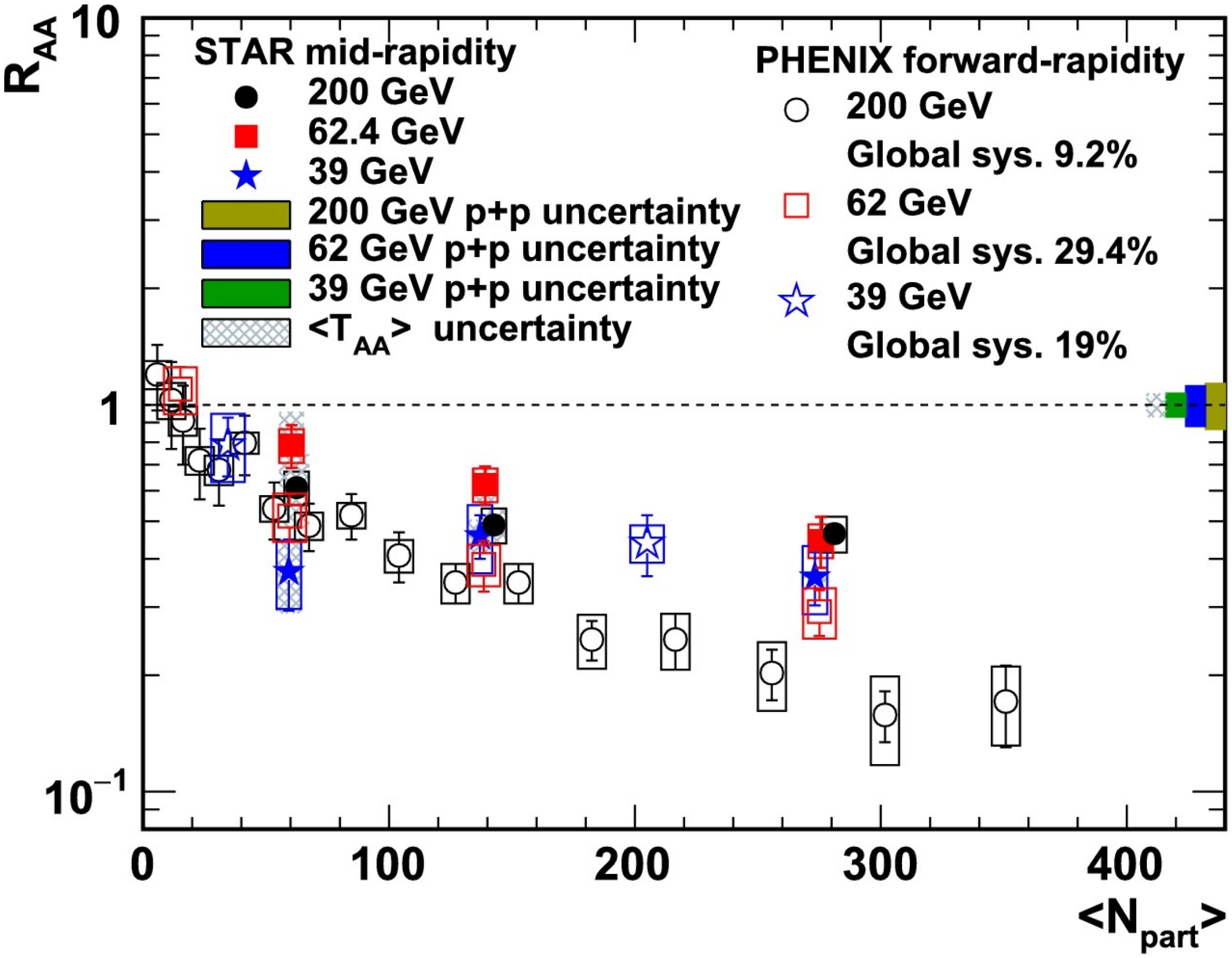}
\includegraphics[width=0.4\hsize]
{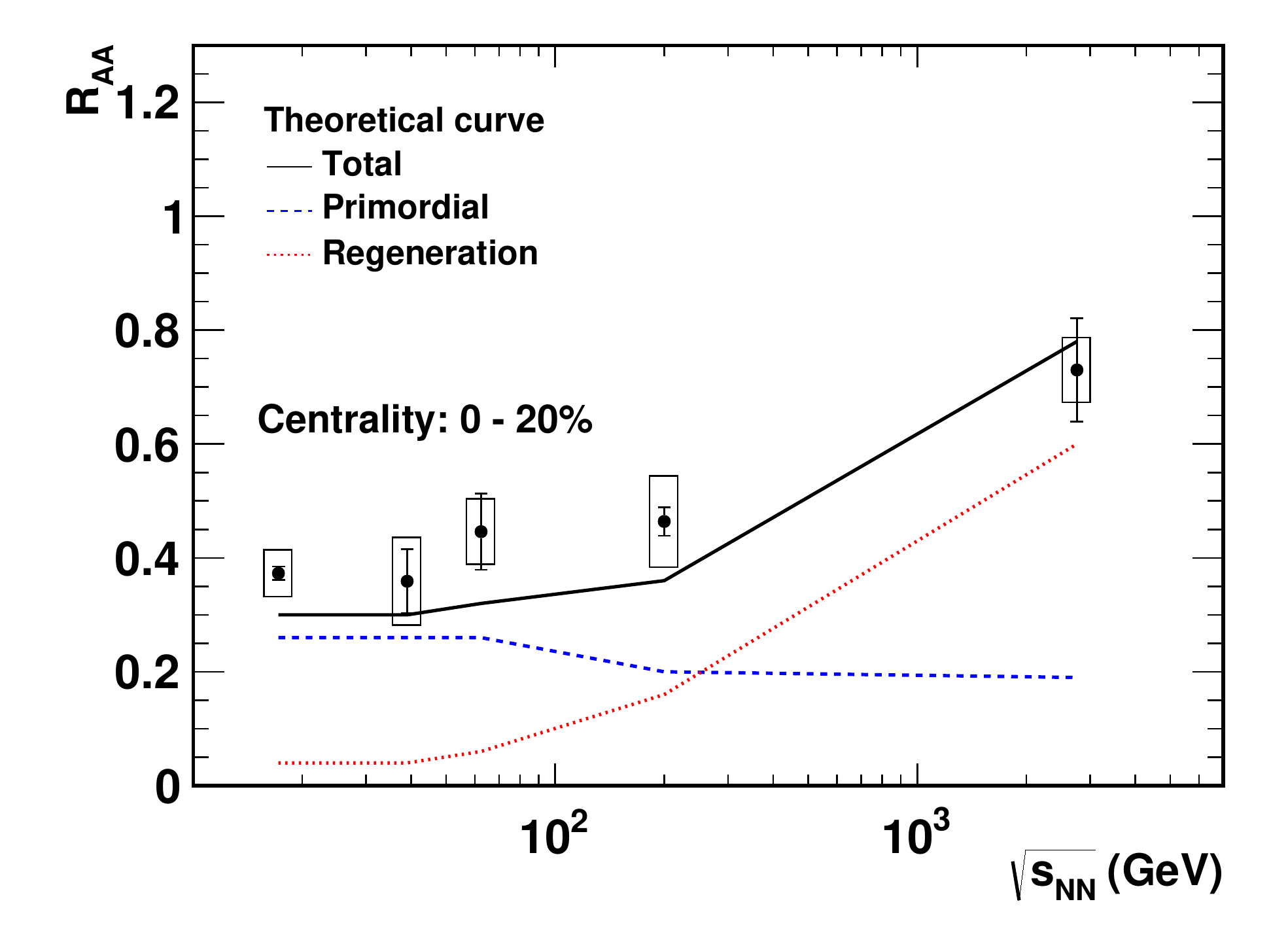}
\caption{ Left: Inclusive \Jpsi \raa as a function of centrality at mid-rapidity and forward rapidity in \auau collisions at 39, 62.4 and 200 GeV. Right:  Inclusive \Jpsi \raa as a function of center-of-mass energy in 0-20\% central heavy-ion collisions (\auau at RHIC and Pb+Pb at SPS and LHC). The figure is taken from~\cite{Jpsi_STAR_BESI}.}
\label{fig:Jpsi_BES}
\end{figure*}

RHIC launched the Beam Energy Scan (BES) program in 2010 to explore the QCD phase diagram. Both STAR and PHENIX collected data in \auau collisions at 62.4 and 39 GeV in 2010 and at  27 and 19 GeV in 2011 in the phase-I of the BES program (BES-I). These center-of-mass beam energies fill the large gap between SPS energy and RHIC top energy. These BES data can be used to study the evolution of the CNM effects, QGP melting and (re)combination from SPS to RHIC. The production cross section of \Jpsi decreases dramatically with decreasing center-of-mass energy and the luminosity of RHIC also decreases quickly with decreasing beam energy. We were only able to measure \Jpsi production in the collisions at 39 and 62.4 GeV. To obtain \raa at these two energies, \Jpsi cross section in \pp collisions is needed. There are several measurements from $p$ + A fixed-target experiments and \pp collider experiment at the Intersection Storage Ring (ISR) near these two energies performed in last century. But unfortunately some data at mid-rapidity are found to be not consistent with each other. At mid-rapidity, STAR uses \Jpsi production cross section derived from world wide experimental data~\cite{Jpsi_pp_systematic} to calculation \Jpsi \raa in \auau collisions at 39 and 62.4 GeV~\cite{Jpsi_STAR_BESI}. At forward-rapidity, PHENIX used the reference data derived based on the data at Fermilab fixed-target experiment, ISR collider experiment and CEM model calculations~\cite{Jpsi_PHENIX_BESI}. The left panel of Fig.~\ref{fig:Jpsi_BES} shows the inclusive \Jpsi \raa as a function on $N_{\textrm{part}}$ in \auau collisions at 39 and 62.4 GeV at both mid- and forward rapidity and compared to that at 200 GeV. It is consistent with no suppression in peripheral collisions but exhibit strong suppression towards central collisions. At mid-rapidity, no significant energy dependence is observed within uncertainties from SPS (\sNN = 17.3 GeV) to RHIC top energy (\sNN =200 GeV). However, at forward rapidity, the suppression seems less in \auau collisions at \sNN = 39 and 62.4 GeV than at 200 GeV. The rapidity dependence of the suppression can be obtained by comparing the STAR data at mid-rapidity (|y|<1) and the PHENIX data at forward rapidity (1.2<|y|<2.2). The suppression is much stronger at forward rapidity than at mid-rapidity in \auau collisions at 200 GeV. However, at 39 and 62.4 GeV, the suppression of \Jpsi shows no significant rapidity dependence within uncertainties. This could be due to the energy and rapidity dependence of the (re)combination contribution, which is larger at higher collision energy and mid-rapidity. In order to understand the collision energy dependence, the inclusive \Jpsi \raa in central heavy-ion collisions is plotted as a function of center-of-mass energy. The \raa is flat at $\sim 0.4$ from \sNN = 17.3 to 200 GeV then dramatically increase to $>0.6$ at LHC. The curves shown in Fig.~\ref{fig:Jpsi_BES} are the theoretical calculations from a transport model~\cite{XingboRalf2010} which implemented CNM effects, QGP melting and (re)combination. The dashed curve represent the \raa of primordially produced J/$\psi$, whose production yield suffers from the QGP melting and CNM effects. The \raa is fairly flat from \sNN = 17.3 to 62 GeV then decreases with center-of-mass energy. The trend is the result of counter-balance of the CNM effects and QGP melting. The nuclear absorption cross section decreases with increasing center-of-mass energy, resulting in increasing \raa with increasing center-of-mass energy. While the QGP melting results in decreasing trend due to the increasing energy density. The QGP melting plays significant role at RHIC top energy. The \raa with only CNM effects is estimated to about 0.6 in central \auau collisions at \sNN = 200 GeV~\cite{XingboRalf2010}. The QGP melting brings down the \raa from 0.6 to 0.2. The dotted curve in Fig.~\ref{fig:Jpsi_BES} shows the \raa for the \Jpsi produced from (re)combination. It is negligible from SPS energy to RHIC BES at \sNN $<\sim 50$ GeV and starts to play a role at higher center-of-mass energy. At RHIC top energy, the contribution of \Jpsi from (re)combination is comparable to the survived primordial J/$\psi$ and becomes dominant at LHC. The solid curve in Fig~\ref{fig:Jpsi_BES} shows the sum of the two components. It can describe the inclusive \Jpsi \raa in central heavy-ion collisions from SPS (CNM effects domain) to LHC ((re)combination domain) within uncertainties. 

It is particularly interesting to study the \Jpsi suppression in heavy-ion collisions at center-of-mass energy around 50 GeV where the (re)combination contribution remains negligible as SPS, but the energy density is higher and the expected CNM effects such as nuclear absorption is smaller than at SPS. The STAR experiment has collected a large data sample of \auau collisions at \sNN = 54 GeV. The statistics is one order of magnitude higher than that of the 62 GeV data sample. This will allow more precise and differential measurement of the \Jpsi suppression at \sNN around 50 GeV. A fixed-target experiment using the LHC beams (AFTER@LHC~\cite{AFTER_2017} or current LHC experiments in fixed-target mode) will be able to collect high statistics heavy-ion collision data at \sNN = 72 GeV with Pb beam of 2.76 TeV per nucleon. The $p$+A data can also be taken at the same energy with Pb beam on proton target to study the CNM effects. 

\subsection{Collision system dependence}

\begin{figure*}[!htb]
\begin{minipage}[b]{0.45\linewidth}
\includegraphics[width=0.75\hsize]
{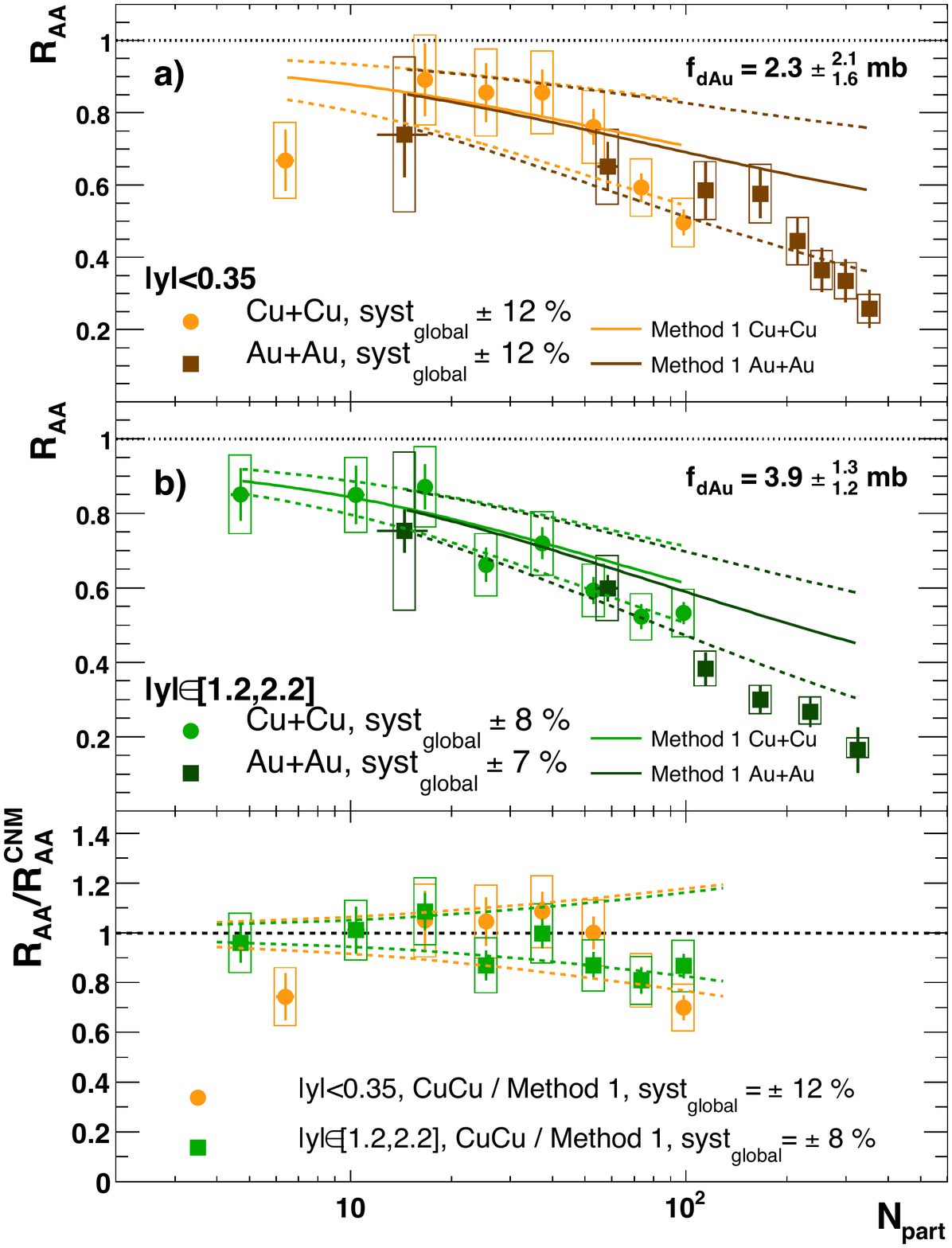}
\caption{Inclusive \Jpsi \raa as a function of $N_{\textrm{part}}$ at mid-rapidity (a) and forward/backward rapidity (b) and the ratio of \raa to the expectation from CNM effects (c) in Cu+Cu collisions at 200 GeV. The figure is taken from~\cite{Jpsi_PHENIX_CuCu}.}
\label{fig:Jpsi_CuCu}
\end{minipage}
\hspace{0.5cm}
\begin{minipage}[b]{0.45\linewidth}
\includegraphics[width=0.75\hsize]
{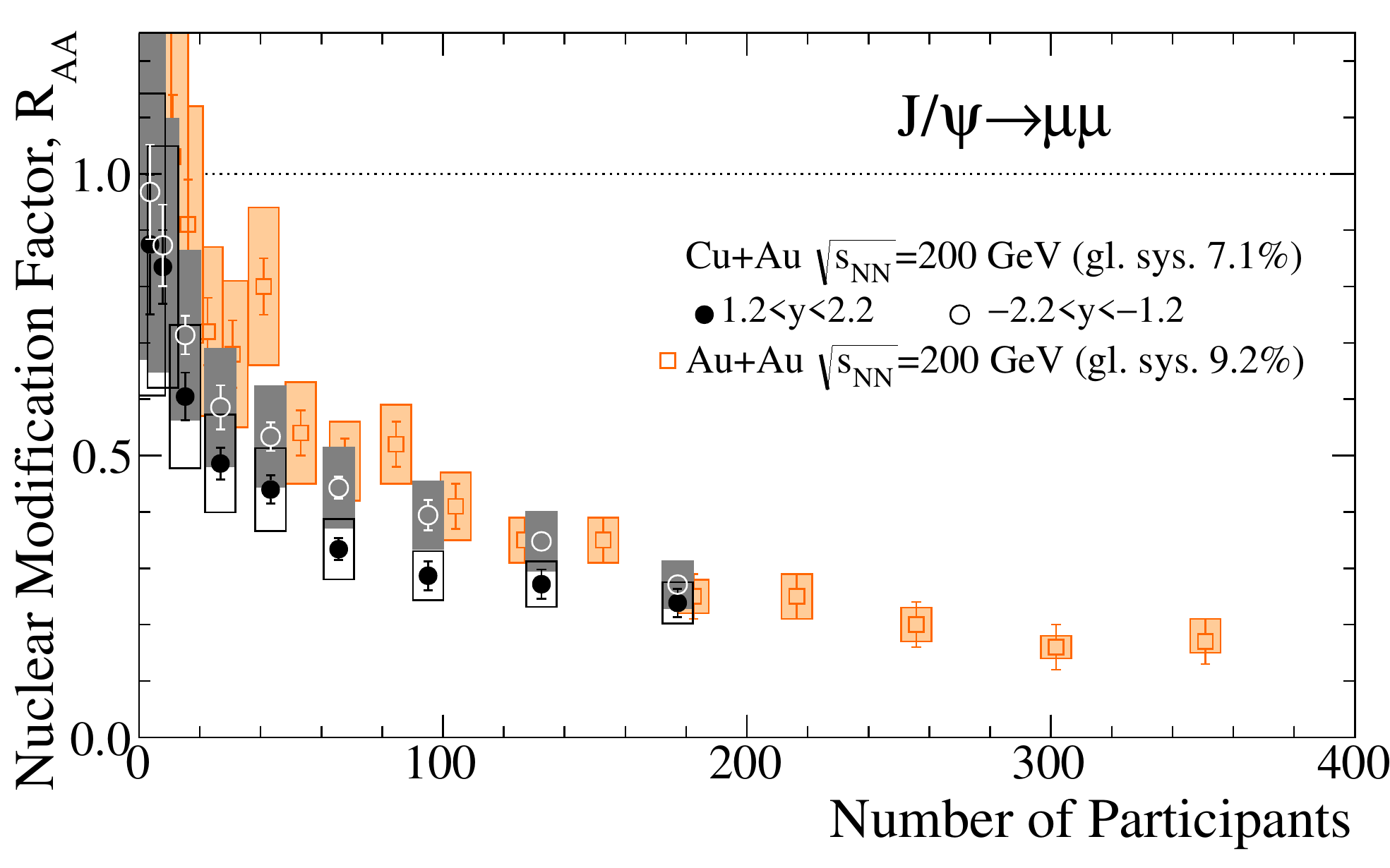}
\includegraphics[width=0.75\hsize,]
{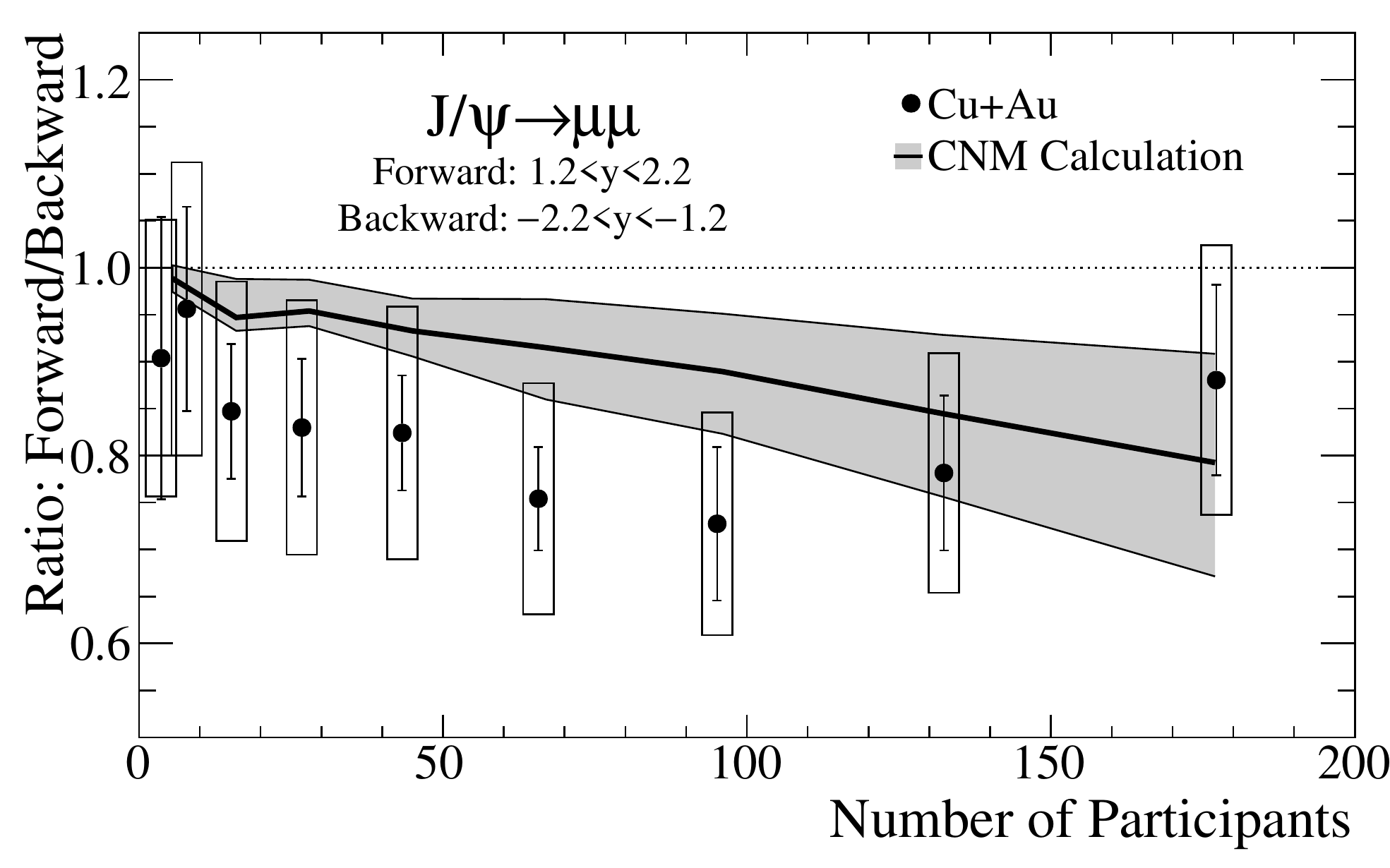}
\caption{ Top: Inclusive \Jpsi \raa as a function of $N_{\textrm{part}}$ at forward (1.2<y<2.2) and backward rapidity (-2.2<y<-1.2) in Cu+Au collisions at 200 GeV. The data in \auau collisions at 200 GeV are also shown for comparison. Bottom: The ratio of \raa in forward and backward rapidity in Cu+Au collisions. The figure is taken from~\cite{Jpsi_PHENIX_CuAu}.}
\label{fig:Jpsi_CuAu}
\end{minipage}
\end{figure*}

The measurements in Pb+Pb collisions at SPS shows anomalous \Jpsi suppression from semi-peripheral to central Pb+Pb collisions ($N_{\textrm{part}}>\sim 100$) at \sNN = 17.3 GeV. At RHIC energy, the energy density is expected to reach the energy density required for QGP formation based on Lattice QCD calculations at $N_{\textrm{part}}$ below 100. However, the \auau data in this QGP transition threshold region is limited. In order to provide crucial information in such an important region, PHENIX has measured \Jpsi suppression in Cu+Cu collisions at \sNN = 200 GeV. Figure~\ref{fig:Jpsi_CuCu}.a and \ref{fig:Jpsi_CuCu}.b shows the inclusive \Jpsi \raa as a function of \npart in \cucu collisions at \sNN = 200 GeV for mid-rapidity (|y|<0.35) and forward/backward rapidity (1.2<|y|<2.2), respectively. The results in \auau collisions at the same energy is also shown for comparison. The \raa in \cucu and \auau are consistent with other within uncertainties at comparable $N_{\textrm{part}}$. The \cucu data covers the \npart range upto 100 with much finer bins than in \auau collisions. The observed suppression has large contribution from CNM effects. In order to extract the possible QGP melting in \cucu collisions, the CNM effects are estimated by projecting the suppression measurement in $d$+Au collision at the same energy with nPDF and nuclear absorption. The nPDF is taken from the shadowing model EKS98~\cite{EKS98} or nDSg~\cite{nDSg}. The nuclear absorption cross section are optimized separately for mid-rapidity and forward rapidity from the fit to $d$+Au data. The estimated CNM effects with EKS98 (method 1) is shown in Fig.~\ref{fig:Jpsi_CuCu}.a and \ref{fig:Jpsi_CuCu}.b as solid lines. The dashed lines depict the calculation by varying the nuclear absorption cross section by $1\sigma$. The predicted CNM \raa show almost no difference in \cucu and \auau collisions at the same $N_{\textrm{part}}$. Figure~\ref{fig:Jpsi_CuCu}.c shows the measured \raa in \cucu collisions divided by the predicted CNM \raa with EKS98 parameterization at both mid- and forward/backward rapidity. At \npart $<\sim 50$, the measured \raa in \cucu collisions is seen to be consistent with the CNM projection within about 15\% uncertainties. At \npart above 50, the centroid of the measured ratios at both mid-rapdity and forward rapidity are smaller than unity. But no strong conclusion can be drawn with the larger uncertainties. More precise measurement in $p(d)$+A collisions and better understanding of how to project the CNM effects in $p(d)$+A collisions to A+A collisions is needed. Nevertheless, the CNM effects dominate \Jpsi production in \cucu collisions and peripheral and semi-peripheral \auau collisions.

RHIC collided Cu+Au collisions at 200 GeV in 2012. The rapidity dependence of \Jpsi suppression in the asymmetric collision system may provide key insights on the balance of CNM effects and hot nuclear matter effects. The parton distribution functions are more strongly modified in heavier Au nucleus than in light nucleus. At forward rapidity (Cu-going  direction), \Jpsi probes gluons at lower Bjorken $x$ in the Au nucleus while higher $x$ in the Cu nucleus. This is reversed at backward rapidity. The shadowing effects are expected to be stronger at forward rapidity than at backward rapidity. On the other hand, the \Jpsi produced at forward rapidity has a large rapidity relative to the Au nucleus thus have shorter proper time. This will result in less nuclear absorption of \Jpsi or energy loss at forward rapidity. Furthermore, the energy density and hadron multiplicity are also asymmetric in Cu+Au collisions. It is higher at backward rapidity (Au-going direction). The asymmetric energy density and hadron multiplicity may result in different CNM and hot matter effects. The breakup of \Jpsi by comovers depends on the density of comovers and is expected to be stronger at backward rapidity than at forward rapidity. The asymmetric hot nuclear matter effects are not so straightforward. The QGP melting effect is stronger at the backward rapidity thus results in smaller \raa at backward rapidity. However, the (re)combination effect is also stronger at the backward rapidity thus results in larger \raa at backward rapidity. 

The upper panel of Fig.~\ref{fig:Jpsi_CuAu} shows the inclusive \Jpsi \raa as a function of \npart at forward (1.2<y<2.2) and backward (-2.2<y<-1.2) rapidity in Cu+Au collisions at \sNN = 200 GeV. The forward and backward rapidity combined \auau results at the same center-of-mass energy is also shown for comparison. The \raa in Cu+Au collisions at backward rapidity (Au-going direction) is similar to that for \auau collisions at similar $N_{\textrm{part}}$. While \raa in Cu+Au collisions at forward rapidity (Cu-going direction) is systematically lower. The difference between forward rapidity and backward rapidity is more clear in the bottom panel of Fig.~\ref{fig:Jpsi_CuAu} in which the ratio of \raa at forward rapidity to that at backward rapidity is shown. The ratios are 20\%-30\% lower than unity and no significant \npart dependence. The grey band depicts the prediction from a simple Glauber model incorporating gluon distribution function modification from the EPS09~\cite{EPS09} parameterization, and a rapidity-independent effective $c\bar{c}$ breakup cross section of 4 mb to account for the nuclear absorption effect. The expected shadowing difference has the right sign as the data and the level of the difference is comparable with the data within uncertainties but systematically smaller than the data especially in peripheral collisions. The QGP melting are expected to result in the forward-to-backward ratio above unity. However, the expectation from (re)combination effect has the same sign as the data and decrease from peripheral collisions. It seems that both QGP melting and (re)combination can not explain the possible difference between the data and the estimated \raa from gluon shadowing effect.

\begin{figure}[!htb]
\includegraphics[width=0.75\hsize]
{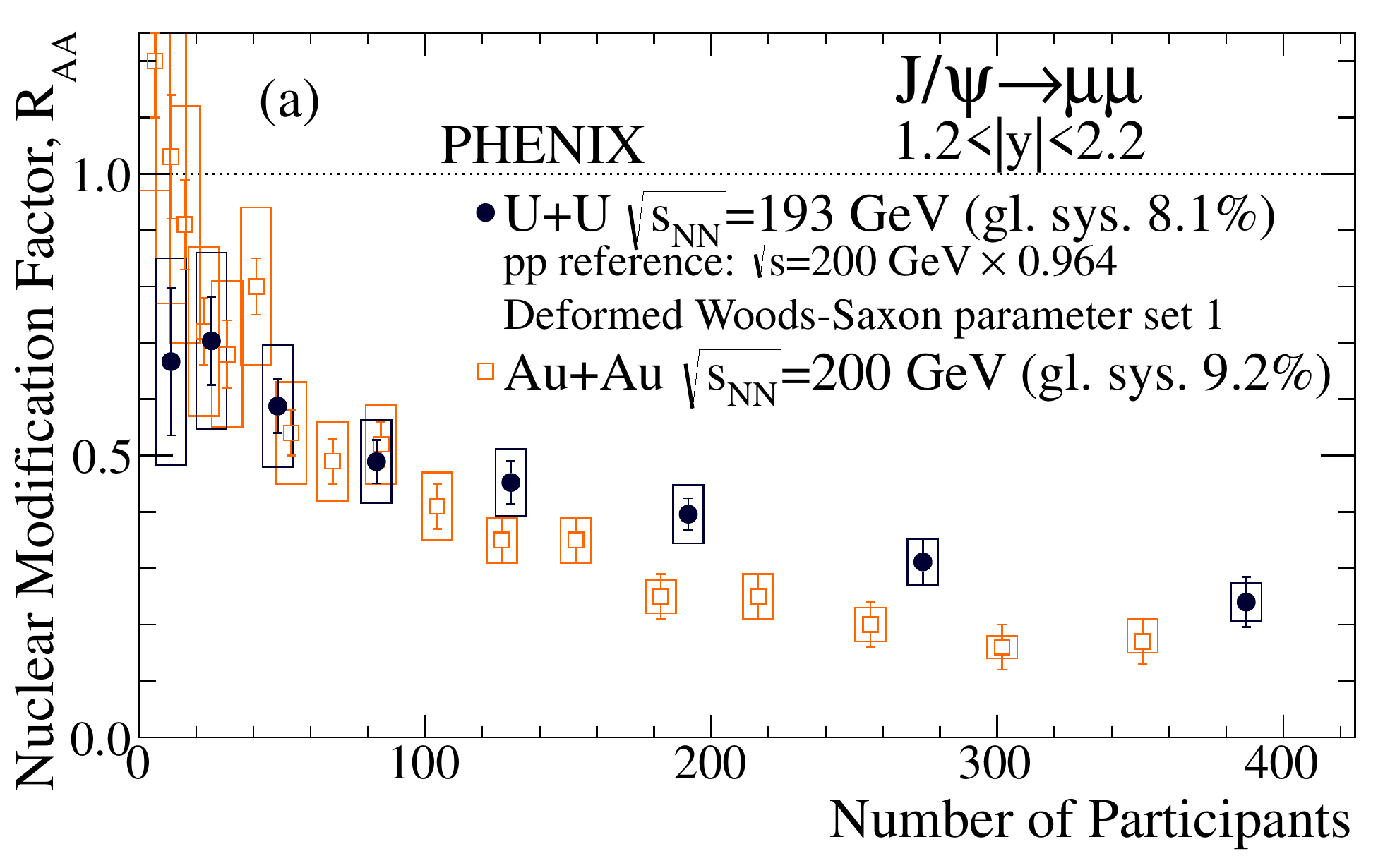}
\includegraphics[width=0.75\hsize]
{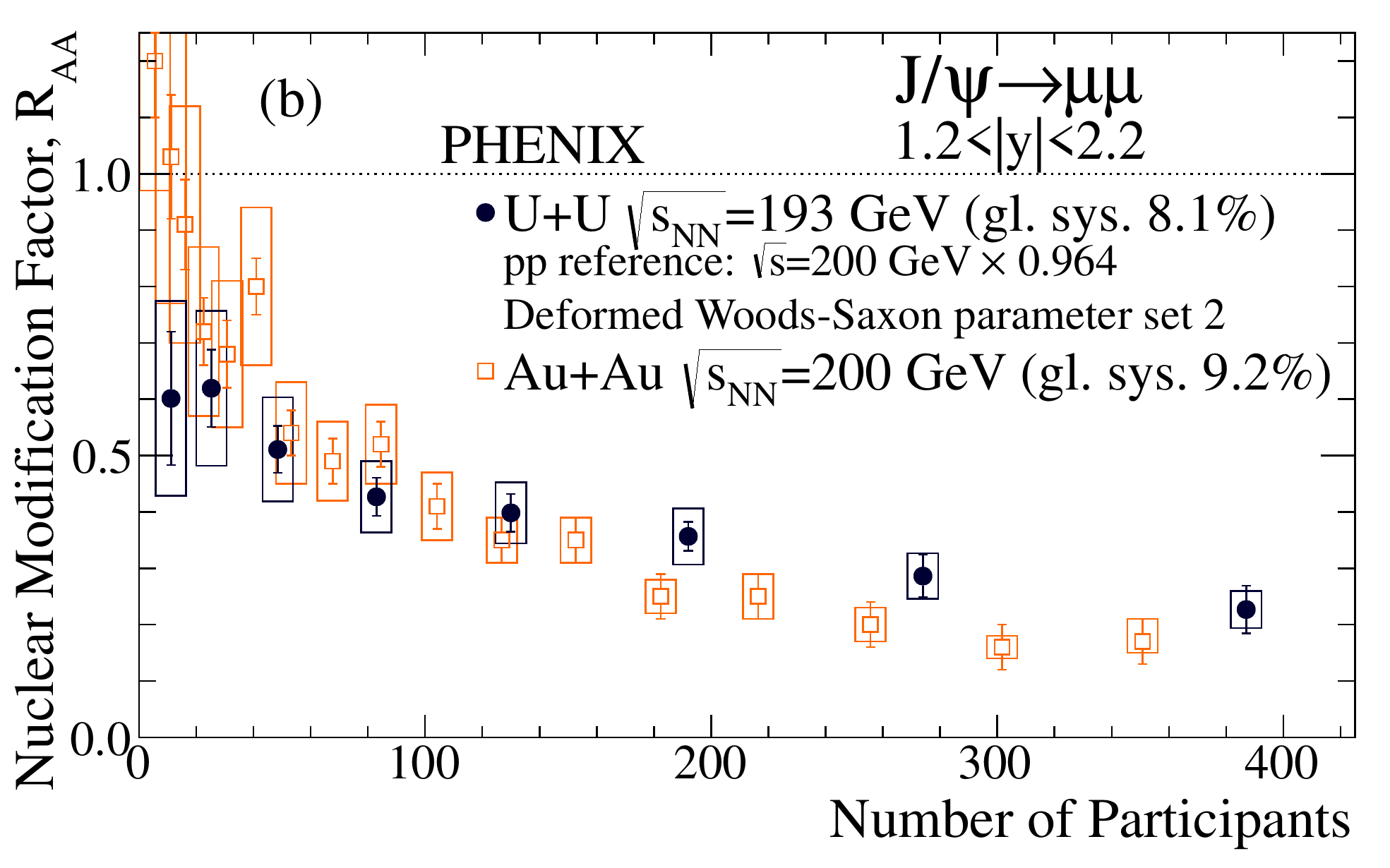}
\caption{Inclusive \Jpsi \raa as a function of $N_{\textrm{part}}$ at forward rapidity (1.2<|y|<2.2) in U+U collisions at \sNN = 193 GeV and compared to \auau collisions at \sNN = 200 GeV. The figure is taken from~\cite{Jpsi_PHENIX_UU}.}
\label{fig:Jpsi_UU}
\end{figure}

The energy density or particle multiplicity dependence of \Jpsi suppression can also be studied in U+U collisions. The energy density in U+U collisions is about 20\% higher than that in \auau collisions with similar number of participants. Figure~\ref{fig:Jpsi_UU} shows the inclusive \Jpsi \raa as a function of \npart in U+U collisions at 193 GeV at forward rapidity and compared to that in \auau collisions~\cite{Jpsi_PHENIX_UU}. The U+U data were taken in 2012. Unlike Au nucleus, U nucleus is deformed and the shape is not well understood. The number of participants and number of binary nucleon-nucleon collisions in U+U collisions depends on the shape of the U nucleus. The U+U results shown in the upper and lower panel of Fig.~\ref{fig:Jpsi_UU} are from two parameterization of deformed Woods-Saxon distribution for U (set 1~\cite{Masui:2009qk} and set 2~\cite{Shou:2014eya}). The parameterization of set 2 has smaller surface diffuseness, resulting in a notably more compact nucleus (and larger number of binary nucleon-nucleon collisions). 

From both parameterizations, the observed \raa in U+U collisions is similar as in \auau collisions with the same number of participants in peripheral and semi-peripheral collisions, but exhibit less suppression than in \auau collisions in central collisions. The CNM effects due to shadowing are expected to be similar in \auau and U+U collisions. The difference in \auau and U+U are likely due to hot nuclear matter effects. The increase of \raa from \auau collisions to U+U collisions is consistent with a picture in which the enhancement due to (re)combination becomes more important than the suppression due to QGP melting. 

\subsection{Transverse momentum dependence}

\begin{figure*}[!htb]
\begin{minipage}[b]{0.69\linewidth}
\includegraphics[width=0.99\hsize]
{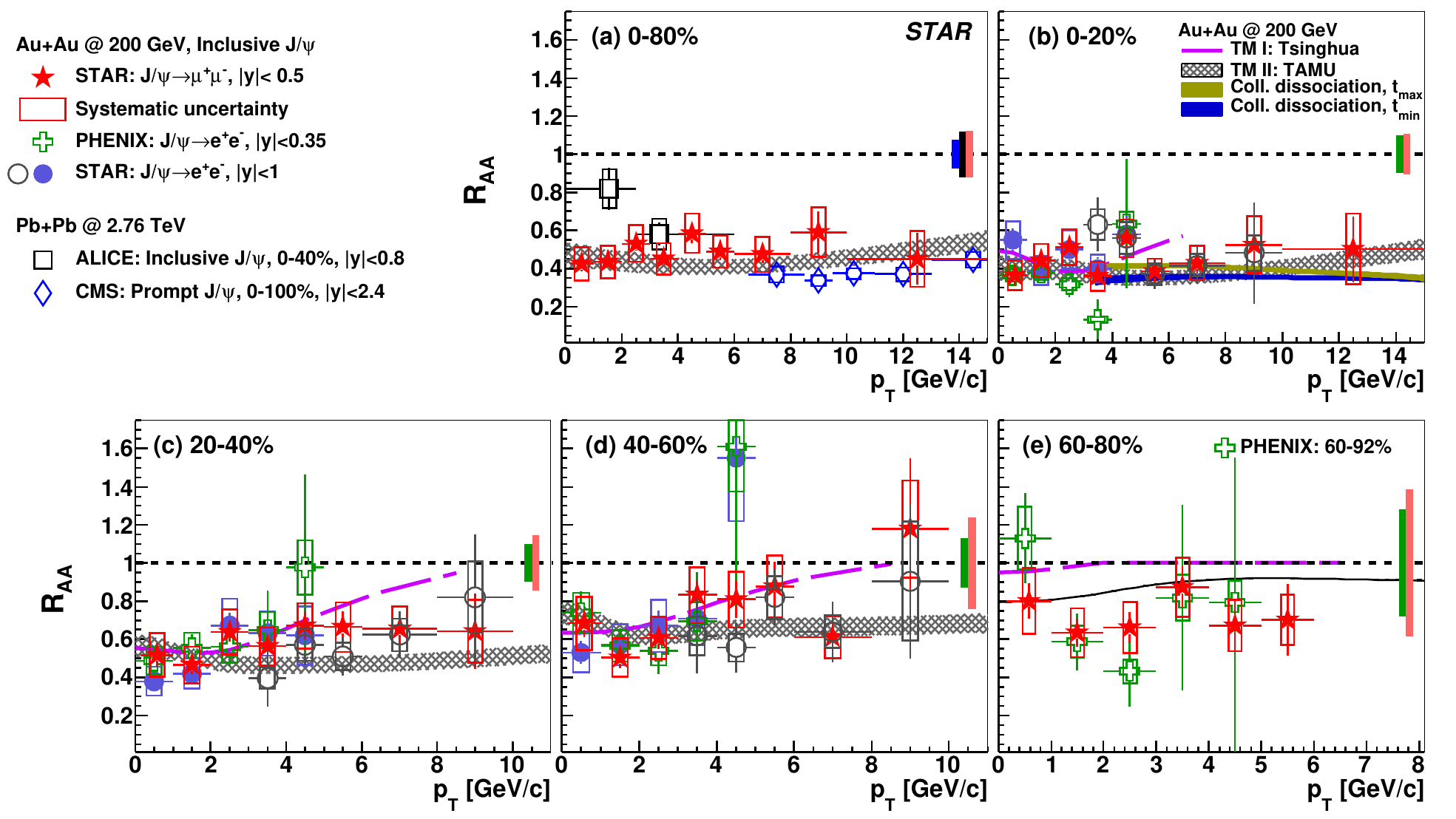}
\caption{Inclusive \Jpsi \raa as a function of \pT in \auau collisions at \sNN = 200 through di-electron and di-muon decay channels. The results from Pb+Pb collisions at \sNN = 2.76 GeV are also shown for comparison. The curves and shaded bands depict theoretical calculations. See text for detail descriptions. The figure is taken from~\cite{Jpsi_AuAu_STAR_PLB2019}.}
\label{fig:Jpsi_vs_pt_STAR}
\end{minipage}
\hspace{0.3 cm}
\begin{minipage}[b]{0.28\linewidth}
\includegraphics[width=0.96\hsize]
{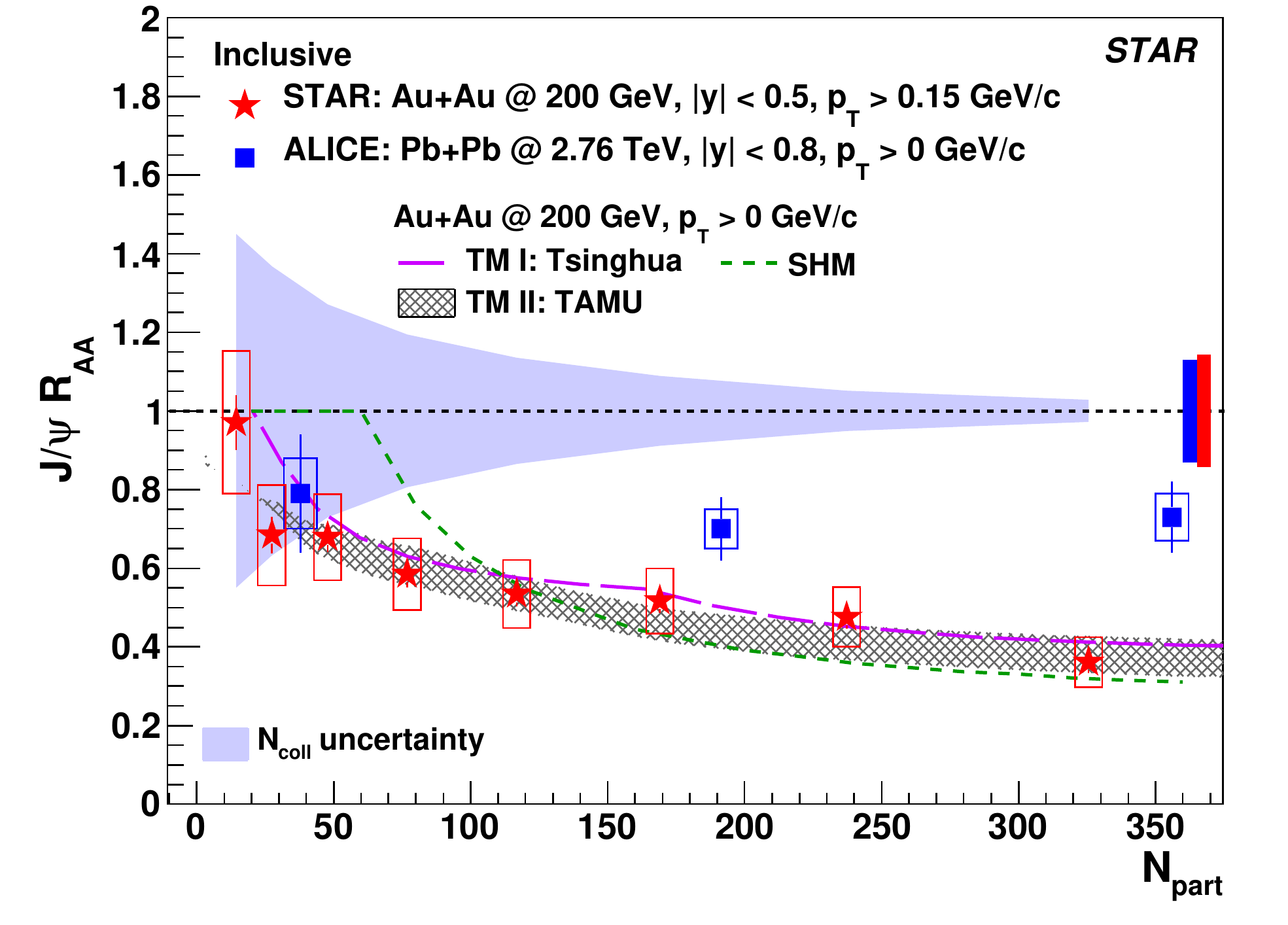}
\includegraphics[width=0.96\hsize,]
{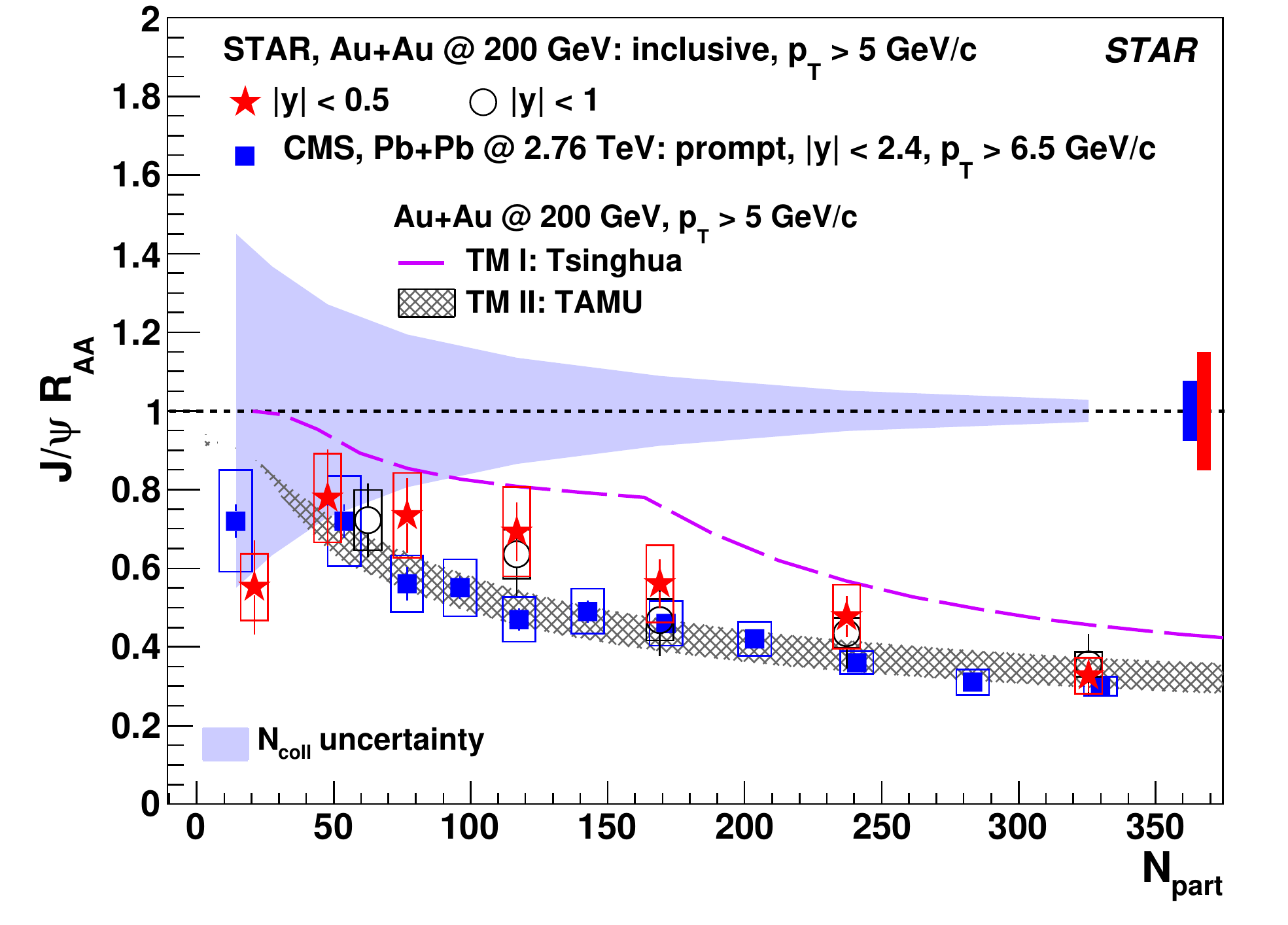}
\caption{Inclusive \Jpsi \raa as a function of $N_{\textrm{part}}$  in Au+Au collisions at 200 GeV for low-\pT and high-\pT  J/$\psi$. The figure is taken from~\cite{Jpsi_AuAu_STAR_PLB2019}.}
\label{fig:Jpsi_vs_Npart_STAR}
\end{minipage}
\end{figure*}

The QGP melting, (re)combination and CNM effects not only have collision energy and collision system dependence, but also depend on the transverse momentum of J/$\psi$. The \raa from CNM effects usually exhibit an increasing trend as a function of \Jpsi $p_T$. The E866~\cite{CNM_E866} and HERA-B~\cite{CNM_HERA-B} experiments found \Jpsi suppression factor $\alpha$, which is obtained by assuming the cross section dependence on nuclear mass, A, to be of the form $\sigma_A = \sigma_N \times A^\alpha$, in fixed-target $p$+A collisions has a clear increasing trend as a function of $p_T$ and cross unity at \pT around $2-3$ GeV/$c$. The increasing trend is usually attributed to multiple scattering of the incident parton before the hard scattering and of the nascent $c\bar{c}$ in the final state. This effect is sometimes also referred to the Cronin effect. At RHIC, the PHENIX Collaboration published \Jpsi suppression as a function of \pT in $d$+Au collisions at \sNN = 200 GeV at both mid- and forward/backward rapidity~\cite{Jpsi_dAu_PHENIX2013} and recently submitted the results in $p$+Al, $p$+Au and $^3$He+Au collisions at \sNN = 200 GeV for publication~\cite{PHENIX_Jpsi_pAlpAuHeAu}. The STAR Collaboration also measured the \pT dependence of \Jpsi suppression in $p$+Au collisions at \sNN = 200 GeV~\cite{Jpsi_pAu_STARQM2017}. In all the collision systems with Au, the suppression of inclusive \Jpsi also shows an increasing trend. The suppression is on the level of about 30\% at low-\pT but consistent with no suppression at \pT above $3-4$ GeV/$c$. A transport model~\cite{XingboRalf2010} predicts \raa of about 0.4 at \pT around zero and lies on unity at \pT from 4.5 to 10 GeV/$c$ at mid-rapidity in 0-20\% central \auau collisions at \sNN = 200 GeV with CNM effects including nuclear absorption and feeddown contribution from $B$.

The (re)combination effect is expected to decrease with $p_T$. This is mainly because the yield of \Jpsi from (re)combination is approximately proportional to the square of the number of charm quarks, which falls fast with $p_T$. The transport models ~\cite{XingboRalf2010, Zhuang2009} show that the contribution of (re)combination is comparable with the primordial \Jpsi at \pT below 1 GeV/$c$ and is negligible at \pT above 5 GeV/$c$. The \pT dependence of QGP melting is not well understand. The formation time effect predicts increasing trend because \Jpsi with higher \pT is more likely to form outside of the medium and less affected by the hot, dense medium. However, the dissociation temperature of \Jpsi may depend on the relative velocity of \Jpsi and the medium and its \pT dependence is model dependent. \Jpsi with higher \pT may have higher or lower dissociation temperature in different models~\cite{Td_velocity_Zhuang, adscft}. A detailed differential measurement of \Jpsi suppression over a broad kinematic range can shed new lights on \Jpsi production mechanism in heavy-ion collisions and the properties of QGP.

The STAR Collaboration attempted to extend the \Jpsi measurement in heavy-ion collisions to \pT beyond 5 GeV/$c$ since 2006~\cite{Jpsi_highpT_PRC2009,Jpsi_STARpp2009HT}. The \Jpsi production at high $p_T$, where the CNM effects and (re)combination is negligible, is found to be consistent with no suppression in \cucu and peripheral \auau collisions but significantly suppressed in (semi-)central \auau collisions at \sNN = 200 GeV. But limited by statistics, no firm conclusion was drawn. 

In 2014 and 2016, the STAR Collaboration collected large samples of \auau collisions at \sNN = 200 GeV utilizing the Muon Telescope Detector (MTD). The MTD detector is designed to trigger on and identify muons and is completed in early 2014. Compare to the previous measurement at mid-rapidity through the di-electron channel~\cite{Jpsi_STAR_BESI, Jpsi_STARpp2009HT, Jpsi_lowpT_CuCuAuAu_PRC2014}, the new data allows to extend the kinematic reach towards high \pT with better precision. Figure~\ref{fig:Jpsi_vs_pt_STAR} shows the inclusive \Jpsi \raa as a function of \pT in \auau collisions at \sNN = 200 GeV from the data taken in 2014. The filled stars represent the new results from STAR through di-muon channel and the open and filled circles depict the previous results through di-electron channel. The results at low \pT from PHENIX are also shown as hollow crosses~\cite{PHENIX_Jpsi_AuAu}. The new results are consistent with previous results in the overlapping kinematic range, but have better precision and covers a wider kinematic range. Within the uncertainties, \Jpsi suppression shows little \pT dependence from $p_T \sim 0$ upto 14 GeV/$c$. For comparison, \Jpsi suppression at mid-rapidity in Pb+Pb collisions at \sNN = 2.76 TeV measured by the ALICE~\cite{Jpsi_ALICE_JHEP2015} and CMS~\cite{Jpsi_CMS_EPJC2017} Collaborations are shown in panel (a). The \pT dependence at LHC and RHIC are significantly different. \Jpsi \raa at RHIC is lower than at LHC at low \pT, but systematical higher than at LHC at high $p_T$. The difference of the \raa is due to different (re)combination contribution in \Jpsi production and initial temperature and lifetime of the QGP in heavy-ion collisions at different collision energy. The shaded bands and dashed lines represent two transport model calculations for \auau collisions at \sNN = 200 GeV from Tsinghua (TM1)~\cite{Zhuang2009} and TAMU (TM2)~\cite{XingboRalf2010} groups.  The CNM effects, QGP meting and (re)combination are taken into account in both models, although the detailed treatments are different. The TM1 model describes the data reasonable well at low $p_T$, but show a stepper increasing trend towards high \pT than data. The TM2 model has better \pT dependence as the data, but the absolute values are systematically lower than the data from intermediate to high $p_T$. The two solid bands covering $3.5<p_T<15$ GeV/$c$ in panel (b) show theoretical calculations using vacuum \Jpsi wave function without any color-screening, and including both radiative energy loss of color-octet $c\bar{c}$ pairs and collisional dissociation of J/$\psi$~\cite{Sharma:2012dy}. The two bands are corresponding to two different values of the \Jpsi formation time. Both of them are consistent with data. All these calculations include feeddown contribution constrained by the measurements in \pp collisions as well as CNM effects constrained by the measurements in $p(d)$+A collisions.

Figure~\ref{fig:Jpsi_vs_Npart_STAR} shows the centrality dependence of \Jpsi suppression in heavy-ion collisions at both RHIC and LHC energies for low-\pT (upper) and high-\pT (lower) J/$\psi$. For low-\pT J/$\psi$, the suppression decreases towards central collisions at RHIC but more flat at LHC. The low-\pT \Jpsi suppression is the interplay of all CNM effects, QGP melting and (re)combination. The much less suppression in central heavy-ion collisions at LHC than at RHIC is likely due to the different fraction of (re)combination at RHIC and LHC, expected from the different charm quark production cross section at these energies.  The high-\pT \Jpsi is suppressed by a factor of 3.1 with a significance of 8.1$\sigma$ in 0-10\% \auau collisions. The CNM effects and (re)combination contribution are expected to be minimal at this \pT range ($p_T>5$ GeV/$c$). The significant suppression of high-\pT \Jpsi in central \auau collisions provides strong evidence for the color-screening effect in QGP. Unlike the low-\pT J/$\psi$, the high-\pT \Jpsi is more suppressed at LHC than at RHIC. This could be because the temperature of the medium created at the LHC is higher than that at RHIC. 

\subsection{Collective flow}


The measurements on collective flow of \Jpsi may also shed light on the relative contribution of primordial \Jpsi and \Jpsi from (re)combination. The primordial \Jpsi is dominantly produced before the QGP formation thus do not have initial collective flow. In non-central collisions, the primordial \Jpsi may have different suppression along different azimuthal angle with respect to the reaction plane due to the different path lengths in azimuthal. But the azimuthal anisotropy should be limited. On the other hand, the \Jpsi produced from the (re)combination of charm quark and its antiquark should inherit the flow of charm quarks and may have considerable flow.


As discussed in Sec.~\ref{sec.IIA}, the $D^0$ $v_2$ shown in Fig.~\ref{fig:v2CompareWithData} is found to follow the mass ordering at low $p_T$, expected from hydrodynamics, and NCQ-scaling as the light and strange hadrons in the intermediated $p_T$, expected from quark coalescence. It is concluded that the charm quarks have gained significant flow in QGP.

The radial flow of $D^0$ meson in heavy-ion collisions is also studed with the precise measurements of  \pT or $m_T$ spectra in \auau collisions at \sNN = 200 GeV~\cite{D0_Spectra_STAR_PRC2019}, as discussed in Sec.~\ref{sec.IIA}. The $T_{\text{eff}}$, slope parameter of an exponential fit to the $m_T$ spectra, as a function of particle mass for light flavor hadrons, strange hadrons and $D^0$ mesons clearly show two different systematic trends. The light flavor hadrons $\pi$, $K$, and $p$ data follow one linear dependence while strange and charm hadrons $\phi$, $\Lambda$, $\Omega$ and $D^0$ data follow another linear dependence. The $T_{\textrm{fo}}$ and $\langle \beta_t \rangle$ of $D^0$, extracted by fitting the \pT spectra with blast-wave (BW)~\cite{BlastWave_PRC1993} or Tsallis blast-wave (TBW)~\cite{Tsallis09} model, is found to group with the multi-strangeness particles $\phi$, $\Xi$ and $\Omega$, showing much smaller  $\langle \beta_t \rangle$ and larger $T_{\textrm{fo}}$ compared to light hadrons $\pi$, $K$ and $p$. This suggests that the $D^0$ flows with the medium and the collectivity obtained is mostly through the partonic rescattering in the QGP phase. If \Jpsi produced from the (re)combination of charm quark and and its antiquark is the dominant process, it should have significant $v_2$ and radial flow.

\begin{figure}[!htb]
\includegraphics[width=0.88\hsize]
{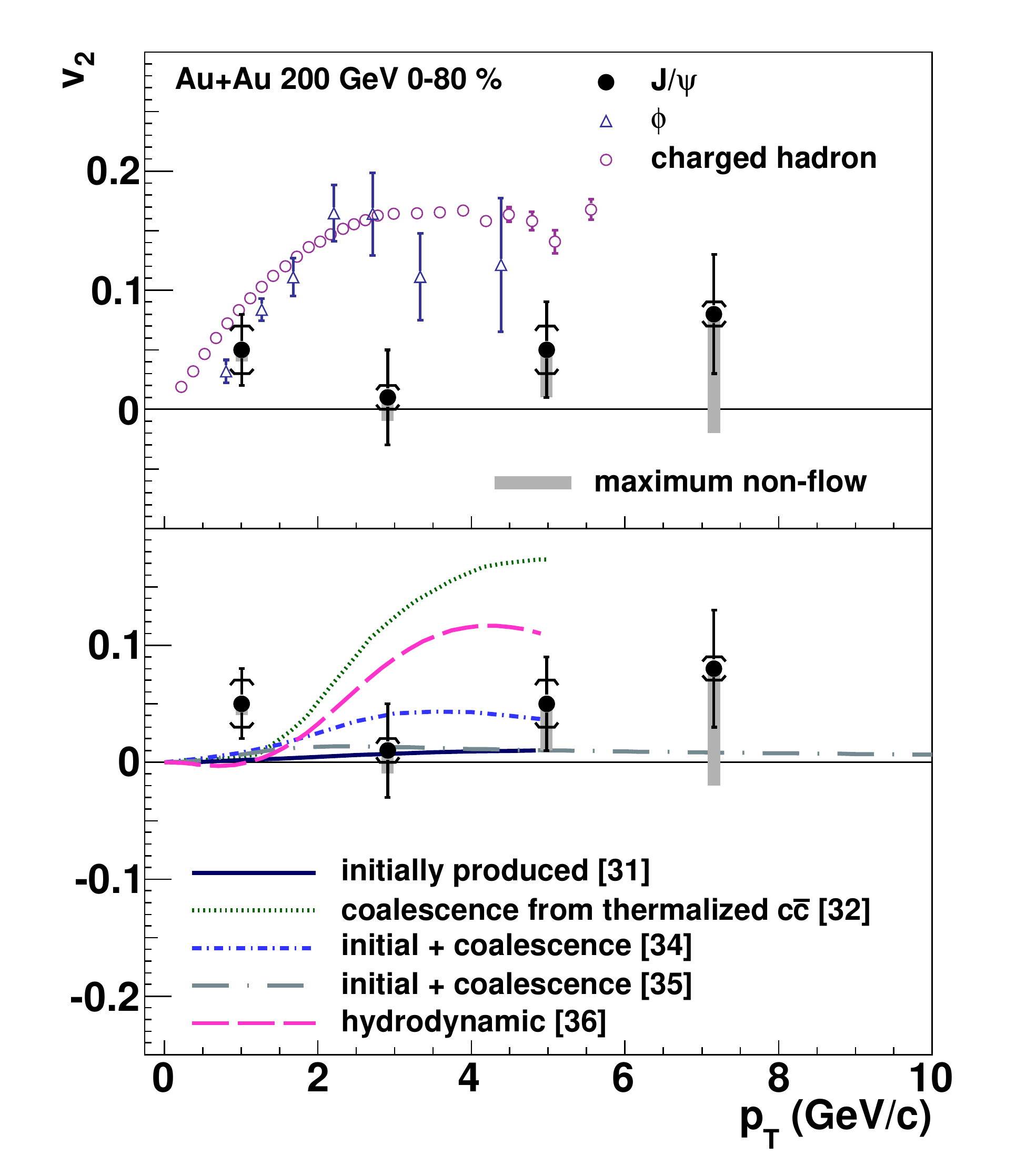}
\caption{ \Jpsi $v_2$ as a function of \pT in 0-80\% \auau collisions at \sNN = 200 GeV, compared to the results for $\phi$ and charged hadrons and theoretical calculations. The figure is taken from~\cite{Jpsi_v2_STAR}.}
\label{fig:Jpsi_v2}
\end{figure}

The STAR Collaboration measured \Jpsi $v_2$ in \auau collisions at \sNN = 200 GeV from the combination of various triggers operating in 2010. The inclusive J/$\psi$s are reconstructed through di-electron channel. The upper panel of Fig.~\ref{fig:Jpsi_v2} shows $v_2$ as a function of \pT for inclusive J/$\psi$, $\phi$ and charged hadrons which are dominated by $\pi$s. The grey box on the \Jpsi data shows the estimated maximum possible range of $v_2$ if the influence of non-flow is corrected. It is estimated using the measurement of \Jpsi-hadron correlation in \pp collisions at the same energy. Unlike the $D^0$, the \Jpsi $v_2$ is significantly lower than that for $\phi$ and charged hadrons at \pT above 2 GeV/$c$.

The lower panel of Fig.~\ref{fig:Jpsi_v2} shows the comparison of the \Jpsi $v_2$ data and theoretical calculations. The solid line shows the calculation for \Jpsi produced from the initial hard scattering. It is non-zero but limited in the \pT range of 0-5 GeV/$c$. Although significant suppression of \Jpsi is observed in \auau collisions at \sNN = 200 GeV, the azimuthally different suppression along the different path lengths in azimuth is limited, beyond the sensitivity of the current measurement. The dotted line shows the prediction for \Jpsi produced by coalescence of fully thermalized charm quarks at the freeze-out ((re)combination). The maximum is similar as that of light and strange hadrons, but shifted to higher \pT due to significant larger mass of charm quark and \Jpsi particle. The prediction for \Jpsi from coalescence is systematic higher than the data at \pT > 2 GeV/$c$. The $\chi^2$/ndf is as large as 16.2/3, corresponding to a small $p$-value of $1.0\times 10^{-3}$. The transport models including contribution from both primordial production and (re)combination predict a much smaller 
$v_2$ and are consistent with the data. The $p$-values is $0.58$ and $0.38$ for the TAMU~\cite{Zhao:2008vu} and Tsinghua~\cite{Liu:2009gx} transport model, respectively. The small $v_2$ in the transport model is due to the fact the $v_2$ of charm quark is small at low $p_T$ and the (re)combination contribution is small at high $p_T$. Although both transport models describe the data reasonably well, there is sizable difference among the models. The TAMU model is more close to the $v_2$ for initially produced J/$\psi$.  The measurement with improved precision will be helpful to distinguish or constrain the two transport models. The hydrodynamics model tuned to describe $v_2$ of light hadrons predicts a \Jpsi $v_2$ that strongly increase with \pT below 4 GeV/$c$ and fails to describe the data. The $\chi^2$/ndf is 7.0/3, corresponding to $p$-value of 0.072~\cite{UllrichHeinz}. Based on the data and model comparisons, it is concluded that the \Jpsi $v_2$ data disfavor the scenario that \Jpsi with \pT $>2$ GeV/$c$ are produced dominantly by coalescence from charm quarks and anti-charm quarks which are thermalized and flow with QGP. 

The \Jpsi radial flow at SPS and RHIC is systematically study in \cite{Tsallis11} with the Tsallis blast-wave (TBW) model. The \pT spectra of light hadrons and strange hadrons in \auau collisions at \sNN = 200 GeV at RHIC as well as in Pb+Pb collisions at \sNN = 17.3 at SPS are fit with the TBW model to extract the radial flow, kinetic freeze-out temperature and the non-extensive parameter. The \pT spectrum for \Jpsi predicted from the TBW with the same set of parameters as light and strange hardons is much softer than the measurement. It overestimates the yield at low \pT and underestimates the yield at high $p_T$. A fit to the \Jpsi \pT spectrum alone shows that the radial flow of \Jpsi at both RHIC and SPS is consistent with zero. This provides another evidence that \Jpsi production at RHIC and SPS is not dominantly from (re)combination of thermalized charm quarks. 

\subsection{J/$\psi$ photoproduction with nuclear overlap}
 J/$\psi$ can also be generated by the intense electromagnetic fields accompanied with the relativistic heavy ions~\cite{UPCreview}. The intense electromagnetic field can be viewed as a spectrum of equivalent photons by the equivalent photon approximation~\cite{KRAUSS1997503}. The quasi-real photon emitted by one nucleus could fluctuate into $c\bar{c}$ pair, scatters off the other nucleus, and emerge as a real J/$\psi$. The coherent nature of these interactions gives the processes distinctive characteristics: the final products consist of a J/$\psi$ with very low transverse momentum, two intact nuclei, and nothing else. Conventionally, these reactions are only visible and studied in Ultra-Peripheral Collisions (UPC), in which the impact parameter ($b$) is larger than twice the nuclear radius ($R_{A}$) to avoid any hadronic interactions. Several results of J/$\psi$ production in UPC are already available at RHIC~\cite{Afanasiev2009321} and LHC~\cite{20131273,Abbas2013,2017489}, which provide valuable insights into the gluon distribution in the colliding nuclei~\cite{Guzey:2013qza}.
 
 Can the coherent photon products also exist in Hadronic Heavy-Ion Collisions (HHIC, $b < 2R_{A}$), where the violent strong interactions occur in the overlap region? The story starts with the measurements from ALICE: significant excesses of J/$\psi$ yield at very low $p_{T} (< 0.3$ GeV/c) have been observed in peripheral Pb+Pb collisions at $\sqrt{s_{\rm{NN}}} =$ 2.76 TeV~\cite{LOW_ALICE}, which can not be explained by the hadronic J$/\psi$ production with the known cold and hot medium effects. STAR made the measurements of di-electron~\cite{PhysRevLett.121.132301} in Au+Au collisions at $\sqrt{s_{\rm{NN}}} =$ 200 GeV, and also observed significant enhancement at very low $p_{T}$ in peripheral collisions. The anomaly excesses observed possess characteristics of coherent photon interaction and can be quantitatively described by the theoretical calculations with coherent photon-nucleus~\cite{PhysRevC.93.044912, PhysRevC.97.044910, Jpsi_photo_Isobar, SHI2018399} and photon-photon~\cite{ZHA2018182,PhysRevC.97.054903,Klusek-Gawenda:2018zfz} production mechanisms, which points to evidence of coherent photon reactions in HHIC. The observed excesses may originate from coherent photon induced interaction, which impose great challenges for the existing models, e.g., how the broken nuclei satisfy the requirement of coherence. Measurements of J/$\psi$ production at very low $p_{T}$ at different collision energies, collision systems, and centralities can shed new light on the origin of the excess.
 
\begin{figure}[!htb]
 	\includegraphics[width=0.88\hsize]
 	{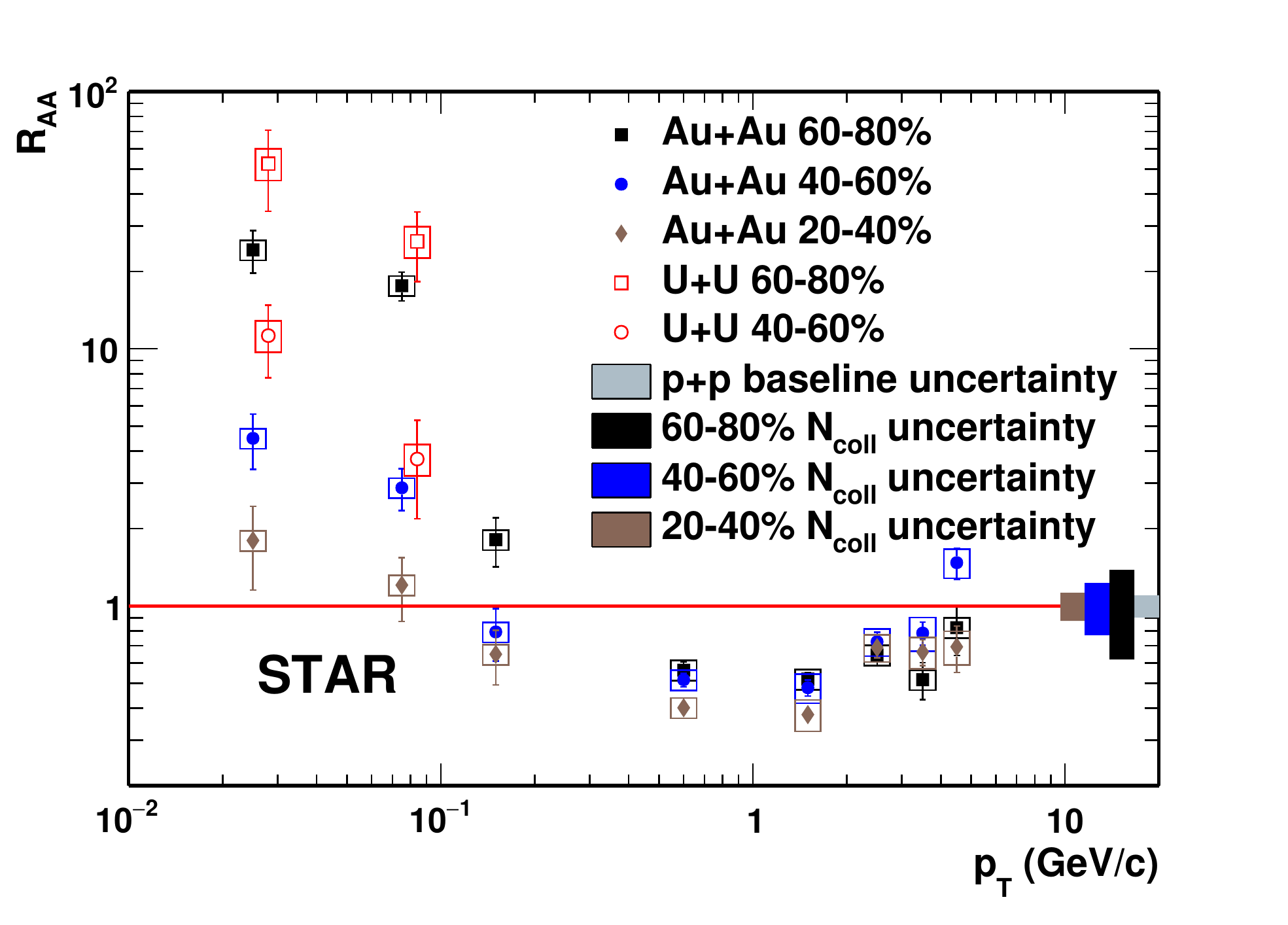}
 	\caption{ The J/$\psi$ $R_{\rm{AA}}$ as a function $p_{T}$ in Au+Au collisions at $\sqrt{s_{\rm{NN}}} =$ 200 GeV and U+U collisions at $\sqrt{s_{\rm{NN}}} =$ 193 GeV. The figure is taken from~\cite{PhysRevLett.123.132302}.}
 	\label{fig:JPSI_LOWPT_RAA}
 \end{figure}

 The STAR Collaboration measured J/$\psi$ production yields at very low $p_{T}$ in Au+Au collisions at $\sqrt{s_{\rm{NN}}} =$ 200 GeV and U+U collisions at $\sqrt{s_{\rm{NN}}} =$ 193 GeV at mid-rapidity via the di-electron decay channel. Figure~\ref{fig:JPSI_LOWPT_RAA} shows the J/$\psi$ \raa as a function of $p_{T}$ in Au+Au collisions and U+U collisions for different centrality classes. Suppression of J/$\psi$ production is observed for $p_{T} >$ 0.2 GeV/c in all collision centrality classes, which is consistent with the previous measurements~\cite{Jpsi_lowpT_CuCuAuAu_PRC2014, Jpsi_STARpp2009HT, Jpsi_STAR_BESI, PHENIX_Jpsi_AuAu} and can be well described by the transport models~\cite{XingboRalf2010, Zhuang2009} incorporating cold and hot medium effects. However, in the extremely low $p_{T}$ range, i.e., $p_T < 0.2$ GeV/c, a large enhancement of $R_{\rm{AA}}$ above unity is observed in peripheral collisions (40-80$\%$) both for Au+Au and U+U collisions. In this $p_{T}$ range, the color screening and CNM effects would suppress J/$\psi$ production, and the only gain effect, which is regeneration, is negligible in peripheral collisions~\cite{Zhuang2009}. The overall effect would lead to $R_{\rm{AA}} <$ 1 for hadronic production, which is far below the current measurement. For $p_T < 0.05$ GeV/c in the 60-80$\%$ centrality class, the $R_{\rm{AA}}$ is $24\pm5 (\rm{stat.}) \pm9 (\rm{syst.})$ for Au+Au collisions and $52\pm18 (\rm{stat.}) \pm16 (\rm{syst.})$ for U+U collisions, significantly deviating from the hadronic \pp reference with $N_{\rm{coll}}$ scaling, which strongly suggests an additional production mechanism.

\begin{figure}[!htb]
	\includegraphics[width=0.88\hsize]
	{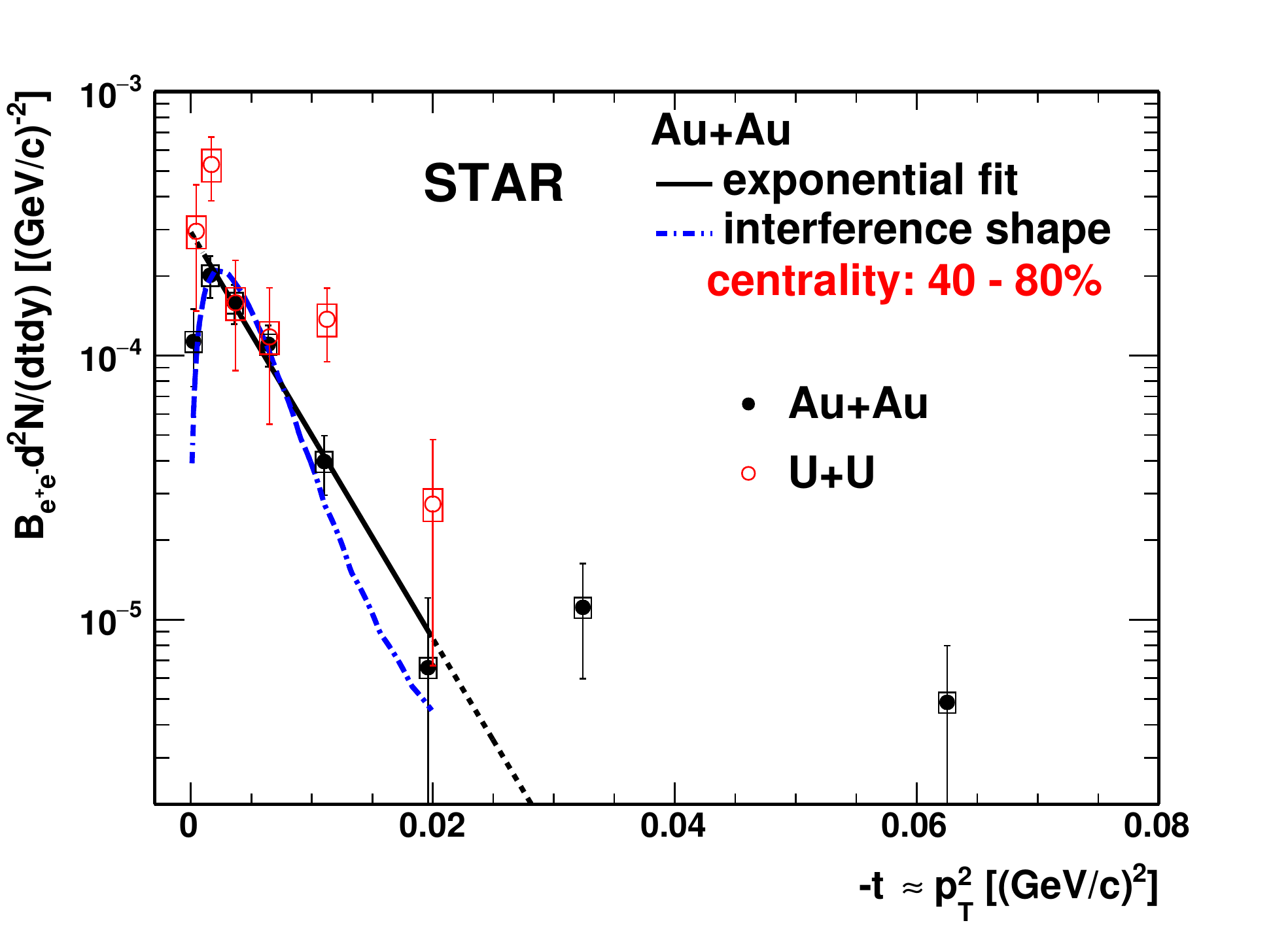}
	\caption{ The J/$\psi$ yield as a function of the negative momentum transfer squared $-t$ ( $-t \sim p_{T}^{2}$ ) for the 40-80$\%$ collision centrality class in Au+Au and U+U collisions. The figure is taken from~\cite{PhysRevLett.123.132302}.}
	\label{fig:JPSI_LOWPT_T}
\end{figure}

Assuming that the excess observed originates from coherent photoproduction, STAR also reported the differential cross section $d\sigma/dt$, where $t$ is the negative momentum transfer squared $-t \sim p_T^2$,  which reveals the distribution of interaction sites and is closely related to the parton distribution in the nucleus. Figure~\ref{fig:JPSI_LOWPT_T} shows the J/$\psi$ yield with the expected hadronic contribution subtracted as a function of $-t$ for the 40-80$\%$ centrality class in Au+Au and U+U collisions in the low $p_{T}$ range. The shape of the $dN/dt$ distribution is very similar to that observed in UPC for $\rho^{0}$ meson~\cite{PhysRevC.77.034910}. An exponential fit has been applied to the distribution in the  $-t$ range of $0.001-0.015~(\text{GeV}/c)^{-2}$ for Au+Au collisions. The slope parameter of this fit can be related to the position of the interaction sites within the target. The extracted slope parameter is $177 \pm 23~(\text{GeV}/c)^{-2}$, which is consistent with that expected for an Au nucleus (199 $(\text{GeV/c})^{-2}$)~\cite{UPC_PT} within uncertainties. As shown in the figure the data point at $-t < 0.001 ~(\text{GeV}/c)^{-2}$ is significantly lower ($3.0 \sigma$) than the extrapolation of the exponential fit. This suppression may be a hint of interference, which has been confirmed by STAR~\cite{PhysRevLett.102.112301} in the UPC case for $\rho^{0}$ meson. The theoretical calculation with interference from~\cite{PhysRevC.97.044910}, shown as the blue curve in the plot, can describe the Au+Au data reasonably well ($\chi^{2}$/ndf = 4.8/4) for $-t < 0.015~(\text{GeV}/c)^{-2}$. It should be aware that there also exists possible contribution from incoherent J/$\psi$ photoproduction. The fitting $-t$ range is chosen to ensure that the coherent production ($\langle -t \rangle \sim 0.005~(\text{GeV}/c)^{-2}$) is dominant over the incoherent production ($\langle -t \rangle \sim 0.250~(\text{GeV}/c)^{-2}$). Due to the different nuclear profile, the $-t$ distribution in U+U collisions is expected to be different from that in Au+Au collisions, however, as shown in the figure, the difference is not observed due to the large uncertainties.

\begin{figure}[!htb]
	\includegraphics[width=0.88\hsize]
	{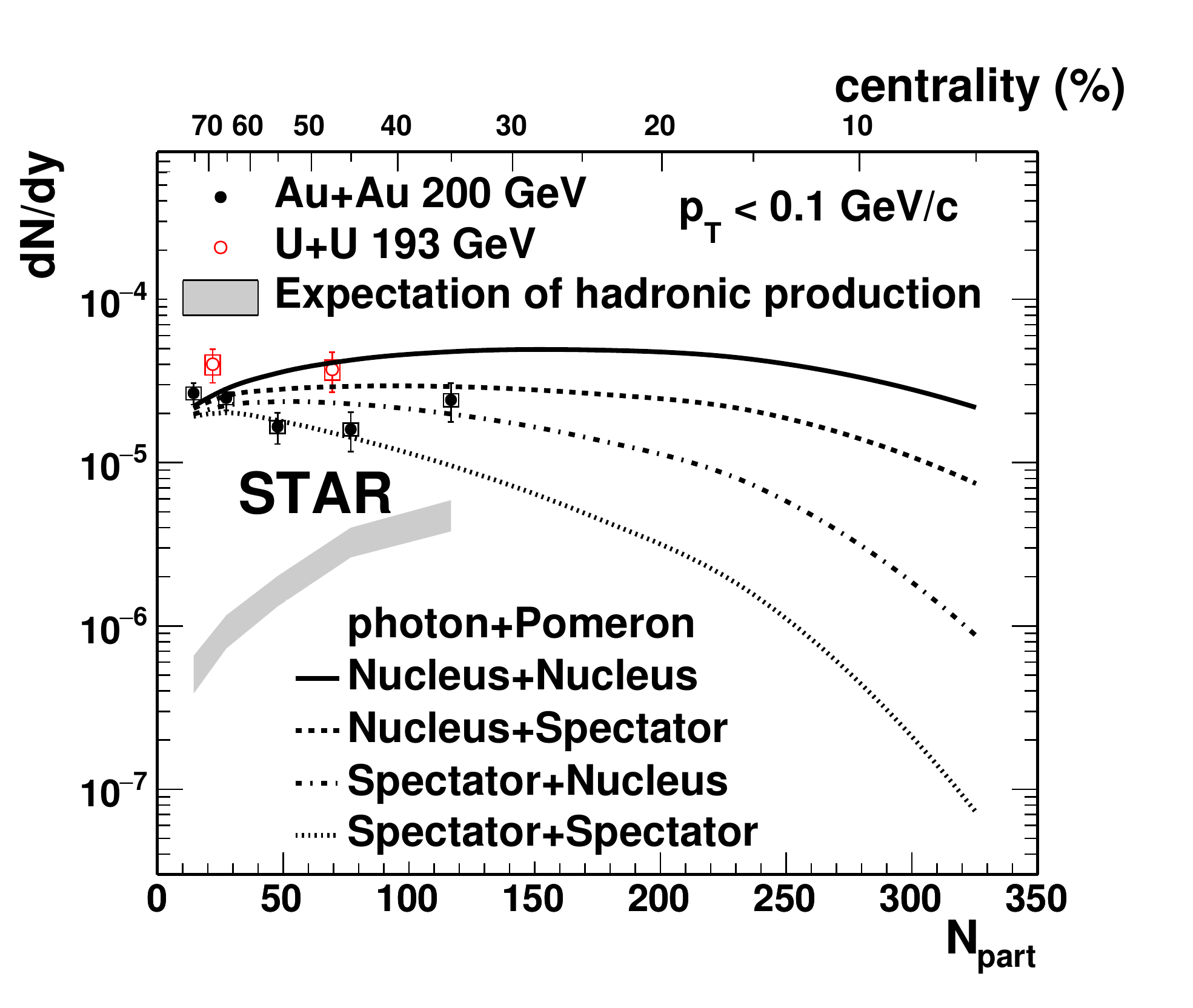}
	\caption{ The $p_{T}$-integrated J/$\psi$ yields ($p_{T} <$ 0.1~GeV/c) as a function of $N_{\rm{part}}$ for 30-80$\%$ Au+Au collisions and 40-80$\%$ U+U collisions. The figure is taken from~\cite{PhysRevLett.123.132302}.}
	\label{fig:JPSI_LOWPT_yield}
\end{figure}

Figure~\ref{fig:JPSI_LOWPT_yield} shows  $p_{T}$-integrated J/$\psi$ yields for $p_{T} <$ 0.1 GeV/$c$ with the expected hadronic contribution subtracted as a function of $N_{\rm{part}}$ for 30-80$\%$ Au+Au and 40-80$\%$ U+U collisions. The expected hadronic contributions in Au+Au collisions are also plotted for comparison. As depicted in the figure, the contribution from hadronic production is not dominant for the low-$p_{T}$ range in the measured centrality classes. Furthermore, the hadronic contribution increases dramatically towards central collisions, while the measured excess shows no sign of significant centrality dependence within uncertainties. With the assumption of coherent photoproduction, the excess in U+U collisions should be larger than that in Au+Au collisions. Indeed the central value of measurements in U+U collisions is larger than that in Au+Au collisions. However, limited by the current experimental precision, the observed difference (2.0$\sigma$) is not significant. The model calculations for Au+Au collisions with the coherent photoproduction assumption~\cite{PhysRevC.97.044910} are also plotted for comparison. In the model calculations, the authors consider either the whole nucleus or only the spectator nucleons as photon and Pomeron emitters, resulting in four configurations for photon emitter + Pomeron emitter: (1) Nucleus + Nucleus; (2) Nucleus + Spectator; (3) Spectator + Nucleus; (4) Spectator + Spectator. All four scenarios can describe the data points in the most peripheral centrality bins (60-80$\%$). However, in more central collisions, the Nucleus + Nucleus scenario significantly overestimates the data, which suggests that there may exist a partial disruption of the coherent production by the violent hadronic interactions in the overlapping region. The measurements in semi-central collisions seem to favor the Nucleus + Spectator or Spectator + Nucleus scenarios. The approach used in the model effectively incorporates the shadowing effect, which can describe the UPC results in the $x$-range probed by the RHIC measurement. However, the coherently produced J$/\psi$ could be modified by hot medium effects, e.g. QGP meltingg, which is not included in the model. More precise measurements toward central collisions and advanced modeling with hot medium effects included are essential to distinguish the different scenarios.

\section{$\Upsilon$ production in medium}

The $b\bar{b}$ cross section is much smaller than that of $c\bar{c}$ at both RHIC and LHC. Based on an FONLL calculation~\cite{FONLL1, FONLL2, FONLL_online}, the number of $b\bar{b}$ pairs per event is estimated to be less than 0.1 in 0-10\% \auau collisions at \sNN = 200 GeV and $2.3 \pm 0.4$ ($4.9 \pm 0.9$) in 0-10\% Pb+Pb collisions at \sNN = 2.76 (5.5) TeV. And due to the large masses of $b$ quarks, it is difficult to reach the thermalization. Thus the contribution of (re)combination is negligible at RHIC for $\Upsilon$. In terms of CNM effects, Ref.~\cite{Y_Rapp_PRC2017} points out that the dissociation of $\Upsilon(1S)$ by comovers is much smaller than that of \Jpsi and can be neglected at RHIC. Thus, compared to J/$\psi$, $\Upsilon$ provides a cleaner probe of QGP melting at least at RHIC. 

The binding energy of radius of $\Upsilon(1S)$, $\Upsilon(2S)$ and $\Upsilon(3S)$ are quite different. Since the dissociation temperature depends on the radius of the quarkonium states. The measurements on the suppression of various $\Upsilon$ states can be used to study the properties of the color screening and the QGP. 

\subsection{$\Upsilon$ production in $p$+Au collisions}

\begin{figure}[!htb]
\includegraphics[width=0.85\hsize]
{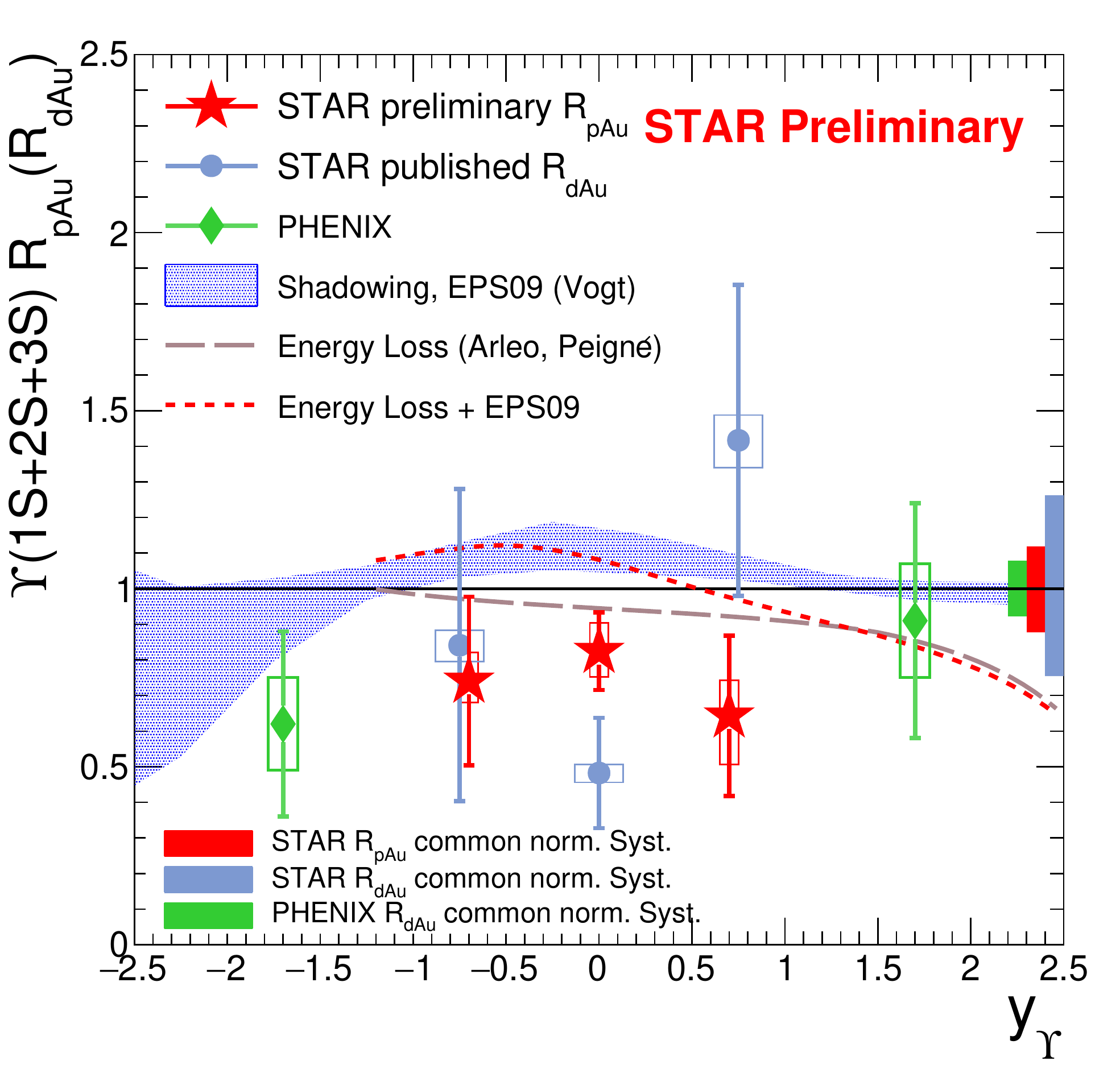}
\caption{ $\Upsilon$ suppression as a function of rapidity in $p(d)$+Au collisions at \sNN = 200 GeV, compared to theoretical calculations. The figure is taken from~\cite{ZhaochenYe_QM17Proceedings}.}
\label{fig:Y_pAudAu}
\end{figure}

The CNM effects on $\Upsilon$ production can be studied in $p$+A or $d$+A collisions. Figure~\ref{fig:Y_pAudAu} shows the suppression of $\Upsilon(1S+2S+3S)$ as a function of rapidity in $p$+Au and $d$+Au collisions at \sNN = 200 GeV. The STAR results are measured at mid-rapidity through di-electron channel~\cite{Upsilon_STAR_dAuAuAu2014, ZhaochenYe_QM17Proceedings} and the PHENIX results are measured at the forward/backward rapidity through di-muon channel~\cite{Upsilon_PHENIX_pp_dAu}. The data point at $|y|<0.5$ in $p$+Au collisions has the highest precision. It is $R_{pAu} = 0.82 \pm 0.10~\textrm{(stat.)} ^{+0.08}_{-0.07}~\textrm{(syst.)} \pm 0.10~\textrm{(global)}$, indicating suppression of $\Upsilon$ due to CNM effects. The shaded area shown in the figure represents the calculation from CEM with the EPS09 nuclear parton distribution function~\cite{Frawley:2008kk}. It predicts enhancement at $y \sim 0$, mainly due to the anti-shadowing effect. The long dashed line depicts the calculation including parton energy loss only~\cite{Upsilon_EPS09andEnergyLoss}. It is more closer to the data. The dashed line shows the calculation including both EPS09 nPDF and parton energy loss~\cite{Upsilon_EPS09andEnergyLoss}. It is closer to the calculation with EPS09 alone at mid-rapidity, while is closer to the calculation with parton energy loss only at forward rapidity. The data is systematically lower than the calculations, particularly at mid-rapidity. This suggests that other CNM effects than the nPDF effect alone are needed to describe the data.

\subsection{$\Upsilon$ production in \auau collisions}

\begin{figure}[!htb]
\includegraphics[width=0.9\hsize]
{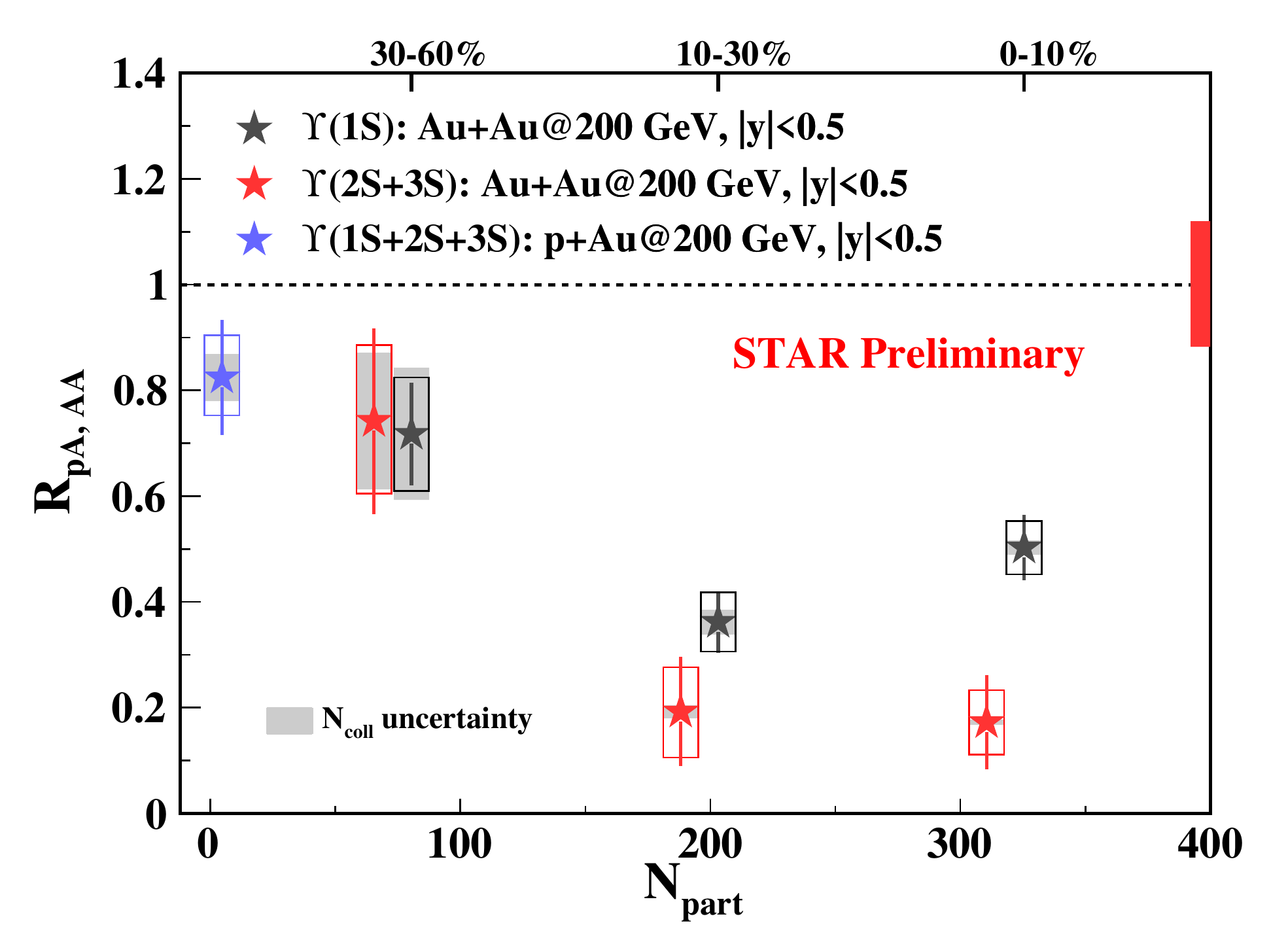}
\caption{\raa as a function of \npart for $\Upsilon(1S)$ and $\Upsilon(2S+3S)$ in \auau collisions at \sNN = 200 GeV. The figure is taken from~\cite{PengfeiWang_QM2018Proceedings}.}
\label{fig:Y_STAR_Only}
\end{figure}

The $\Upsilon$ measurements at RHIC is very challenging due to the small production cross section. The STAR Collaboration collected large data samples in 2014 and 2016 for $\Upsilon$ study in \auau collisions using di-muon trigger utilizing the MTD detector completely installed in early 2014. Thanks to the large statistics and good momentum resolution in \auau collisions, the separation of $\Upsilon(1S)$ and $\Upsilon(2S+3S)$ from the invariant mass spectrum of di-muon is possible. Figure~\ref{fig:Y_STAR_Only} shows \raa as a function of \npart for $\Upsilon(1S)$ and $\Upsilon(2S+3S)$ in \auau collisions at \sNN = 200 GeV. The suppression of $\Upsilon(1S+2S+3S)$ in $p$+Au collisions at \sNN = 200 GeV is also shown for comparison. The suppression for both $\Upsilon(1S)$ and $\Upsilon(2S+3S)$ increases towards central collisions. In central collisions, significant suppression for both $\Upsilon(1S)$ and $\Upsilon(2S+3S)$ is observed. The observed suppression for inclusive $\Upsilon(1S)$ could be due to the suppression of the feeddown contribution of higher bottomonium states. The fraction of direct $\Upsilon(1S)$ in the inclusive $\Upsilon(1S)$ is estimated to be $(71 \pm 5)\%$ at low-\pT and $(45.5 \pm 8.5)\%$ at high $p_T$ in \pp collisions~\cite{quarkonium_feeddown}. It is not clear whether the direct $\Upsilon(1S)$ is suppressed with current precision. The suppression for $\Upsilon(2S+3S)$ is found to be larger than $\Upsilon(1S)$ in (semi-)central collisions, consisting with the ``sequential'' suppression picture.

\subsection{Comparison between RHIC and LHC}

\begin{figure}[!htb]
\includegraphics[width=0.85\hsize]
{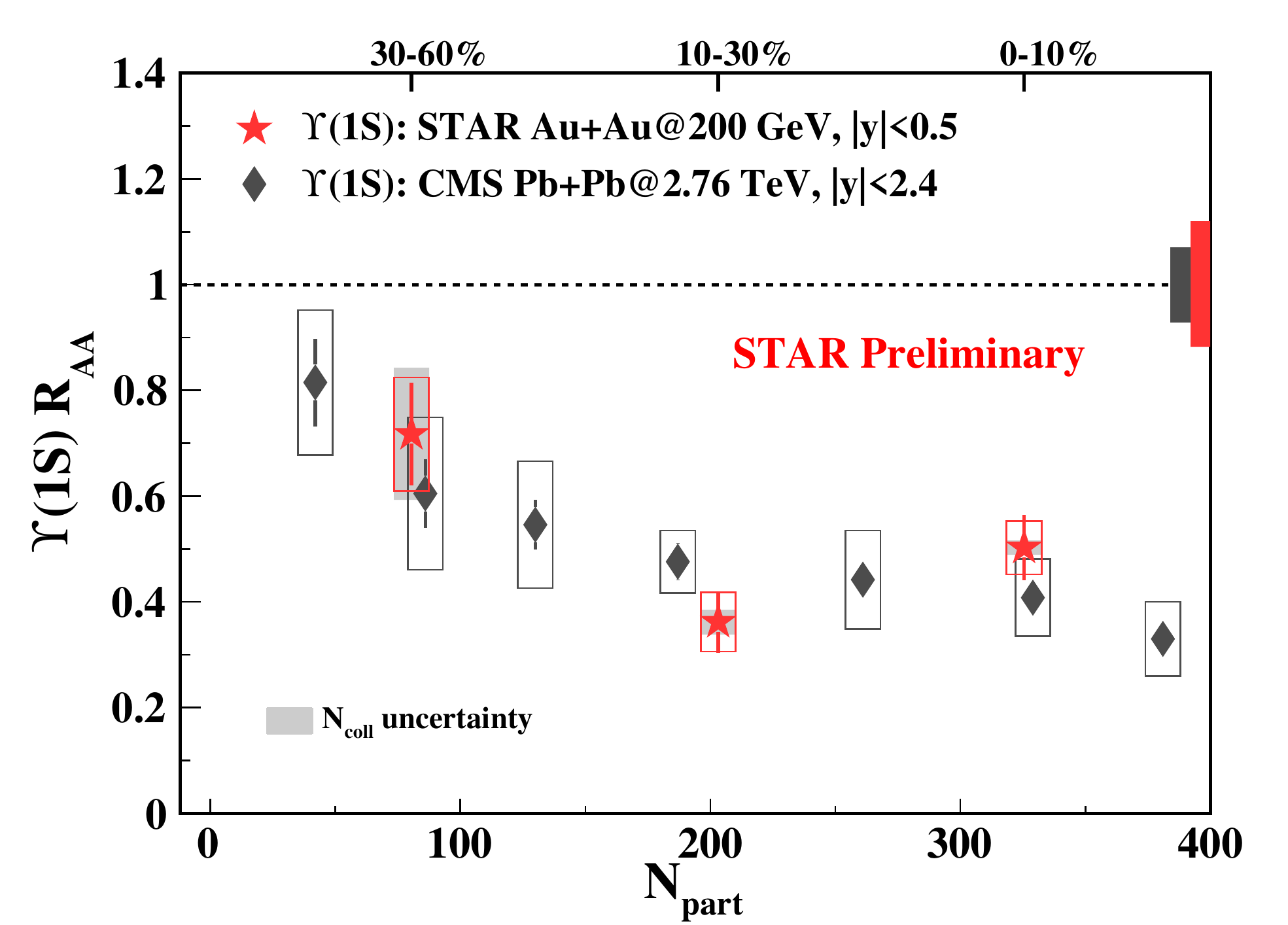}
\includegraphics[width=0.85\hsize]
{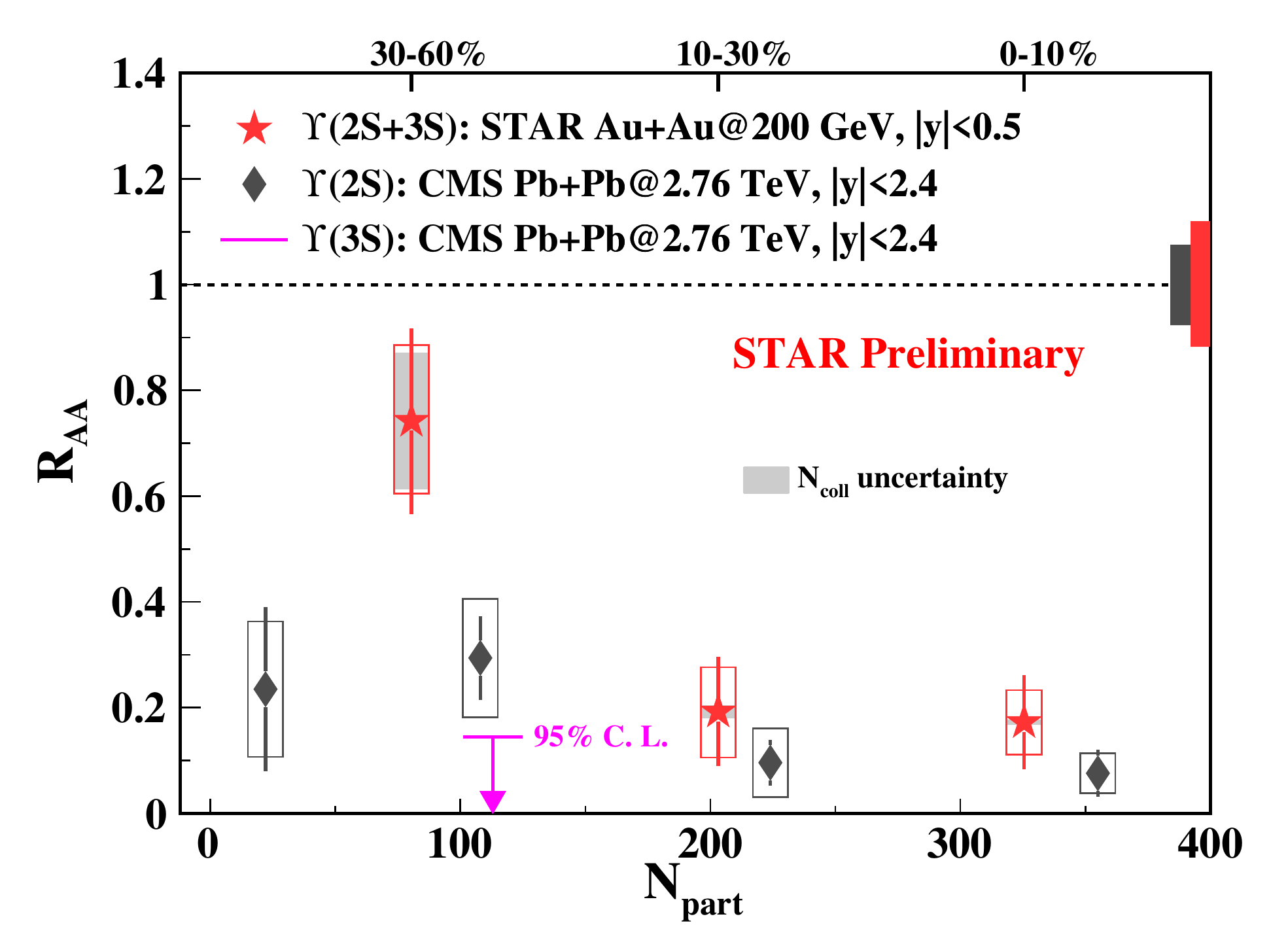}
\caption{ \raa as a function of \npart for  $\Upsilon(2S+3S)$ in \auau collisions at \sNN = 200 GeV and for $\Upsilon(2S)$ and  $\Upsilon(3S)$ in Pb+Pb collisions at \sNN = 2.76 TeV. The figure is taken from~\cite{PengfeiWang_QM2018Proceedings}.}
\label{fig:Y_STAR_vs_CMS}
\end{figure}

The results at RHIC are compared to the results at LHC in Pb+Pb collisions at \sNN = 2.76 TeV measured by the CMS Collaboration~\cite{CMS_Upsilon_PLB2017}. Both measurements are done at mid-rapidity through the di-muon channel. 

The upper panel of Fig.~\ref{fig:Y_STAR_vs_CMS} is for $\Upsilon(1S)$. The suppression for $\Upsilon(1S)$ at RHIC and CMS is similar from peripheral to central heavy-ion collisions although the center-of-mass energies differ by one order of magnitude. It is plausible that the inclusive $\Upsilon(1S)$ suppression arises mainly from the CNM effects and the suppression of the feeddown from excited bottomonium states while the direct $\Upsilon(1S)$ remains unaffected by the deconfined medium at both RHIC and LHC.

The lower panel of Fig.~\ref{fig:Y_STAR_vs_CMS} shows the comparison of \raa as a function of \npart for $\Upsilon(2S+3S)$ in \auau collisions at \sNN = 200 GeV and for $\Upsilon(2S)$ and  $\Upsilon(3S)$ in Pb+Pb collisions at \sNN = 2.76 TeV. The $\Upsilon(2S+3S)$ seems less suppressed at RHIC than at LHC, especially in the peripheral collisions. This could be due to the different temperature profile of the medium produced in the heavy-ion collisions at RHIC and LHC. It is expected that the initial temperature is higher in central collisions and at higher center-of-mass energy. If the initial temperature reached in the heavy-ion collisions at LHC and RHIC are both well above the dissociation temperature of $\Upsilon(2S)$ and  $\Upsilon(3S)$, no significant difference is expected at RHIC and LHC. However, if the initial temperature is close to the dissociation temperature of $\Upsilon(2S)$, the suppression of inclusive $\Upsilon(2S)$ will be sensitive to the temperature profile of the medium and could result in the different behavior at RHIC and LHC. 

\subsection{Comparison between experiment and theory}

\begin{figure}[!htb]
\includegraphics[width=0.85\hsize]
{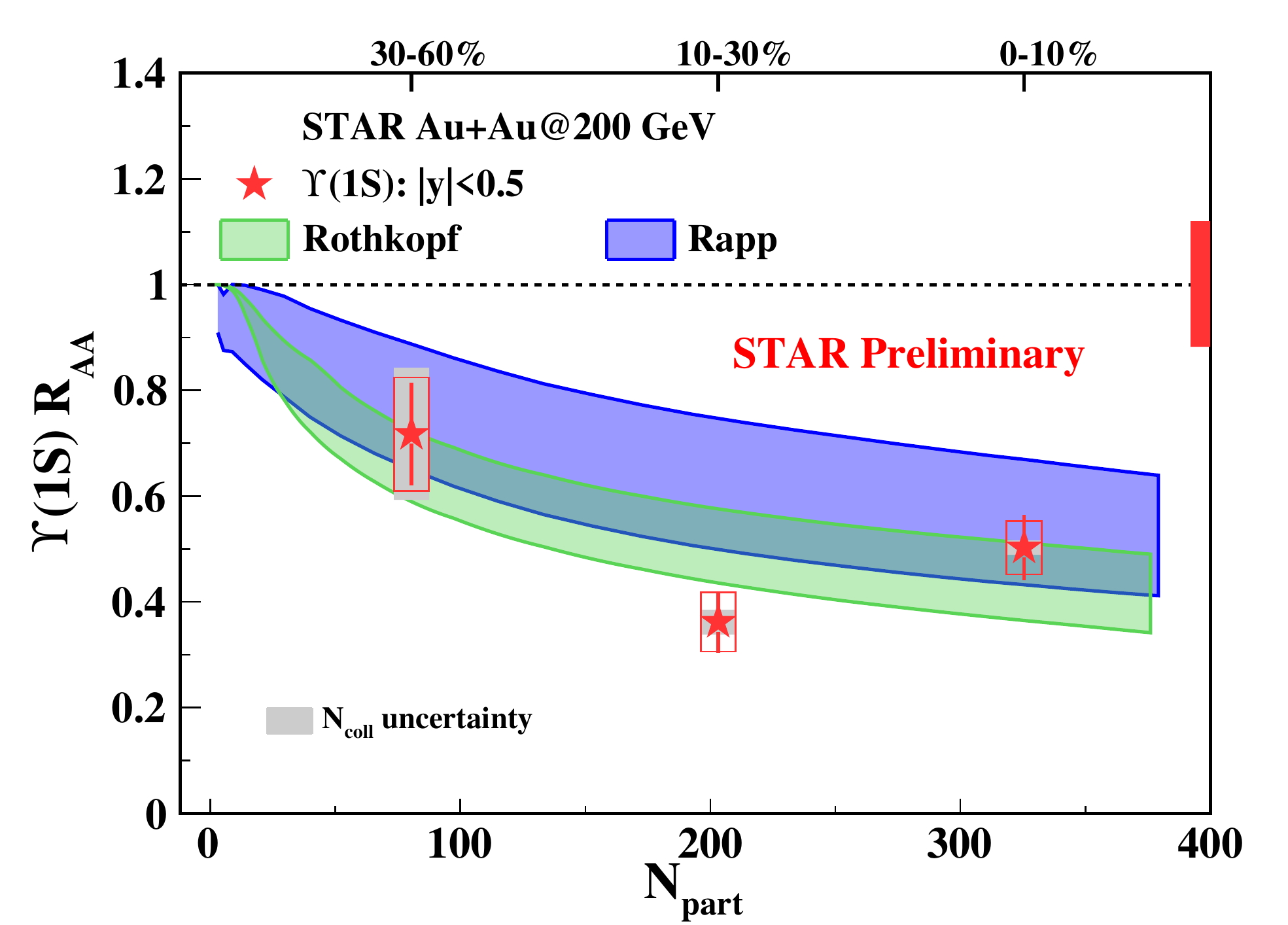}
\includegraphics[width=0.85\hsize]
{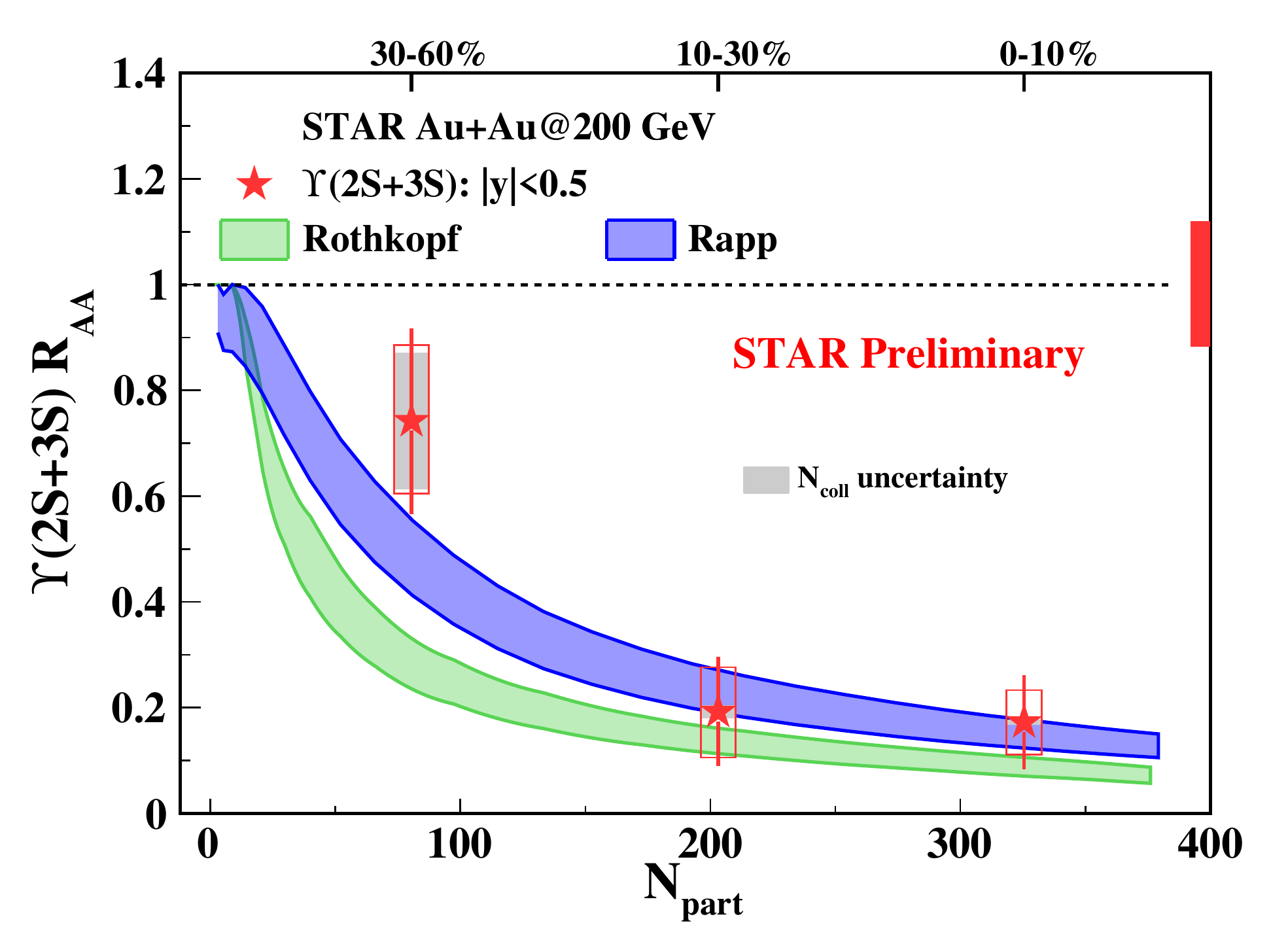}
\caption{ \Jpsi $v_2$ as a function of \pT in 0-80\% \auau collisions at \sNN = 200 GeV, compared to the results for $\phi$ and charged hadrons and theoretical calculatioßns. The figure is taken from~\cite{Jpsi_v2_STAR}.}
\label{fig:Y_STAR_vs_Theory}
\end{figure}

For better understanding of the $\Upsilon$ production and constraining the temperature of the medium produced in heavy-ion collision at RHIC, the $\Upsilon$ suppression data is compared to two theoretical calculations. In the TAMU transport model (Rapp)~\cite{Y_Rapp_PRC2017}, the QGP melting and (re)combination of the $\Upsilon$ mesons are controlled by a kinetic-rate equation. The binding energies in the medium are predicted by thermodynamic microscopic T-matrix calculations using the internal energy from lattice QCD as potential. The space-time evolution of the fireball is dictated by a lattice-QCD-based equation of state. The initial temperature of the fireball is about 310 MeV in the most central \auau collisions. The CNM effects are also considered in this calculation. The model by Rothkopf and his collaborators~\cite{Krouppa:2017jlg} uses a lattice QCD vetted complex value heavy-quark potential coupled with a QGP background following anisotropic hydrodynamic evolution. The initial temperature used is about 440 MeV in the most central collisions. No (re)combination and CNM effects are included in Rothkopf calculation. 

Figure~\ref{fig:Y_STAR_vs_Theory} shows the comparison of the STAR measurements on $\Upsilon(1S)$ and $\Upsilon(2S+3S)$ and the corresponding theoretical calculations from the two models mentioned above. Both model calculations are consistent with the data for the ground and excited $\Upsilon$ states within experimental and theoretical uncertainties. The precision of the data and the theoretical calculations are to be improved in order to extract the temperature reached in the heavy-ion collisions via the systematic study of quarkonium suppression.

\section{Summary}
This paper presents a review of recent experiment measurements of open heavy flavor and quarkonium production at RHIC. Heavy quarks, due to their large masses, are expected to behave different from light flavors when interacting with the nuclear matter created in high energy heavy-ion collisions, such as the production mechanisms, hadronization, thermalization, interactions with medium and so on. Taking advantage of the silicon vertex detector technology development, precise measurement is achieved and will provide better constraints on theoretical calculations. In the following, we summarize few key points for open heavy flavor and quarkonium production.

{\it Open heavy flavor production}: Open charm hadrons $p_{T}$  spectra, including $D^0$, $D_s$ and $\Lambda_c$ in various centrality bins at midrapidity $|y|<1$ in \sNN = 200 GeV Au+Au collisions are presented. Thermal parameters, $T_{eff}$, $T_{kin}$ and radial flow velocity extracted from $D^0$ $p_{T}$  spectra show that $D^0$ freezes out early than light hadrons. The strong suppression of $D^0$ nuclear modification factor $R_{AA}$ at high $p_{T}$  and large elliptic flow $v_2$ follow the similar level of light flavor hadrons, indicating strong interactions between charm and medium and charm quark may be thermalized as light quarks. The enhancement of $\Lambda_c/D^0$ and $D_{s}/D^{0}$ ratios in Au+Au collisions compared to that in $p$+$p$ collisions provide important data for understanding the charm quark hadronization mechanisms. The comparison to various models with charm-quark coalescence suggest that coalescence mechanism plays an important role in charm-quark hadronization in the presence of QGP. 

The open bottom hadrons are indirectly measured via their decay products, $B{\rightarrow}J/\psi$, $B{\rightarrow}D^0$ and $b{\rightarrow}e$, by the STAR experiment at mid-rapidity in $\sqrt{s_{\rm NN}} = $ 200 GeV Au+Au collisions. The suppression observed for non-prompt $J/\psi$ and non-prompt $D^0$/e at high $p_{T}$  indicates bottom-medium interactions and bottom energy loss. The less suppression of $b{\rightarrow}e$ compared with $c{\rightarrow}e$ and no suppression of $B{\rightarrow}D$ at 4 GeV/c indicating less energy loss of bottom quark due to its extremely large mass and consistent with flavor dependent parton energy loss mechanisms. 

{\it Quarkonium production}: The production of \Jpsi in \pp and heavy-ion collisions are intensively studied at RHIC. In \pp collisions, the measured \pT spectra and polarization (spin alignment) is found to be consistent with the theoretical calculations from (I)CEM and NRQCD. The CEM calcualtion and NRQCD calculations with different Long Distance Matrix Elements predict different polarization for J/$\psi$, but the current data can not tell the difference. The measurement of \Jpsi polarization with high statistics will be helpful to constraint models or LDMEs in NRQCD.

The collision energy, collision size and \pT dependence of \Jpsi production in heavy-ion collisions are measured at RHIC and compared to SPS and LHC. At low $p_T$, it is consistent with the picture that QGP melting, (re)combination and CNM effects play important roles. Their relative contribution varies with the collision energy, collision centrality, system size and the kinematic variables of J/$\psi$. Based on a transport model calculation, low-\pT \Jpsi suppression in central \auau collisions is about 0.6, QGP melting further suppresses it to 0.2, but (re)combination enhances it back to about 0.4. At high $p_T$, where the contribution from CNM effects and (re)combination is negligible, significant suppression is observed, providing strong evidence of QGP melting at RHIC. 

The collectivity of J/$\psi$ is studied in \auau collisions via radial and elliptic flow. The results disfavor that \Jpsi at RHIC is dominantly produced via (re)combination of thermalized charm quarks. 

$\Upsilon$, a cleaner probe of QGP meting, is found to be suppressed significantly in central \auau collisions. Sequential suppression (stronger suppression of $\Upsilon(3S)$ and $\Upsilon(2S)$ than $\Upsilon(1S)$) observed at LHC is confirmed at RHIC, which provided another strong evidence of QGP melting. The comparisons of the RHIC data to LHC data and theoretical calculations are useful to extract the properties of QGP. 

In around 2023, sPHENIX will start high luminosity runs with high-speed silicon vertex detector (MVTX), which is based on a state-of-art monolithic active pixel sensors (MAPS) technology. At least 100 times more statistics will be collected aiming for dedicated bottom measurement via hadronic decay channels. Such as precision measurement of nuclear modification factors and flows for B-mesons and b-tagged jets~\cite{Roland:2019cwl}. 

With the high statistics Zr+Zr and Ru+Ru (3B events for each collision system) data taken in 2018, the \Jpsi \raa and elliptic flow will be measured with good precision to deepen our understanding of the interplay of QGP melting, (re)combination and CNM effects on \Jpsi production. The $Z$ dependence of \Jpsi photoproduction can be also studied with the isobaric collision data and the comparison to Au+Au collisions. The Cu+Au collision data taken by STAR in 2012 is recently been fully produced, \Jpsi production via di-electron decay channel will be measured. The STAR forward upgrade program, including finished inner Time Projection Chamber, Endcap Time-of-Flight upgrades and ongoing Forward Tracking System and Forward Calorimeter System upgrades will extend the rapidity coverage of quarkonium measurement at STAR upto $y=4$. Many unique physics opportunities with quarkonium in \pp, $p(d)$+A and A+A collisions at very forward rapidity will be enabled. With sPHENIX in high luminosity A+A runs, the precision of $\Upsilon$ measurement are expected to be significantly improved.

These facility upgrades will deepen our understanding of the interaction of heavy flavor and quarkonium with the hot dense medium created in heavy-ion collisions at RHIC.

\bibliographystyle{elsarticle-num}
\bibliography{HF_Onium_review.bib}




\end{document}